\documentclass[a4paper,preprintnumbers,showpacs,onecolumn,superscriptaddress,nofootinbib,amsmath,amssymb,notitlepage]{revtex4-1}

\usepackage{enumitem}
\usepackage{times}
\usepackage{dcolumn}
\usepackage{mathrsfs}
\usepackage{comment}
\usepackage{hyperref}
\usepackage{amsmath}
\usepackage{mathtools}
\usepackage{bbm}
\usepackage{bm}
\usepackage{amsfonts}
\usepackage{amssymb}
\usepackage{mathrsfs}
\usepackage{wasysym}
\usepackage{graphicx}
\usepackage[]{subfigure}
\usepackage{filecontents}
\usepackage[dvipsnames]{xcolor}

\hypersetup{
    bookmarks=true,         
    pdftoolbar=true,        
    pdfmenubar=true,        
    pdffitwindow=true,     
    pdfstartview={FitH},     
    pdftitle={My title},     
    pdfauthor={author},      
    pdfsubject={Subject},    
    pdfcreator={Creator},    
    pdfproducer={Producer},  
    pdfkeywords={keyword1} 
    pdfnewwindow=true,       
    colorlinks=true ,        
    linkcolor=Blue  ,        
    citecolor=Blue,      
    filecolor=green,       
    urlcolor=Blue            
}

\newcommand{\arcsinh}{\operatorname{arcsinh}}

\newcommand{\sech}{\operatorname{sech}}

\newcommand{\ed}{\mathrm{d}}

\newcommand{\tL}{\tilde{\Lambda}}

\newcommand{\emit}{\mathrm{emit}}
\newcommand{\obs}{\mathrm{obs}}
\newcommand{\prop}{\mathrm{prop}}
\newcommand{\e}{\mathrm{e}}
\newcommand{\oo}{\mathrm{o}}
\newcommand{\f}{\mathrm{f}}

\begin{document}

\title{Observational signatures of a static $f(R)$ black hole with thin accretion disk}

\author{Mohsen Fathi }
\email{mohsen.fathi@usach.cl}
\affiliation{Departamento de F\'{i}sica, Universidad de Santiago de Chile,
Avenida V\'{i}ctor Jara 3493, Estaci\'{o}n Central, 9170124, Santiago, Chile}

\author{Norman Cruz}
\email{norman.cruz@usach.cl}
\affiliation{Departamento de F\'{i}sica, Universidad de Santiago de Chile,
Avenida V\'{i}ctor Jara 3493, Estaci\'{o}n Central, 9170124, Santiago, Chile}


\begin{abstract}
{In this study, we focus on a static spherically symmetric $f(R)$ black hole spacetime characterized by a linear dark matter-related parameter. Our investigation delves into understanding the influence of different assumed values of this parameter on the observable characteristics of the black hole. To fulfill this task, we investigate the light deflection angles, which are inferred from direct analytical calculations of null geodesics.} To examine the black hole's properties further, we assume an optically thin accretion disk and explore various emission profiles. Additionally, we investigate the shadow cast by the illuminated black hole when affected by the disk. Furthermore, we simulate the brightness of an infalling spherical accretion in the context of silhouette imaging for the black hole. Our findings indicate that, except for some specific cases, the observed brightness of the accretion disk predominantly arises from direct emission, rather than lensing and photon rings. Moreover, we reveal that the linear dark parameter of the black hole significantly influences the shadow size and brightness. Our discussion covers both analytical and numerical approaches, and we utilize ray-tracing methods to produce accurate visualizations.

 \bigskip

{\noindent{\textit{keywords}}: Black holes, $f(R)$ gravity, thin accretion, shadow}\\

\noindent{PACS numbers}: 04.20.Fy, 04.20.Jb, 04.25.-g   
\end{abstract}

\maketitle


\section{Introduction and Motivation}\label{sec:intro}

Since the foundational concepts of black holes were established by Schwarzschild \cite{Schwarzschild:1916} and Finkelstein \cite{Finkelstein:1958}, the search for identifying these enigmatic objects has been on an exhilarating journey. From the initial observational evidence gathered for Cygnus X-1 in 1971 \cite{webster_cygnus_1972, bolton_identification_1972} to the recent shadow images of M87* \cite{Akiyama:2019} and Sgr A* \cite{Akiyama:2022} captured by the Event Horizon Telescope (EHT), the pursuit of understanding black holes has relentlessly advanced. By comparing theoretical predictions with observed shadows, valuable insights into the behavior of light in extremely gravitating systems can be obtained. The EHT results also unveiled the presence of a magnetic field around M87*, possibly connected to the formation of jets emanating from the black hole \cite{2021ApJ...910L..12E,2021ApJ...910L..13E,2021PhRvD.103j4047K}. Shadow images can provide information about the geometric structure near the black hole's horizon \cite{PhysRevD.97.084020} and its physical characteristics \cite{Takahashi_2004}. However, despite these compelling evidences that support the general theory of relativity, there are limitations in the cosmological context where general relativity falls short. These limitations include issues with flat galactic rotation curves, anti-lensing, the accelerated expansion of the universe, observed anisotropies in the cosmic microwave background radiation, and the coincidence problem \cite{Rubin1980, Massey2010, Bolejko2013, Riess:1998, Perlmutter:1999, Astier:2012}.

Many scientists believe that the phenomena described above are attributed to the dark side of the universe, which remains inadequately explained. For instance, introducing the cosmological constant term to Einstein's field equations as a nonzero vacuum energy can account for the acceleration of the universe. However, the reason behind the small value of the cosmological constant remains elusive. To address unresolved cosmological problems, such as the late-time acceleration of the universe, some researchers turn to modified theories of gravity to mimic the effects of dark matter and dark energy and provide an effective time-varying equation of state. In these models, the Einstein-Hilbert action is generalized or extended to explain the dynamics of the universe on cosmic, galactic, or astrophysical scales. One of the most intuitive extensions of general relativity is the $f(R)$ theory, where the Einstein-Hilbert action is replaced with a generic function $f(R)$ \cite{RevModPhys.82.451, de_felice_fr_2010, CAPOZZIELLO2011167}. As a result, $f(R)$ theories of gravity have garnered significant interest and have been thoroughly scrutinized for their consistency \cite{Navarro_2007, NOJIRI2007238, JE-AN:2011, PhysRevD.78.063504, Capozziello_2008, cognola_class_2008, nojiri_modified_2008} (see also the reviews in Refs.~\cite{nojiri_unified_2011, nojiri_modified_2017}). Within the context of $f(R)$ gravity, researchers are also interested in black hole solutions. One primary solution derived from the Palatini formalism of $f(R)$ theory is the Schwarzschild-(anti-)de Sitter metric, which includes an effective cosmological constant. However, this solution faces incompatibilities with primary general relativistic tests since the cosmological constant appears to have no significant role in solar system scales \cite{KAGRAMANOVA2006465}. Nonetheless, this problem can be circumvented by appropriately manipulating the action, rendering the cosmological constant effective on cosmological scales while negligible on solar system scales \cite{starobinsky_disappearing_2007, PhysRevD.76.064004}. To address these issues properly, a suitable $f(R)$ action model has been proposed in Refs. \cite{refId0, refId0_1, RAHVAR:2008}, which is consistent with both galactic and cosmological scales. In Ref. \cite{Saffari:2008}, this model was further elaborated using a generic function in the gravitational action to ensure its compatibility with solar system tests, galactic rotation curves, and the late-time acceleration of the universe. In this context, the authors also propose a static spherically symmetric black hole solution, which is of interest in this paper, particularly concerning light propagation in its geometry and its shadow.

Indeed, the theoretical constraining of black hole shadows based on observational data has been a subject of special interest to scientists, leading to numerous publications dedicated to this area (see, for example, Refs.~\cite{Abdujabbarov:2012bn, Yumoto:2012kz, Atamurotov:2013sca, Zakharov:2014lqa, Papnoi:2014aaa, Johannsen:2015hib, Johannsen:2015mdd, Moffat:2015kva, Giddings:2016btb, Cunha:2016bjh, Tsukamoto:2017fxq, Hennigar:2018hza, Cunha:2018acu, Allahyari:2019jqz, Abdikamalov:2019ztb, Ovgun:2019jdo, Shaikh:2019fpu, bambi_testing_2019, vagnozzi_hunting_2019, Kumar:2020hgm, Li2020, Ovgun:2020gjz, khodadi_black_2020, Zhong:2021mty, Zuluaga:2021vjc, Stashko:2021lad, Rahaman:2021kge, Ovgun:2021ttv, Pantig:2022whj, Bisnovatyi-Kogan:2022ujt, Kazempour:2022asl, Pantig:2022ely, roy_superradiance_2022, chen_superradiant_2022, vagnozzi_horizon-scale_2022,omwoyo_remarks_2022}). However, the recent silhouette imaging by the EHT has added even greater importance to the need for reliable methods to visualize black holes with accretion disks as their illumination sources. This interest was initially ignited by Luminet in 1979 when he calculated the radiation emitted from a thin accretion disk surrounding a Schwarzschild black hole and proposed a ray-traced image of the disk \cite{Luminet:1979nyg}. Generally, this type of accretion is based on the Shakura-Sunyaev \cite{Shakura:1973}, Novikov-Thorne \cite{Novikov:1973}, and Page-Thorne \cite{page_disk-accretion_1974} models, where the disk is assumed to be thin, geometrically and optically. These assumptions, along with the growing interest in black hole imaging, led to the development of a new method for simulating higher-order light rings for black holes with thin accretion disks, proposed in Ref. \cite{Gralla:2019}. Since then, this method has been applied in various publications (e.g., Refs. \cite{Guerrero_2021, li_observational_2021, okyay_nonlinear_2022, PhysRevD.105.084057, hu_observational_2022, PhysRevD.106.044070, Guo_2022, WANG2022116026, PhysRevD.105.064031, uniyal_probing_2023, uniyal_nonlinearly_2023}), and it also plays a significant role in our paper.

We structure our discussion as follows: In Sec. \ref{sec:f(R)_theBH}, we provide a concise overview of the $f(R)$ black hole solution and introduce its cosmological parameters. Moving on to Sec. \ref{sec:lightprop}, we initiate our investigation by studying the causal structure of the spacetime. We then apply a Lagrangian formalism to derive the equations of motion for massless particles (light rays). By calculating the critical impact parameter of photon trajectories, we determine the points at which the orbits become unstable. {As a result, we are able to calculate the theoretical size of the black hole shadow, which directly correlates with the critical impact parameter.} In the same section, we proceed to find the turning points of light ray trajectories as they approach the black hole. We obtain exact analytical solutions for the angular equation of motion for deflecting and critical trajectories. In fact, gravitational lensing serves as a remarkable tool in examining black hole solutions in the strong-field regime \cite{Virbhadra:1998, Virbhadra:2000, Bozza:2001, Virbhadra:2002, Bozza:2002, Hasse:2002, Virbhadra:2008, Virbhadra:2009, He:2020, Adler:2022, Virbhadra:2022}. Additionally, weak lensing is of great importance to astrophysicists and cosmologists, allowing them to estimate matter distribution profiles within galaxies and other observable regions of the universe \cite{Hoekstra:2003, Sheldon:2004, Mandelbaum:2006, Gavazzi:2007, Parker:2007}, thus providing insights into dark matter and dark energy properties. Hence, we use the above analytical solutions as instruments in finding the lens equation and the deflection angles, which we calculate analytically using the Weierstrassian elliptic function. In Sec. \ref{sec:thinModel}, we construct a thin accretion disk for the black hole based on the Novikov-Thorne model. We calculate the dynamical characteristics of accreting particles in stable orbits and derive the radial profiles of the disk's radiation flux and temperature. Applying the method from Ref. \cite{Gralla:2019}, we visualize the light rings and accretion disk of the $f(R)$ black hole for three different disk emission profiles. Additionally, we calculate the thickness of the rings, which is inferred from the observed effective intensity profiles. Finally, we assume a spherically symmetric infalling accretion for the black hole and calculate the observed disk emission, concluding this section by simulating the black hole shadow under this condition. Our paper concludes with Sec. \ref{sec:conclusion}, where we summarize our findings and discuss their implications. Throughout this work, we use the signature convention $(- + +\,+)$, and primes on functions denote differentiation with respect to the radial coordinate. We apply the geometrized unit system, where $G=c=1$.

\section{A particular model of $f(R)$ gravity and its black hole solution}\label{sec:f(R)_theBH}

The gravitational action of the theory can be written in its most generic form as
\begin{equation}
\mathcal{S} = \frac{1}{2\kappa}\int\ed x^4\sqrt{-g} ~f(R)+\mathcal{S}_m, 
    \label{eq:action}
\end{equation}
in which, $\kappa$ is a coupling constant, $f(R)$ is a function of the Ricci scalar of the spacetime with the metric determinant $g$, and $\mathcal{S}_m$ is a matter field action. Accordingly, the field equations are derived as
\begin{equation}
F(R) R_{\mu\nu}-\frac{1}{2}g_{\mu\nu} f(R)-\left(\nabla_\mu\nabla_\nu-g_{\mu\nu}\square\right)F(R)=\kappa T_{\mu\nu},
    \label{eq:field_f(R)}
\end{equation}
by varying the action $\mathcal{S}$ with respect to the metric, where $F(R)=\frac{\ed f(R)}{\ed R}$, $\square=\nabla_\lambda\nabla^\lambda$ and $T_{\mu\nu}$ is the energy-momentum tensor. Adopting a generic spherically symmetric line element 
\begin{equation}
\ed s^2 = -B(r)\ed t^2 + A(r) \ed r^2+r^2\left(\ed\theta^2+\sin^2\theta\ed\phi^2\right),
    \label{eq:metric_0}
\end{equation}
in the usual Schwarzschild coordinates $x^\mu=(t,r,\theta,\phi)$, the field equation \eqref{eq:field_f(R)} can be recast as \cite{Saffari:2008}
\begin{eqnarray}
&& 2 F(r)\frac{X'(r)}{X(r)}+rF'(r)\frac{X'(r)}{X(r)}-2rF''(r)=0,\label{eq:field_f(R)_X1}\\
&& B''(r)+\left[\frac{F'(r)}{F(r)}-\frac{1}{2}\frac{X'(r)}{X(r)}\right]B'(r)-\frac{2}{r}\left[\frac{F'(r)}{F(r)}-\frac{1}{2}\frac{X'(r)}{X(r)}\right]B(r)-\frac{2}{r^2}B(r)+\frac{2}{r^2}X(r)=0,
    \label{eq:field_f(R)_X2}
\end{eqnarray}
for the vacuum space (i.e. $T_{\mu\nu}=0$), in which $X(r)=B(r) A(r)$, and based on the $r$-dependence of the Ricci scalar, we have taken into account $F=F(R)\equiv F(r)$. Obviously, general relativity is recovered for $F(R)=1$, for which Eq.~\eqref{eq:field_f(R)_X1} results in $X(r)=1$ (or $B(r)=A(r)^{-1}$), and from Eq.~\eqref{eq:field_f(R)_X2}, the Schwarzschild solution is obtained. We adopt the ansatz $F(r)=(1+r/d)^{-\alpha}$ \cite{Saffari:2008}, where $\alpha$ and $d$ are the free parameters of the action. Here $\alpha$ is dimensionless whereas $[d] = \mathrm{m}$ and is of galactic size. Applying this ansatz, Eq. \eqref{eq:field_f(R)_X1} results in the solution
\begin{equation}
X(r) = X_0 \left(1+\frac{r}{d}\right)^{-2(1+\alpha)}\left(1+\frac{2-\alpha}{2}\frac{r}{d}\right)^{{4(1+\alpha)}/{(2-\alpha)}},
    \label{eq:X(r)_gen}
\end{equation}
where $X_0$ represents an integration constant. When $\alpha=0$, we obtain $X_0=1$, which corresponds to the Schwarzschild solution. Solving Eq. \eqref{eq:field_f(R)_X2} allows us to determine $B(r)$. 
By substituting the expression from Eq.~\eqref{eq:X(r)_gen} into Eq.~\eqref{eq:field_f(R)_X2}, and considering terms up to the first-order of the free parameters, the lapse function can be obtained as \cite{Saffari:2008}
\begin{equation}
    B(r) = 1-\frac{2M}{r}+\beta r-\frac{1}{3}\Lambda r^2,
    \label{eq:lapse_0}
\end{equation}
in which $\beta = {\alpha}/{d}$. Here, $\Lambda$ represents the cosmological constant having the value $|\Lambda|\leq10^{-52}~\mathrm{m}^{-2}$ \cite{Carmeli:2001}\footnote{Unless otherwise specified, this value of $\Lambda$ is assumed throughout the subsequent sections of the paper.}. It is worth noting that the above expression resembles the Mannheim-Kazanas vacuum solution in fourth-order Weyl conformal gravity \cite{mannheim_exact_1989}. In that solution, the linear term $\beta r$ acts as an additional potential that compensates for the flat galactic rotation curves. Similarly, the model presented in Eq. \eqref{eq:lapse_0} suggests that small values of $\alpha$ can yield flat galactic rotation curves for galaxies in a similar manner. The same solution has been further explored in various aspects, including investigations on the quasinormal modes \cite{Aragon:2020xtm, Malekmakan:2022kfd}, dynamics of the accretion disk \cite{Soroushfar:2020kgb}, thermodynamic geometry \cite{Upadhyay:2018bqy}, and particle collision in the black hole's exterior \cite{Majeed:2017txa}. {The lapse function \eqref{eq:lapse_0} can be seen as a linear combination of solutions from two distinct $f(R)$ gravity models, each pertinent to a different length scale. In the solar system scale ($r\ll d$), the model is derived as \cite{Saffari:2008}
\begin{equation}
    f(R) = R+R_0\ln\frac{R}{R_c},
    \label{eq:reducedModel}
\end{equation}
where $R_0={6\alpha^2}/{d^2}$, and $R_c$ is an integration constant. The solution to the above model covers the first three terms in the lapse function \eqref{eq:lapse_0}. On the other hand, in the galactic scales ($r>d$), the $f(R)$ theory is based on the model \cite{Saffari:2008}
\begin{equation}
f(R) = R_c^{-{\alpha}/{2}}\left(R+\Lambda\right)^{1+{\alpha}/{2}}.
    \label{eq:reducedModel_1}
\end{equation}
For cosmological scales ($\alpha\ll1$), this model reduces to $f(R)=R+\Lambda$, which is equivalent to the Einstein-Hilbert action with a cosmological constant. Therefore, in the context of the exact solution, this model contributes the final term in the lapse function \eqref{eq:lapse_0}. It is worth noting that in Ref. \cite{Saffari:2008}, the generic model
\begin{equation}
f(R) = R+\Lambda+\frac{R+\Lambda}{{R}/{R_0}+{2}/{\alpha}}\ln\frac{R+\Lambda}{R_c},
\label{eq:model_f(R)}
\end{equation}
has been proposed, which aligns with both of the above models under different scale criteria. In the regime of strong curvature, characterized by $R\gg\Lambda$ and ${R}/{R_0}\gg {2}/{\alpha}$, we obtain the model \eqref{eq:reducedModel}. Conversely, in cosmological scales, where $R\simeq R_0\simeq\Lambda$ and $\alpha\ll1$, it simplifies to $f(R)=R+\Lambda$.
} Consequently, the small-valued parameter $\beta$ in Eq. \eqref{eq:lapse_0} can effectively capture both large and small-scale phenomena in the universe.
{In the following section, we will assign presumed values to this model parameter within a favorable range. These values will serve as the basis for our subsequent analysis and simulations in the discussion.} 

We begin by studying the black hole exterior geometry and its causal structure. For the sake of convenience in the calculations and demonstrations, we adimensionalize the parameters by introducing the quantities
\begin{equation}
\tilde{r}\rightarrow \frac{r}{M},\quad \tilde{\beta}\rightarrow\beta M,\quad \tL\rightarrow \frac{1}{3}\Lambda M^2.
  \label{eq:adimensionalize}
\end{equation}
In the forthcoming sections, however, we remove the "tilde" overscript from the dimensionless parameters in Eq. \eqref{eq:adimensionalize}, which is equivalent to letting $M=1$.

\section{Propagation of light and unstable photon orbits}\label{sec:lightprop}

The spacetime's causal structure, characterized by the lapse function \eqref{eq:lapse_0}, can be understood by studying the hypersurfaces where $B(r) = 0$, corresponding to the black hole horizons, resulting in a cubic equation with the solutions
\begin{eqnarray}
&& r_{1} =\frac{\beta}{3\Lambda}-\frac{4}{\Lambda}\sqrt{\frac{g_2}{3}}\cos\left(
\frac{1}{3}\arccos\left(
\frac{3g_3}{g_2}\sqrt{\frac{3}{g_2}}\right)-\frac{4\pi}{3}
\right),\label{eq:r1} \\
&& r_2 =\frac{\beta}{3\Lambda}-\frac{4}{\Lambda}\sqrt{\frac{g_2}{3}}\cos\left(
\frac{1}{3}\arccos\left(
\frac{3g_3}{g_2}\sqrt{\frac{3}{g_2}}\right)-\frac{2\pi}{3}
\right),\label{eq:r2}\\
&& r_{3} =\frac{\beta}{3\Lambda}-\frac{4}{\Lambda}\sqrt{\frac{g_2}{3}}\cos\left(
\frac{1}{3}\arccos\left(
\frac{3g_3}{g_2}\sqrt{\frac{3}{g_2}}\right)
\right),\label{eq:r3}
\end{eqnarray}
where
\begin{subequations}
\begin{align}
    & g_2 = \frac{1}{12}\left(\beta^2+3\Lambda\right),\label{eq:g2}\\
    & g_3 = -\frac{1}{16}\left(
    \frac{2\beta^3}{27}+\frac{\beta\Lambda}{3}-2\Lambda^2\right).\label{eq:g3}
\end{align}
\end{subequations}
The existence of real values for the above radii, however, depends on the sign of the polynomial's discriminant, i.e. $\Delta = g_2^3-27 g_3^2 =\frac{\Lambda^2}{256}[\beta^2(1+8\beta)+4(1+9\beta)-108\Lambda^2]$, which is always of positive values for $\beta, \Lambda\ll1$. So, we infer that all the solutions \eqref{eq:r1}--\eqref{eq:r3} are real-valued. It is straightforward to verify that $r_1>r_2>0$ and $r_3<0$. Hence we identify $r_{++}=r_1$, as the cosmological horizon of the black hole where the infinite blueshift happens, and $r_+=r_2$, as its event horizon, where the infinite redshift happens.  This way, the lapse function can be rewritten as
\begin{equation}
    B(r) = \frac{\Lambda}{r}\left(r_{++}-r\right)\left(r-r_+\right)\left(r-r_3\right).
   \label{eq:lapse_1}
\end{equation}

\subsection{Motion of mass-less particles}\label{subsec:massless_motion}

\begin{figure}[t]
    \centering
    \includegraphics[width=7cm]{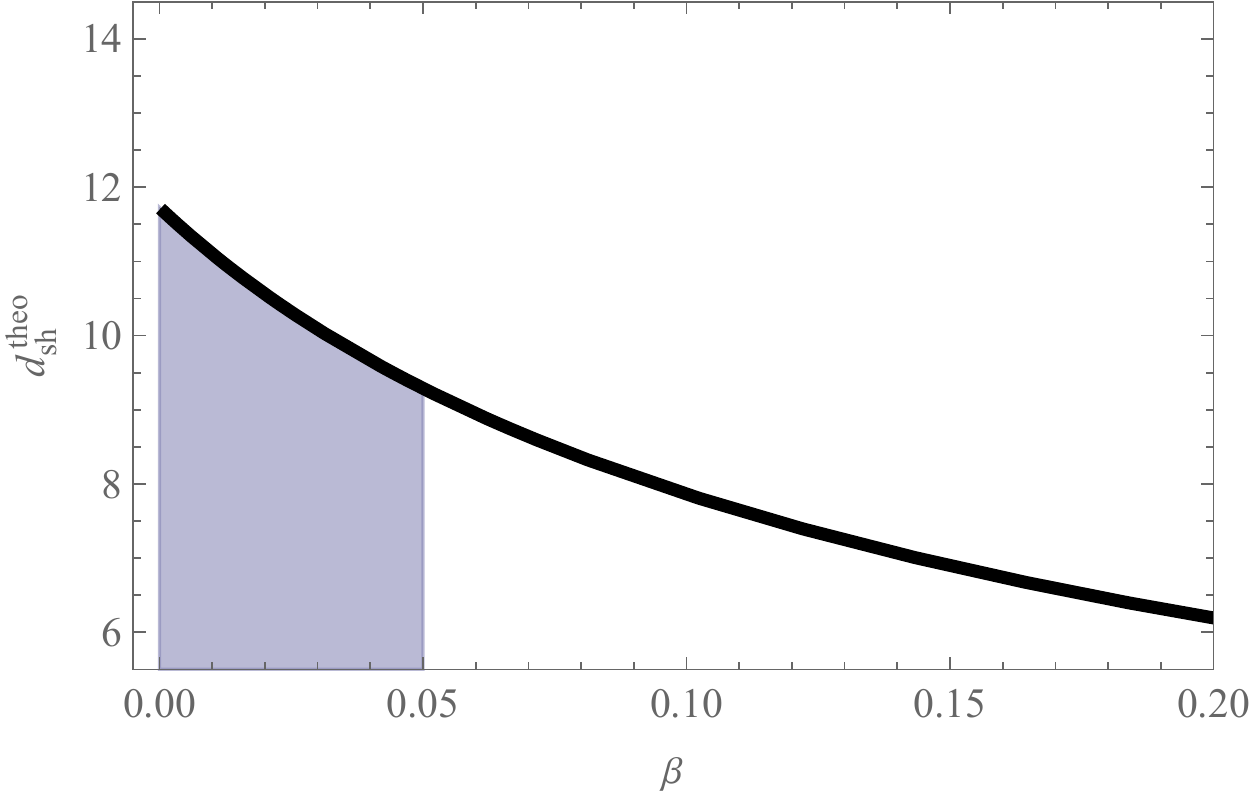}~(a)\qquad
    \includegraphics[width=7cm]{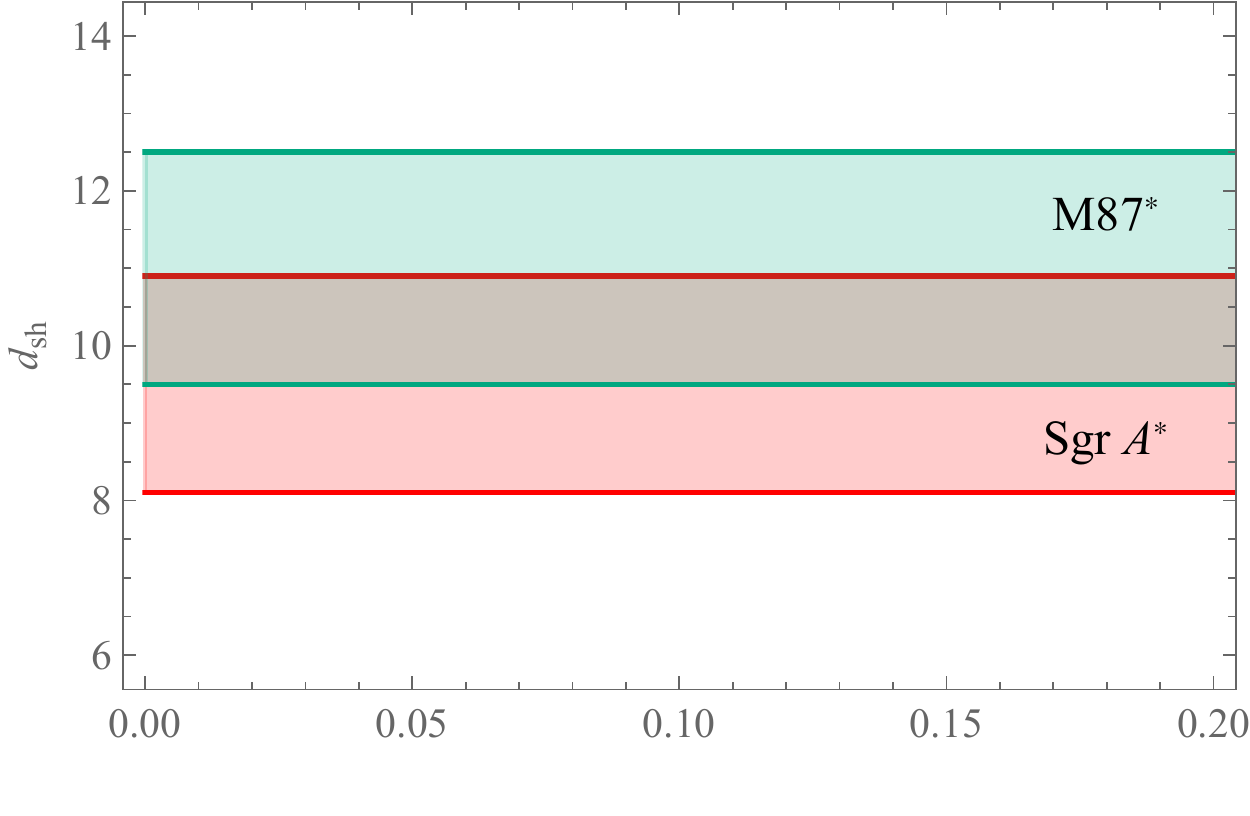}~(b)
    \caption{{The diagrams show (a) the $\beta$-profile of $d_{\mathrm{sh}}^{\mathrm{theo}}$ and (b) the observed shadow diameters $d_{\mathrm{sh}}$ for M87* (green region) and Sgr A* (red region), each with $1\sigma$ uncertainties. The shaded region in panel (a) represents the assumed range for the $\beta$-parameter, limiting it within the domain $0 \leq \beta < 0.05$.}}
    \label{fig:betaConfinement}
\end{figure}

The motion of test particles can be described by the Lagrangian
\begin{eqnarray}
    2\mathscr{L}&=&g_{\mu\nu}\dot{x}^\mu\dot{x}^\nu\nonumber\\
    &=&-B(r)\dot t^2+\frac{\dot{r}^2}{B(r)}+r^2\dot{\theta}^2+r^2\sin^2\theta\dot{\phi}^2, 
    \label{eq:lagrangian}
\end{eqnarray}
{where "dot" stands for differentiation with respect to the affine parameter $\tau$ of the geodesic curves.} Enjoying the spherical symmetry of the spacetime, we confine the motion of particles to the equatorial plane (i.e. $\theta={\pi}/{2}$) without loss of generality. One can then define the conjugate momenta
\begin{equation}
\Pi_\mu=\frac{\partial\mathscr{L}}{\partial{\dot{x}^\mu}},
    \label{eq:conj_moment}
\end{equation}
which provides the two constants of motion
\begin{subequations}
\begin{align}
    & \Pi_t=-B(r)\dot{t}=-E,\label{eq:E}\\
    & \Pi_\phi=r^2\dot{\phi}=L,\label{eq:L}
\end{align}
\end{subequations}
in accordance with the Killing symmetries of the spacetime, and we name them, respectively, as the energy and the angular momentum of the test particles. These two quantities allow us to define the impact parameter $b\equiv {L}/{E}$. This parameter corresponds to the vertical distance between the tangent to the geodesic curves and the line passing the black hole singularity and is of importance in the identification of possible photon trajectories. In fact, the motion of photons can be described by the equation $\mathscr{L}=0$ which characterizes the null geodesics. Thus, by means of Eq. \eqref{eq:lagrangian}, the equations of motion are obtained as
\begin{eqnarray}
    && \dot{r}^2 = E^2-V(r),\label{eq:dotr}\\
   &&  \left(\frac{\ed r}{\ed\phi}\right)^2 = \frac{r^4}{b^2}\left[1-\frac{V(r)}{E^2}\right],\label{eq:drdphi}
\end{eqnarray}
in which 
\begin{equation}
V(r) = L^2\frac{B(r)}{r^2},
    \label{eq:V(r)}
\end{equation}
represents the effective potential for photons. Then, the turning points $r_t$ in the orbits correspond to $\dot r=0$, which is encountered when $V(r_t)=E^2$. This potential has a maximum at
\begin{equation}
r_p=\frac{\beta_0-1}{\beta},
    \label{eq:rp_0}
\end{equation}
given that $\beta_0=\sqrt{1+6\beta}$, which is where the photon orbits become unstable. Hence, this maximum corresponds to the radius of the photon sphere. As it is observed, this radius is independent of $\Lambda$, and decreases as $\beta$ increases. One can also verify that 
\begin{equation}
\lim_{\beta\rightarrow0}r_p=3,
    \label{eq:rp-Schw}
\end{equation}
which is the radius of unstable photon orbits for a Schwarzschild-de Sitter black hole. It is straightforward to calculate the critical value of the impact parameter which is obtained as
\begin{equation}
b_p = \frac{\sqrt{3}\left(\beta_0-1\right)}{\sqrt{\beta^2\left(2\beta_0-1\right)-18\beta\Lambda+6\left(\beta_0-1\right)\Lambda}},
    \label{eq:bp}
\end{equation}
{which is crucial in determining the size of the black hole's apparent shape in the observer's sky.} Historically, various definitions have been proposed in the literature to describe the visual appearance of black holes. These include the \textit{escape cone} by Synge \cite{Synge:1966}, and the \textit{cone of gravitational radiation capture} by Zeldovich and Novikov \cite{zeldovich_relativistic_1966}. 
Bardeen, Chandrasekhar, and Luminet popularized the terms \textit{optical appearance of black holes} and \textit{black hole image} to describe the visual characteristics of black holes \cite{Bardeen:1973b,Chandrasekhar:1998,Luminet:1979nyg,luminet_seeing_2018}. These terms are widely used in the field. In addition, it is worth noting that the term "black hole shadow," which is also utilized in this study, was introduced by Falcke, Melia, and Agol \cite{Falcke_2000}. This term refers to the dark region observed within the apparent boundary of the black hole. The apparent boundary is occasionally referred to as the \textit{photon ring} \cite{johannsen_testing_2010,johnson_universal_2020}, although in this study, the term holds a broader significance (see Ref.~\cite{perlick_calculating_2022} for further historical insights on this matter). In this context, the shadow of black holes is created by the presence of light rings, which represent the lensed images of their luminous background. The radius of these rings can be determined by the critical impact parameter  (i.e. $R_{\mathrm{sh}}=b_p$). Specifically, for the $f(R)$ black hole being examined, the expression for $R_{\mathrm{sh}}$ is given by Eq. \eqref{eq:bp}. This way, the theoretical shadow diameter for this black hole is given by $d_{\mathrm{sh}}^{\mathrm{theo}}=2R_{\mathrm{sh}}=2b_p$.
{In Fig. \ref{fig:betaConfinement}(a), the $\beta$-profile of this theoretical diameter has been plotted. It is worth noting that one can calculate the diameter of the observed shadows in the recent EHT images of M87* and Sgr A*. To do so, let us consider the relation \cite{Bambi:2019tjh}
\begin{equation}
    d_\text{sh} = \frac{D \theta_*}{\gamma},    \label{eq:dsh}
\end{equation}
which calculates the shadow diameter as observed by an observer positioned at a distance $D$ (in parsecs) from the black hole,} where $\gamma$ is the mass ratio of the black hole and the Sun, which is  $\gamma = (6.5 \pm 0.90) \times 10^9$ for M87* at the distance $D=16.8\,\mathrm{Mpc}$ \cite{Akiyama:2019}, and is $\gamma = (4.3 \pm 0.013) \times 10^6$ for Sgr A* at $D=8.127\,\mathrm{kpc}$ \cite{Akiyama:2022}. In Eq.~\eqref{eq:dsh}, $\theta_*$ is the angular diameter of the shadow, which has been measured as $\theta_*=42 \pm 3 \,\mathrm{\mu as}$ for M87*, and $\theta_*=48.7 \pm 7\, \mathrm{\mu as}$ for Sgr A*. This way, one can calculate the shadow diameters as $d_{\mathrm{sh}}^{\mathrm{M87^*}} = 11 \pm 1.5$ and $d_{\mathrm{sh}}^{\mathrm{SgrA^*}} = 9.5 \pm 1.4$.  {In Fig. \ref{fig:betaConfinement}(b), these values are displayed within the $1\sigma$ uncertainties for both black holes. However, these observed diameters do not correspond to the calculated theoretical shadow diameters $d_{\mathrm{sh}}^{\mathrm{theo}}$ in the same physical sense. The theoretical diameter relates to the photon ring and, in general, the photon ring's brightness is predominantly influenced by direct disk emission (further elaborated in Subsec. \ref{subsec:EmissionandRings}). Additionally, uncertainties stemming from the physical characteristics of the accretion disk and its emission profiles introduce further complexities. Consequently, attempting to constrain black hole parameters by comparing the theoretical shadow diameter to observations from the EHT is not viable. Therefore, such analysis does not offer a means to constrain the model parameter $\beta$ for the $f(R)$ black hole.}
{Instead, for the numerical studies in the rest of the paper, we consider the values of the $\beta$-parameter within a specific range, marked in the shaded region of the $\beta$-profile of the theoretical shadow diameter (as seen in Fig. \ref{fig:betaConfinement}(a)). This segment confines the parameter within the range $0 \leq \beta < 0.05$, which we adhere to in the subsequent discussions in this paper.}

\begin{figure}[t]
    \centering
    \includegraphics[width=7cm]{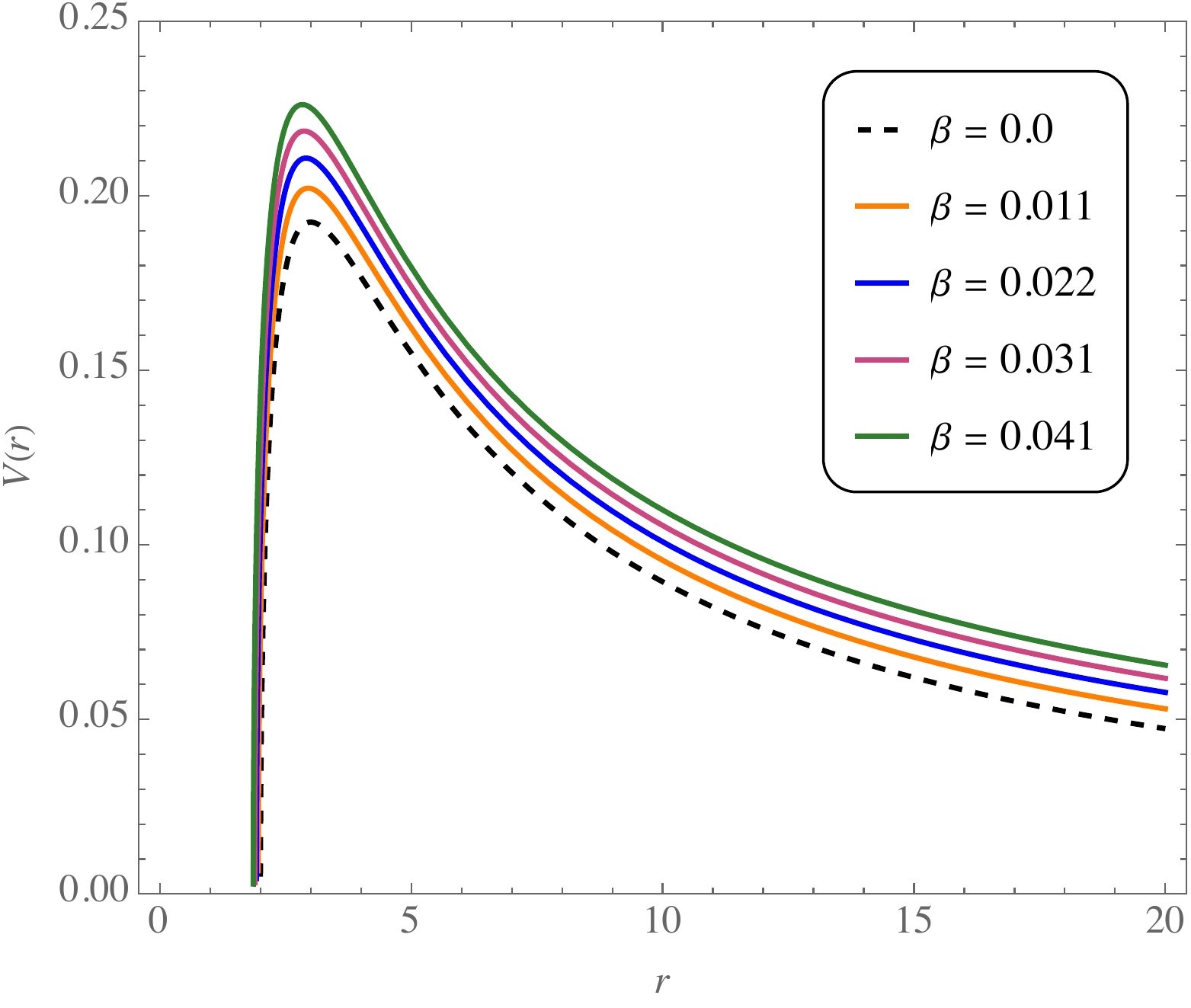}~(a)\qquad\qquad
    \includegraphics[width=7cm]{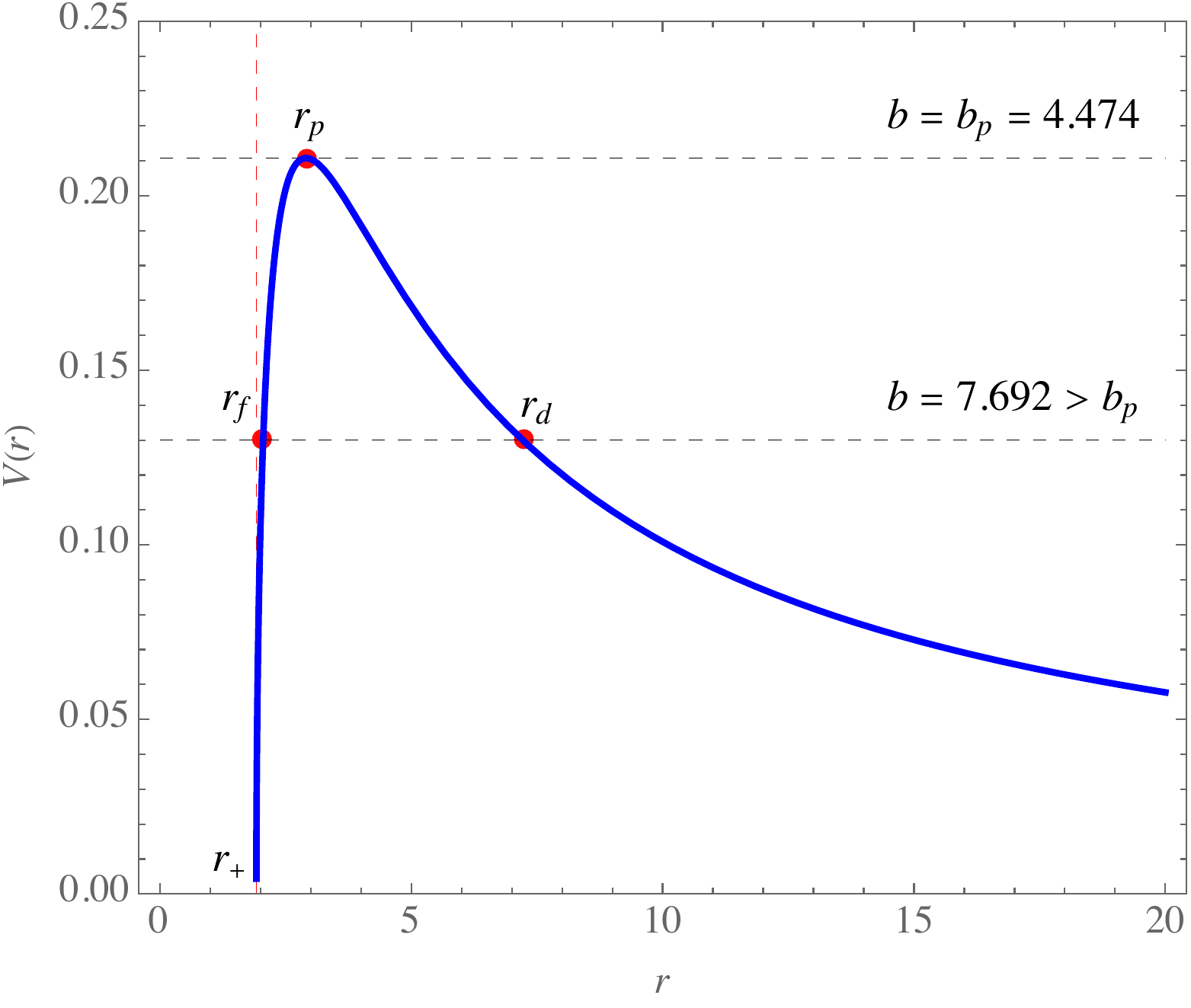}~(b)
    \caption{In panel (a), the radial profile of the effective potential is shown for various values of $\beta$, {within the {assumed range}}. In panel (b), by adopting $\beta=0.022$, 
    a typical effective potential has been shown together with the turning points and their corresponding impact parameters. }
    \label{fig:V(r)}
\end{figure}

Hence, it is now possible to visualize the behavior of the effective potential \eqref{eq:V(r)}, which is shown in Fig. \ref{fig:V(r)}.
In the left panel, five radial profiles of the effective potential have been plotted which correspond to different values of the $\beta$-parameter within the {assumed range}. As it is inferred from the diagram, the potential possesses one maximum, at which, unstable orbits can occur. This is shown in more detail in the right panel of Fig.~\ref{fig:V(r)}, where the orbits are categorized in accordance with the values of the impact parameter $b$. When $b>b_p$, the photons may approach from either of the turning points $r_d$ (where they are recessively deflected by the black hole) or $r_f$ (where they are deflected towards the event horizon). In fact, the turning points can be obtained analytically for the spacetime of the $f(R)$ black hole, by solving the equation $({\ed r}/{\ed\phi})^2= 0$. Applying the Eqs.~\eqref{eq:drdphi} and \eqref{eq:V(r)}, this results in 
\begin{equation}
\left(\frac{\ed r}{\ed\phi}\right)^2=\mathcal{P}_4(r)\equiv r\left(\frac{r^3}{\lambda^2}-\beta r^2-r+2\right) = 0,
    \label{eq:drdphi_new}
\end{equation}
which beside the trivial solution at $r=0$, has one negative root and the two positive roots $r_d=x_d^{-1}$ and $r_f=x_f^{-1}$, where
\begin{eqnarray}
    && x_f = \frac{1}{6}-2\sqrt{\frac{\bar{g}_2}{3}}\sin\left(\frac{1}{3}\arcsin\left(\frac{3\bar{g}_3}{\bar{g}_2}\sqrt{\frac{3}{\bar{g}_2}}
    \right)-\frac{2\pi}{3}\right),\label{eq:xf}\\
    && x_d = \frac{1}{6}-2\sqrt{\frac{\bar{g}_2}{3}}\sin\left(\frac{1}{3}\arcsin\left(\frac{3\bar{g}_3}{\bar{g}_2}\sqrt{\frac{3}{\bar{g}_2}}
    \right)\right)\label{eq:xd},
\end{eqnarray}
with ${\lambda^{-2}}={b^{-2}}+\Lambda$, and
\begin{subequations}\label{eq:gb2,gb3}
\begin{align}
    & \bar{g}_2=\frac{1}{4}\left(\frac{1}{3}+2\beta\right),\\
    & \bar{g}_3=-\frac{1}{4}\left(\frac{1}{\lambda^2}-\frac{\beta}{6}-\frac{1}{54}\right).
\end{align}
\end{subequations}
For the case of $b=b_p$, the photons encounter the turning point $r_p$ (see the right panel of Fig.~\ref{fig:V(r)}), which has been determined in Eq.~\eqref{eq:rp_0}. At this stage, the photons travel on unstable (or critical) orbits, which form the black hole shadow. In Table. \ref{tab:Table1}, the turning points have been given for different values of $\beta$.
\begin{table}[t]
\centering
\begin{tabular}{c |  c  c  c  c  c} 
\hline

 $\beta$ & $0.0$ & 0.011 & 0.022 & 0.031 & 0.041\\ [0.3ex] 
 \hline\hline
 $r_+$ & 2.0001 & 1.9563 & 1.9196 & 1.8880 & 1.8583\\
 
 $r_p$ & 3.0000 & 2.9503 & 2.9077 & 2.8704 & 2.8350\\

 $b_p$ & 5.1968 & 4.9468 & 4.7445 & 4.5762 & 4.4228\\

 $r_d$ & 6.3692 & 6.8091 & 7.2156 & 7.5957 & 7.9813\\

 $r_f$ & 2.1797 & 2.1076 & 2.0544 & 2.0103 & 1.9699\\
 [0.5ex]
 \hline
\end{tabular}
\caption{The turning points of photonic trajectories together with their corresponding values of $b_p$.}
\label{tab:Table1}
\end{table}
As it is inferred from the table, increase in the $\beta$-parameter leads to a smaller black hole (decrease in $r_+$), a wider effective potential (increase in the distance between $r_d$ and $r_f$), and a lower potential maximum. So unstable orbits are less likely to happen for larger $\beta$ (decrease in $b_p$), and the black hole shadow decreases in size. {In such cases, the light rays detected by a distant observer are mostly dominated by the direct emission, which is a simply lensed image of the black hole's emitting disk or its luminous background. This will be discussed in more details in the forthcoming sections.}

Now that the turning points have been obtained and analyzed, we proceed with the determination of the exact solutions for the aforementioned possible photon orbits around the $f(R)$ black hole. In fact, the null geodesics for this black hole have been studied in their most general form in Ref.~\cite{soroushfar_analytical_2015}. In what follows, however, we base our exact solutions on the analytically known turning points, which helps us investigate, separately, the deflecting and critical trajectories that are of importance for the purpose of this paper.


\subsubsection{Deflecting trajectories}\label{subsub:deflection}

As mentioned above, photonic trajectories become deflected at the turning points $r_d$ and $r_f$ which leads to different fates for the photons. Accordingly, the possible deflecting trajectories can be ramified as the orbit of the first kind (OFK) at $r_d$, and the orbit of the second kind (OSK) at $r_f$. Since both of these turning points have been identified analytically, hence, by applying the change of variable $z\doteq{r_i}/{r}$ ($r_i=r_d, r_f$), one can rewrite the differential equation \eqref{eq:drdphi_new} as
\begin{equation}
\left(\frac{\ed z}{\ed\phi}\right)^2=\mathcal{P}_3(z) \equiv \frac{2}{r_i}z^3-z^2-r_i\beta z+\frac{r_i^2}{\lambda^2}.
    \label{eq:dzdphi}
\end{equation}
A further change of variable $u\doteq\frac{1}{2}({z}/{r_i}-{1}/{6})$ provides us with the Weierstrassian differential equation 
\begin{equation}
\left(\frac{\ed u}{\ed\phi}\right)^2=\tilde{\mathcal{P}}_3(u)\equiv4u^3-\tilde{g}_2 u-\tilde{g}_3,
    \label{eq:dudphi}
\end{equation}
in which 
\begin{subequations}
\begin{align}
    & \tilde{g}_2=\frac{1}{12}(1+6\beta),\\
    & \tilde{g}_3 = -\frac{1}{216}\left(\frac{54}{\lambda^2}-9\beta-1\right),
\end{align}
\label{eq:tildeg2g3}
\end{subequations}
are known as the Weierstrass invariants. This leads to the integrals
\begin{eqnarray}
    && \phi-\phi_0=\int_{u_d}^{u}\frac{\ed u'}{\sqrt{\tilde{\mathcal{P}}_3(u')}}~~(\mathrm{with}~ u_d<u),\label{eq:phi(u)_OFK}\\
    && \phi-\phi_0=\int_u^{u_f}\frac{\ed u'}{\sqrt{\tilde{\mathcal{P}}_3(u')}}~~(\mathrm{with}~ u_f>u),\label{eq:{eq:phi(u)_OSK}}
\end{eqnarray}
respectively, for the OFK and OSK, in which $\phi_0$ is the initial azimuth angle, and $u_{i}=\frac{1}{2}({1}/{r_{i}}-{1}/{6})$. Taking into account the applied changes of variables, the above integrals yield
\begin{equation}
r(\phi) = \frac{6}{1+12\wp\left(\omega_d-(\phi-\phi_0)\right)},
    \label{eq:sol_OFK}
\end{equation}
for the OFK and 
\begin{equation}
r(\phi) = \frac{6}{1+12\wp\left(\omega_f+(\phi-\phi_0)\right)},
    \label{eq:sol_OSK}
\end{equation}
for the OSK, where $\wp(x)\equiv\wp(x;\tilde{g}_2,\tilde{g}_3)$ is the  $\wp$-Weierstrassian elliptic function \cite{byrd_handbook_1971}, and we have defined
\begin{equation}
  \omega_i=\wp^{-1}\left(\frac{1}{2r_i}-\frac{1}{12}\right).
\end{equation}


\subsubsection{Deflection angle}\label{subsubsec:strongdeflection}

\begin{figure}[h]
    \centering
    \includegraphics[width=7cm]{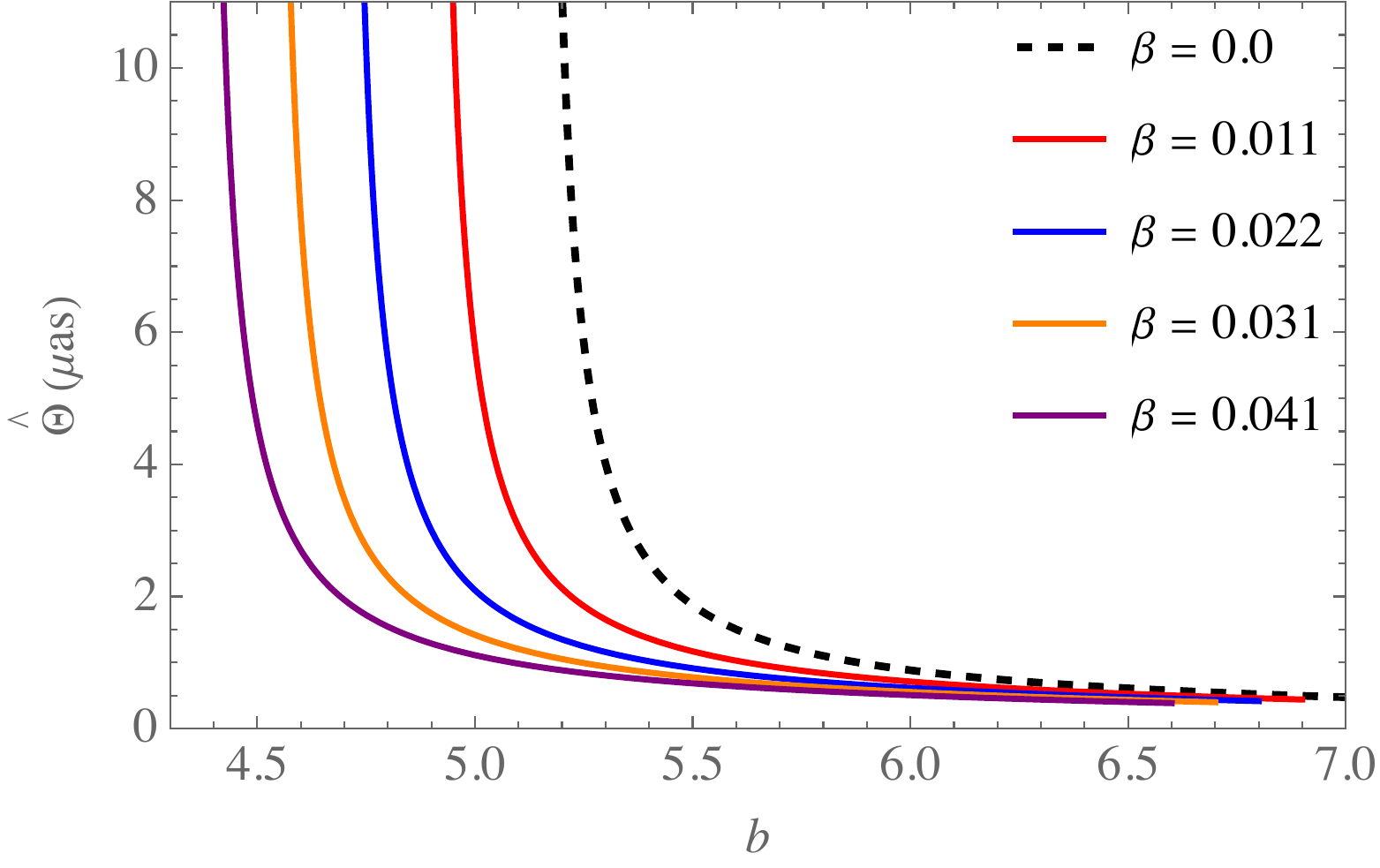}
    \caption{The plot of the deflection angle $\hat{\Theta}$ (in $\mathrm{\mu as}$) versus the changes in the impact parameter $b$, plotted for the different values of the $\beta$-parameter, and $r_\mathbb{O}=10^5$. }
    \label{fig:hatTheta}
\end{figure}

The OFK is in fact related to the gravitational lensing that is caused by the black hole and is, in part, responsible for the formation of the black hole shadow. Hence, by having at hand the integral equation \eqref{eq:phi(u)_OFK}, one can calculate the deflection angle $\hat\Theta$ that an observer $\mathbb{O}$ at the radial position $r_\mathbb{O}$ from the black hole (the lens) measures. Accordingly, we have \cite{misner_gravitation_2017}
\begin{eqnarray}
\hat\Theta&=&2\Delta\phi-\pi\nonumber\\
&=&2\int_{u_\mathbb{O}}^{u_d}\frac{\ed u'}{\sqrt{\tilde{\mathcal{P}}_3(u)}}-\pi =2\left[ \wp^{-1}(u_\mathbb{O})-\wp^{-1}(u_d)\right]-\pi,
    \label{eq:hatTheta_0}
\end{eqnarray}
with the Weierstrass invariants given in Eqs.~\eqref{eq:tildeg2g3}. Using the analytical expression for $r_d$, we have plotted the behavior of $\hat\Theta$ in terms of the impact parameter $b$ in Fig.~\ref{fig:hatTheta}.
As it can be inferred from the diagram, there is no significant sensitivity in the behavior of the deflection angle, given the small changes in the $\beta$-parameter. However, in general, $\hat{\Theta}$ decreases for a given $b$, when the $\beta$-parameter increases from 0 to its maximum value. In this sense, the Schwarzschild-de Sitter spacetime causes the highest deflection angle. Naturally, by approaching the black hole (smaller $b$) the deflection angle increases until it diverges at $b_p$ for each of the cases. So, strong lensing occurs in the near-horizon regions whereas by receding the black hole, the light deflection process will correspond to weak lensing.


\subsubsection{Critical trajectories}\label{subsub:critical}

\begin{figure}[t]
    \centering
    \includegraphics[width=5.5cm]{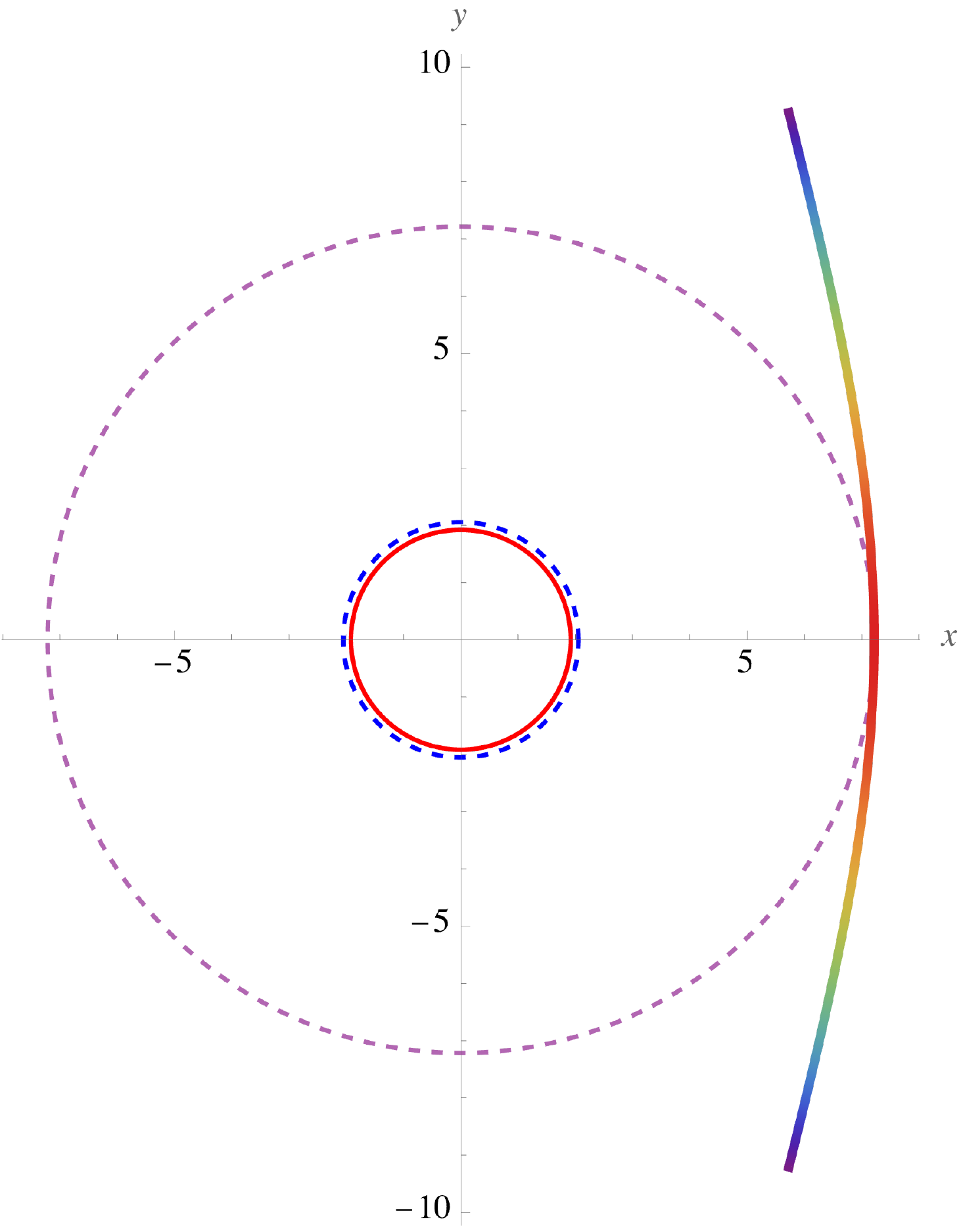}~(a)\qquad\qquad\qquad
    \includegraphics[width=5.5cm]{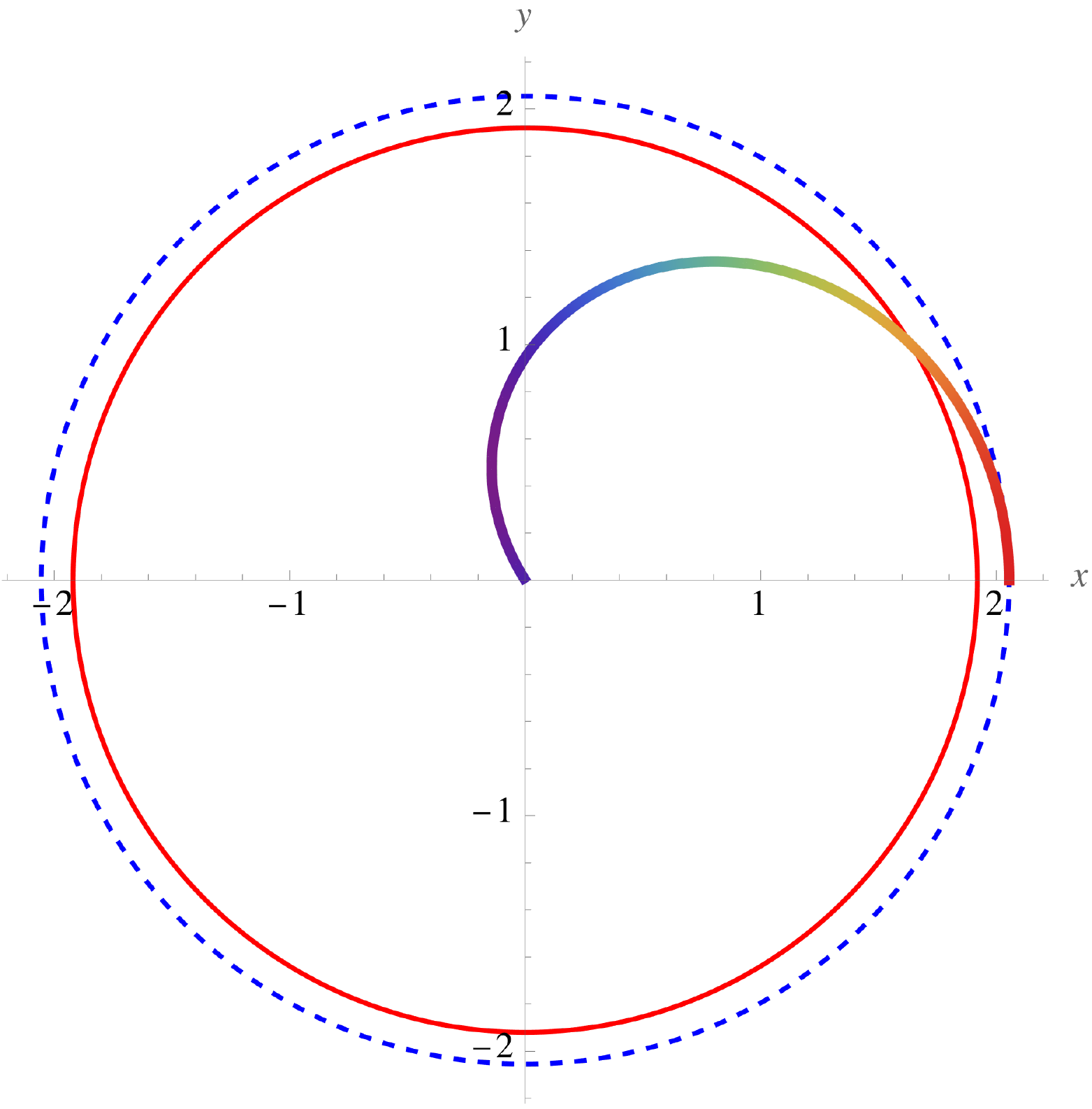}~(b)
    \includegraphics[width=6cm]{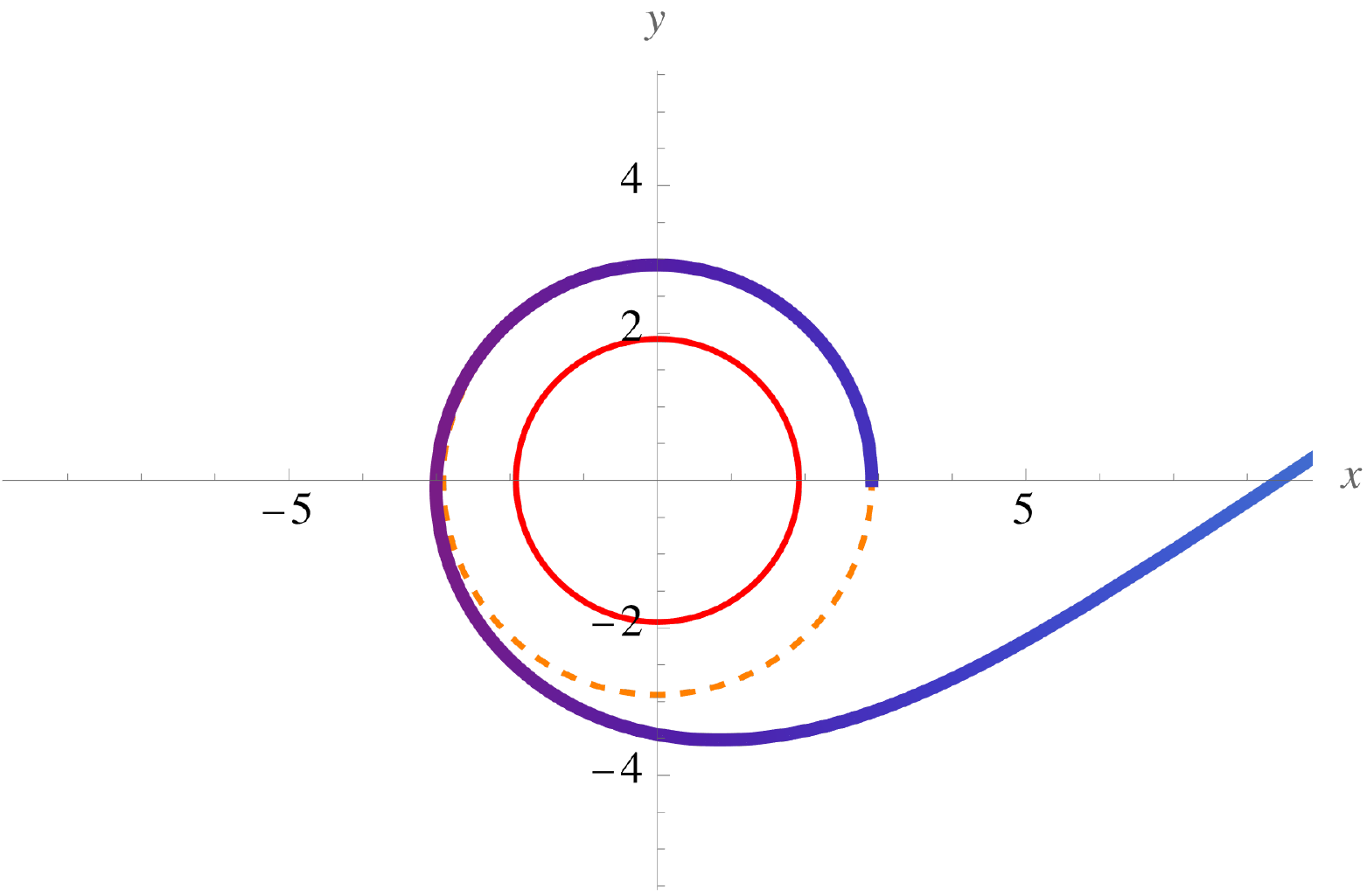}~(c)\qquad\qquad\qquad
    \includegraphics[width=5.3cm]{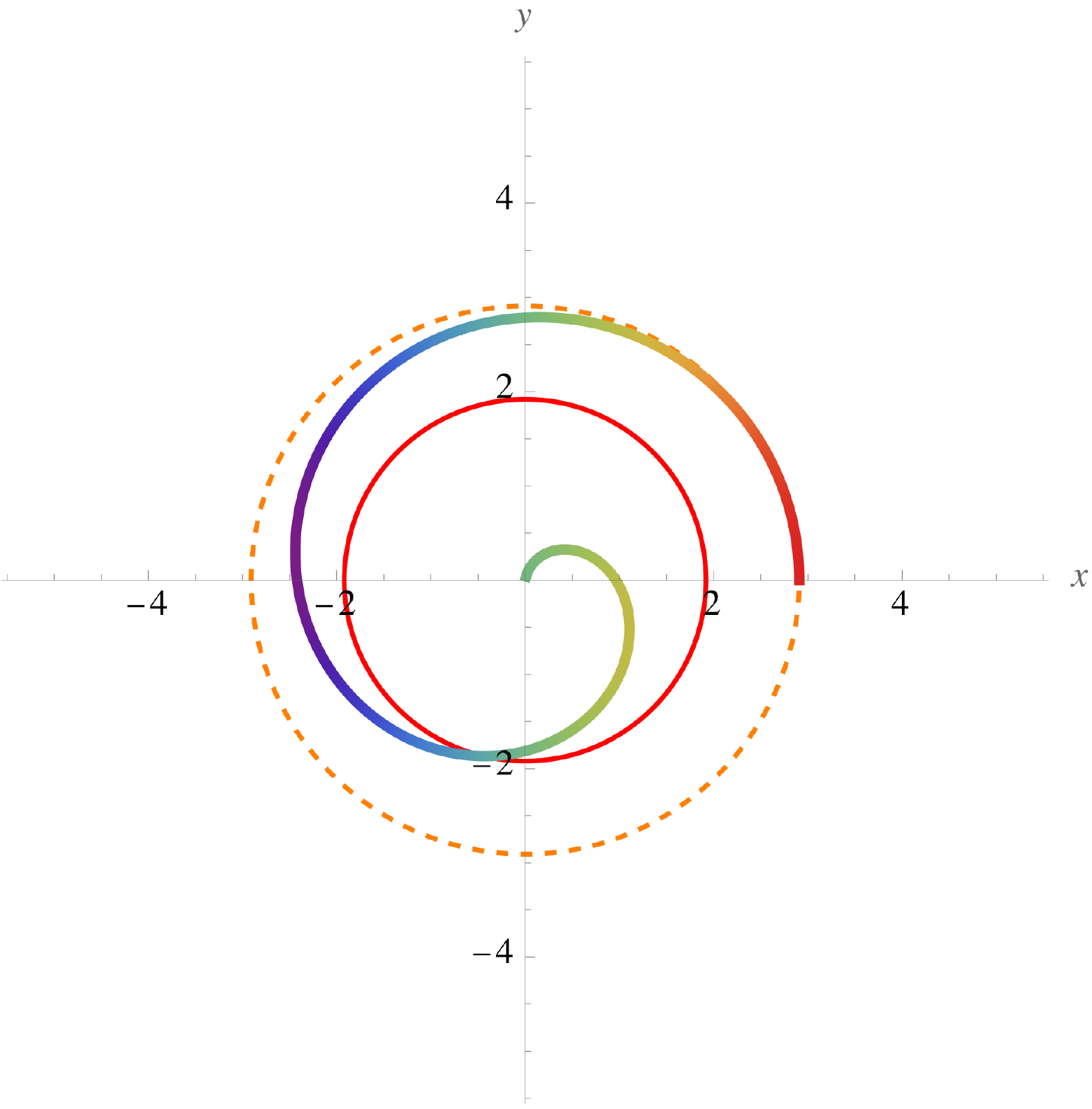}~(d)
    \caption{Examples of (a) OFK, (b) OSK, (c) COFK and (d) COSK, plotted for $\beta=0.022.$ In panels (a) and (b), the red circle at the center denotes $r_+$, while the blue and purple dashed circles correspond to $r_f$ and $r_d$.  In panels (c) and (d), the exterior dashed circles indicate $r_p$.}
    \label{fig:orbits_0}
\end{figure}

Similar to the deflecting trajectories, unstable orbits can also lead to different fates, which we name after as the critical orbits of the first kind (COFK) which happen when photons approach $r_p$ from an initial distance $r_p<r_\mathrm{in}<r_{++}$, and the critical orbits of the second kind (COSK) which correspond to photons approaching $r_p$ from $r_+<r_\mathrm{in}<r_{p}$. When ${\lambda^{-2}}\rightarrow {\lambda_p^{-2}}={b_p^{-2}}+\Lambda$, the point $r=r_p$ is a double root of $\mathcal{P}_4(r)$ in Eq.~\eqref{eq:drdphi_new}. The differential equation of motion can then be factorized as
\begin{equation}
\frac{\ed r}{\ed\phi}\equiv\mathcal{P}_4^{p}(r)=\left|r-r_p\right|\sqrt{\frac{r^2}{\lambda_p^2}+\left(\frac{r_p}{\lambda_p^2}+\chi_1\right)r+\frac{r_p^2}{\lambda_p^2}+\chi_1r_p+\chi_0},
    \label{eq:P4p(r)}
\end{equation}
by means of the method of synthetic division, in which
\begin{subequations}\label{eq:chi0chi1}
\begin{align}
    & \chi_0=r_p\left(\frac{r_p}{\lambda_p^2}-\beta\right)-1,\\
    & \chi_1 = \frac{r_p}{\lambda_p^2}-\beta.
\end{align}
\end{subequations}
One can therefore obtain the exact solutions to Eq.~\eqref{eq:P4p(r)}, by means of direct integration and applying the inversion. This yields the two solutions $r_{\mathrm{I}}(\phi)$ for the COFK and $r_{\mathrm{II}}(\phi)$ for the COSK, which are given as
\begin{multline}
r_{\substack{\mathrm{I}\\\mathrm{II}}}(\phi) = \frac{1}{\mathcal{A}^2+8r_p^2+(\mathcal{A}^2-4\mathcal{B})\cosh\Phi}\Big[
2r_p\left(\mathcal{A}^2-4\mathcal{B}\right)\cosh^2\Phi\pm\left[r_p(r_p+\mathcal{A})+\mathcal{B}\right]\\
\times\sqrt{(\mathcal{A}^2-4\mathcal{B})\sech^2\Phi\tanh^2\Phi}\big(\cosh(2\Phi)\mp2\mathcal{A}\big)\Big],
    \label{eq:rphi_COFSK}
\end{multline}
where $\Phi = \lambda_p^{-1}(\phi-\phi_0)\sqrt{r_p(\mathcal{A}+r_p)+\mathcal{B}}$, and 
\begin{subequations}
\begin{align}
& \mathcal{A} = r_p+\chi_1\lambda_p^2,\\
& \mathcal{B} = r_p^2+\lambda_p^2\left(\chi_0+\chi_1 r_p\right).
    \label{eq:mathcalAB}
\end{align}
\end{subequations}
In Fig.~\ref{fig:orbits_0}, some examples of the possible photon orbits have been shown.
As can be inferred from the diagrams, the boundary of the black hole shadow is based on these four types of orbits. Together, these photon orbits are able to produce the bright ring surrounding the dark shadow in the observer's sky, which is produced as a result of the strong gravitational lensing in the near-horizon regions. 

As previously mentioned, the lensing phenomenon and the existence of critical photon orbits play a crucial role in shaping the black hole shadow and forming the light rings. However, the formation of these features is highly dependent on how the black hole is illuminated. In the case of astrophysical black holes, illumination is often provided by an accretion disk, which is also taken into account in this study. In the following section, we begin by developing a thin accretion model and subsequently delve into both analytical and numerical examinations of the emission process from this disk, which contributes to the formation of the rings.

\section{Thin accretion disk model and emission from the black hole}\label{sec:thinModel}

\begin{figure}[h]
    \centering
    \includegraphics[width=7cm]{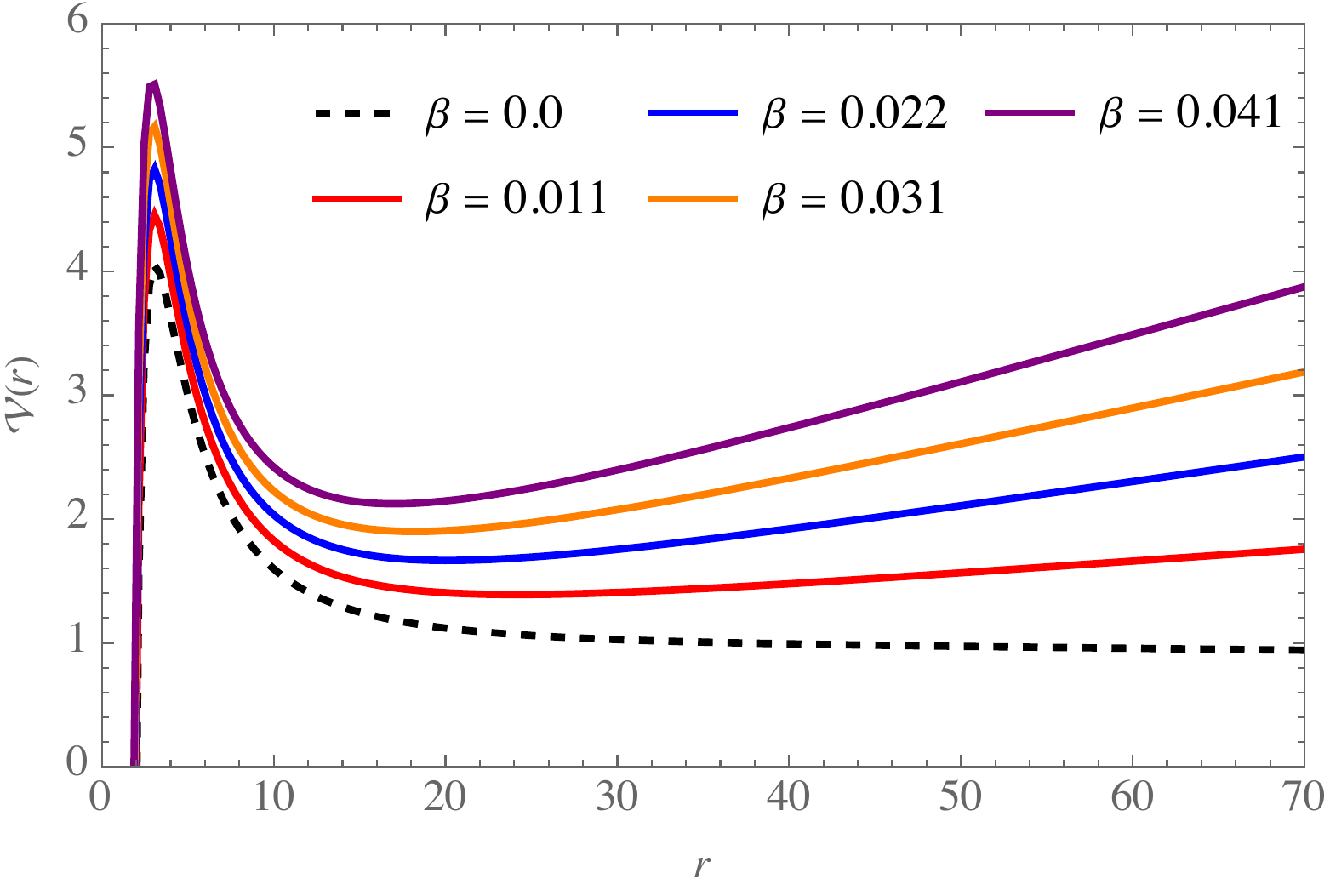}~(a)\qquad\qquad
     \includegraphics[width=7cm]{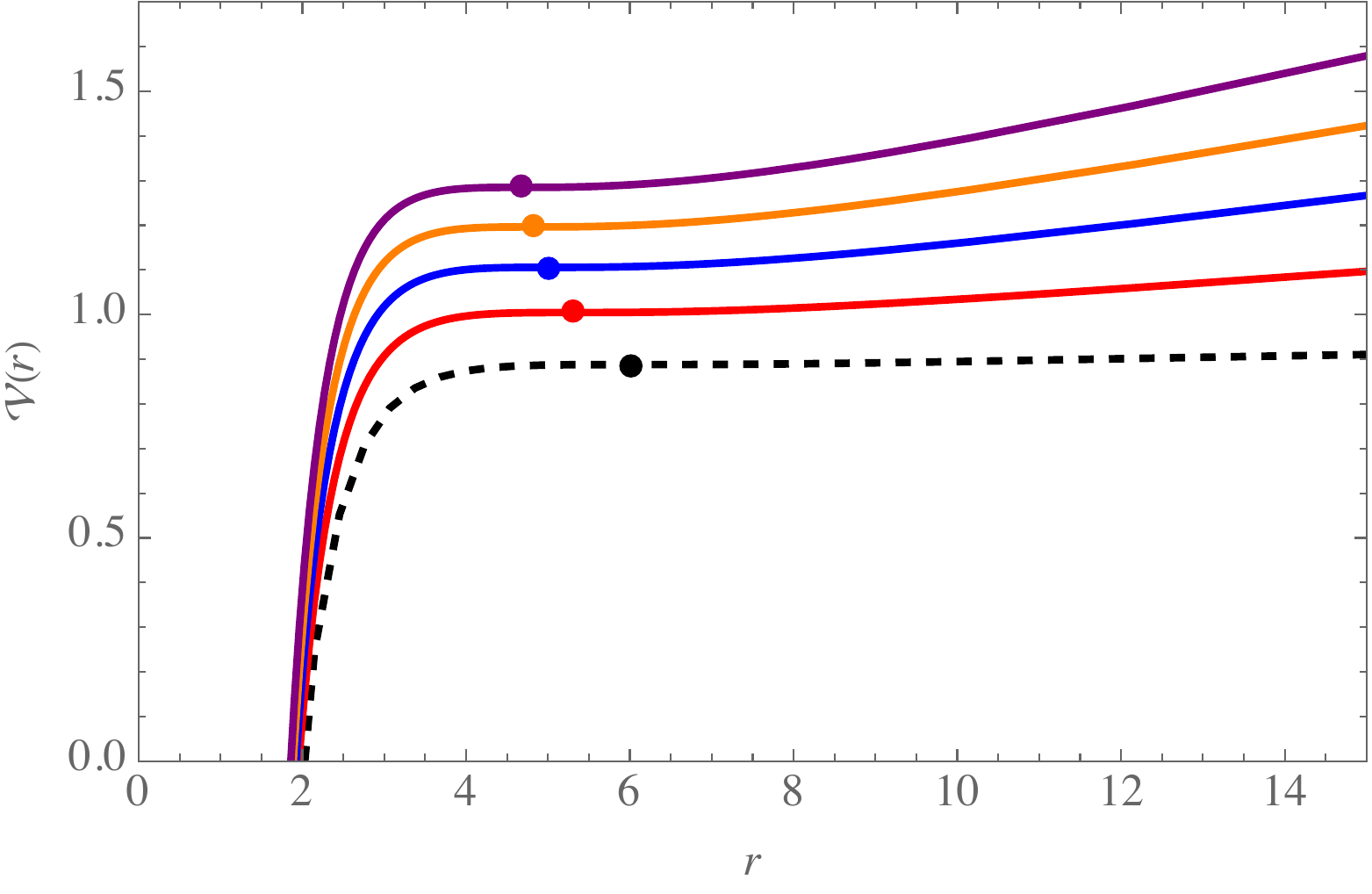}~(b)
    \caption{(a) The radial profile of the effective potential $\mathcal{V}(r)$ plotted for different {values of the $\beta$-parameter} and $\mathcal{L}=10$. (b) The position of ISCO (shown by a point on each of the curves) for the same values of $\beta$. From bottom to top, the corresponding values of the angular momentum are $\mathcal{L} = 3.46, 3.64, 3.74, 3.81$ and $3.86$.}
    \label{fig:Veff_massive}
\end{figure}

\begin{figure}[h]
    \centering
    \includegraphics[width=5.47cm]{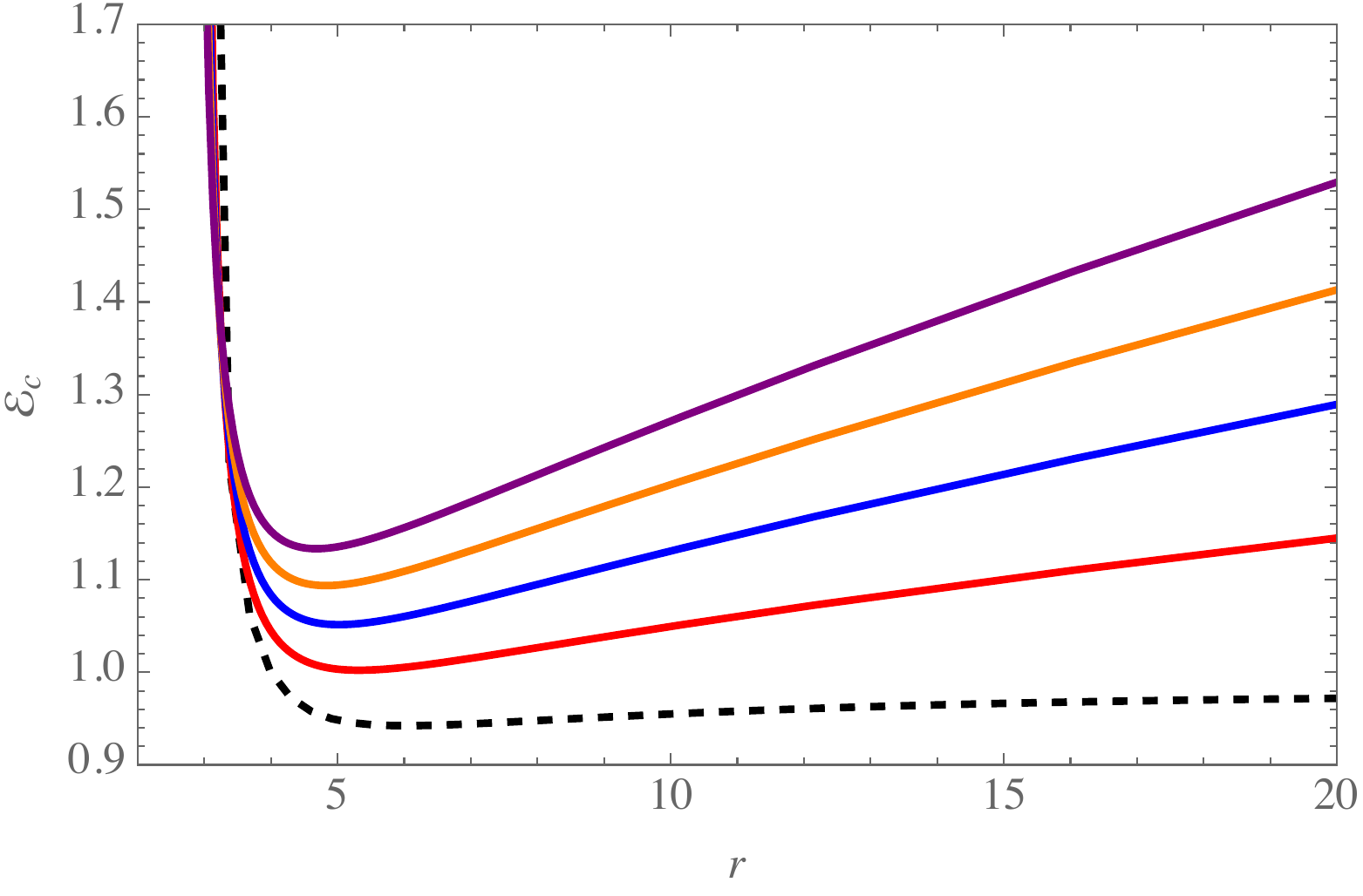}~(a)
     \includegraphics[width=5.47cm]{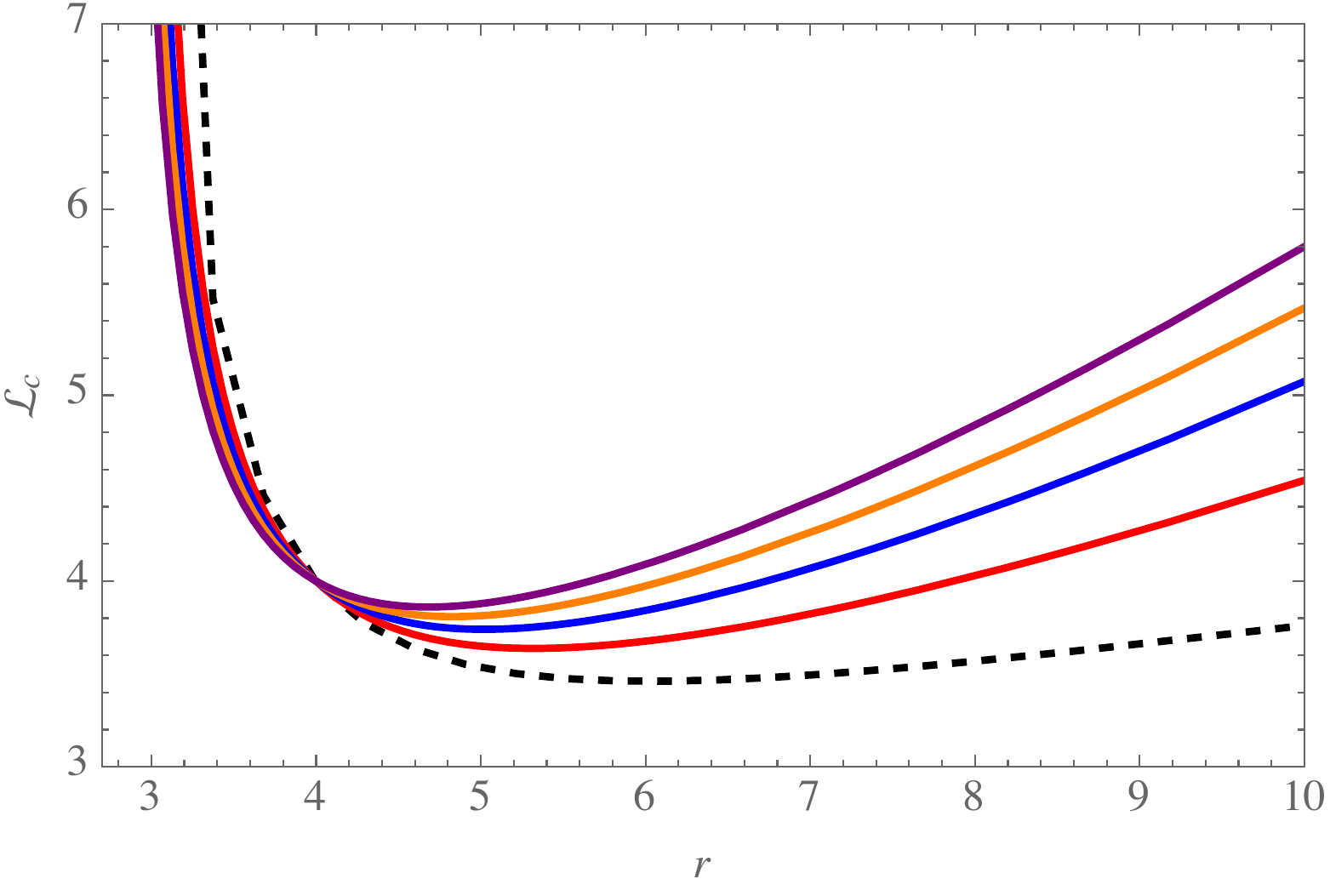}~(b)
     \includegraphics[width=5.47cm]{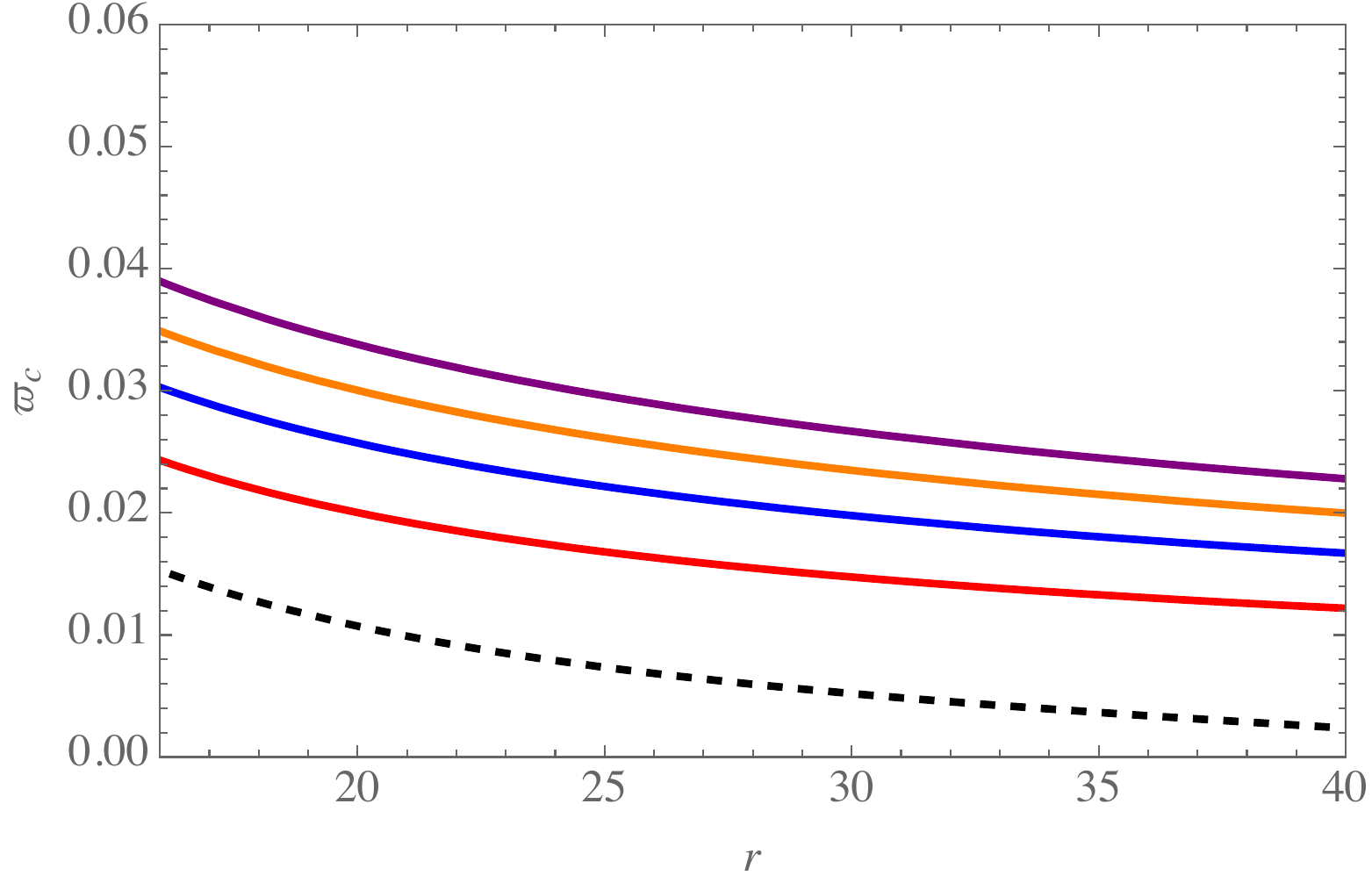}~(c)
    \caption{Radial profiles of (a) $\mathcal{E}_c$, (b) $\mathcal{L}_c$ and (c) $\varpi_c$, for the {adopted values of the $\beta$-parameter} and the same color-coding as in Fig.~\ref{fig:Veff_massive}.}
    \label{fig:ELvarpi_c}
\end{figure}

In this section, we study the observational signatures of the black hole in the case of the existence of a thin accretion disk. We assume that the accretion process is explained by a generalized version of the well-known Shakura-Sunyaev model \cite{Shakura:1973}, proposed by Novikov and Thorne in Ref.~\cite{Novikov:1973}. To proceed with applying this model, let us return to the Lagrangian \eqref{eq:lagrangian}, which is now specified as $2\mathscr{L}=-1$, for massive particles of energy $\mathcal{E}$ and angular momentum $\mathcal{L}$ that constitute the accretion disk. This way, one can rewrite the equations of motion \eqref{eq:dotr} and \eqref{eq:drdphi} as
\begin{eqnarray}
&& \dot{r}=\mathcal{E}^2-\mathcal{V}(r),  \label{eq:rdot_massive}\\
&& \left(\frac{\ed r}{\ed\phi}\right)^2=\frac{\mathscr{P}_6(r)}{\mathcal{L}^2},
\end{eqnarray}
in which $\mathscr{P}_6(r) = r\left[\Lambda r^5-\beta r^4-(1-\mathcal{E}^2-\mathcal{L}^2\Lambda)r^3+(2-\mathcal{L}^2\beta)r^2-\mathcal{L}^2 r-2\mathcal{L}^2\right]$, and 
\begin{equation}
\mathcal{V}(r)=B(r)\left(1+\frac{\mathcal{L}^2}{r^2}\right),
    \label{eq:Veff_massive}
\end{equation}
is the effective potential for massive particles orbiting the black hole in the equatorial plan. The left panel of Fig.~\ref{fig:Veff_massive} shows a typical radial profile of $\mathcal{V}(r)$ which has been plotted for the {adopted values of $\beta$}.
According to the diagram, the effective potential possesses a minimum which allows for stable circular
orbits for the particles. The latter is a necessary condition for the formation of accretion disks in the context of the innermost stable circular orbit (ISCO), whose corresponding radius, $r_c$, can be obtained by the conditions $\mathscr{P}_6(r)=0=\mathscr{P}_6'(r)$.
Note that, although this system of equations cannot be solved by means of common radicals, however, the ISCO radius obeys the following relation \cite{PhysRevD.105.023024}
\begin{equation}
r_c = \frac{3B(r_c) B'(r_c)}{2B'(r_c)^2-B(r_c) B''(r_c)},
    \label{eq:rc}
\end{equation}
in the spacetime geometry of the $f(R)$ black hole.
This radius corresponds to the inner edge of the accretion disk and as we move away from the black hole, particles appear to moving on Keplerian bound orbits. In the right panel of Fig.~\ref{fig:Veff_massive}, the position of the ISCO has been indicated for each of the cases. As it can be inferred, an increase in the $\beta$-parameter decreases $r_c$, and hence, affects the structure of the accretion disk. Furthermore, the above conditions make it possible to obtain the analytical expressions 
\begin{eqnarray}
  && \mathcal{E}_c(r) = \frac{B(r)}{\sqrt{B(r)-r^2 \varpi_c(r)^2}} = 
  \frac{\sqrt{2 \Lambda}   (r_{++}-r) (r-r_{+}) (r-r_{3})}{r \sqrt{ \frac{3 r_{++} r_{+} r_{3}}{r}+r (r_{++}+r_{+}+r_{3})-2 \left[r_{++} (r_{+}+r_{3})+r_{+} r_{3}\right]}}
  ,\label{eq:EISCO}\\
  && \mathcal{L}_c(r) = \frac{r^2\varpi_c(r)}{\sqrt{B(r)-r^2 \varpi_c(r)^2}}
  =\frac{r^{\frac{3}{2}} \sqrt{  \left(-\frac{r_{++} r_{+} r_{3}}{r^2}-2 r+r_{++}+r_{+}+r_{3}\right)}}{\sqrt{\frac{3 r_{++} r_{+} r_{3}}{r}+r (r_{++}+r_{+}+r_{3})-2 \left[r_{++} (r_{+}+r_{3})+r_{+} r_{3}\right]}}
  , \label{eq:LISCO}
\end{eqnarray}
for the energy and angular momentum of particles residing in the ISCO, in which
\begin{equation}
\varpi_c(r)=\frac{\ed\phi}{\ed t}=\sqrt{\frac{B'(r)}{2r}}
=\sqrt{\frac{\Lambda}{2}}  \left(\frac{r_{++}+r_{+}+r_{3}}{r}-\frac{r_{++} r_{+} r_{3}}{r^3}-2\right)^{\frac{1}{2}},
    \label{eq:varpi_c}
\end{equation}
is the angular velocity of orbiting particles, and we have used the expression given in Eq.~\eqref{eq:lapse_1}. In Fig.~\ref{fig:ELvarpi_c}, the radial profile of the above quantities has been plotted for given values of the $\beta$-parameter. It is observed that by increasing $\beta$, all the quantities increase which is a consequence of the relevant changes in the effective potential.
Note that, on the ISCO, one can recast the characteristic polynomial as $\mathscr{P}_6(r) = \Lambda r (r-r_c)^3(r-r_4)(r-r_5)$, where $r_4>0$ and $r_5<0$ are the remaining two real roots of the characteristic polynomial, which can be expressed in terms of $r_c$ by means of the method of synthetic division.

\begin{figure}[h]
    \centering
    \includegraphics[width=5.495cm]{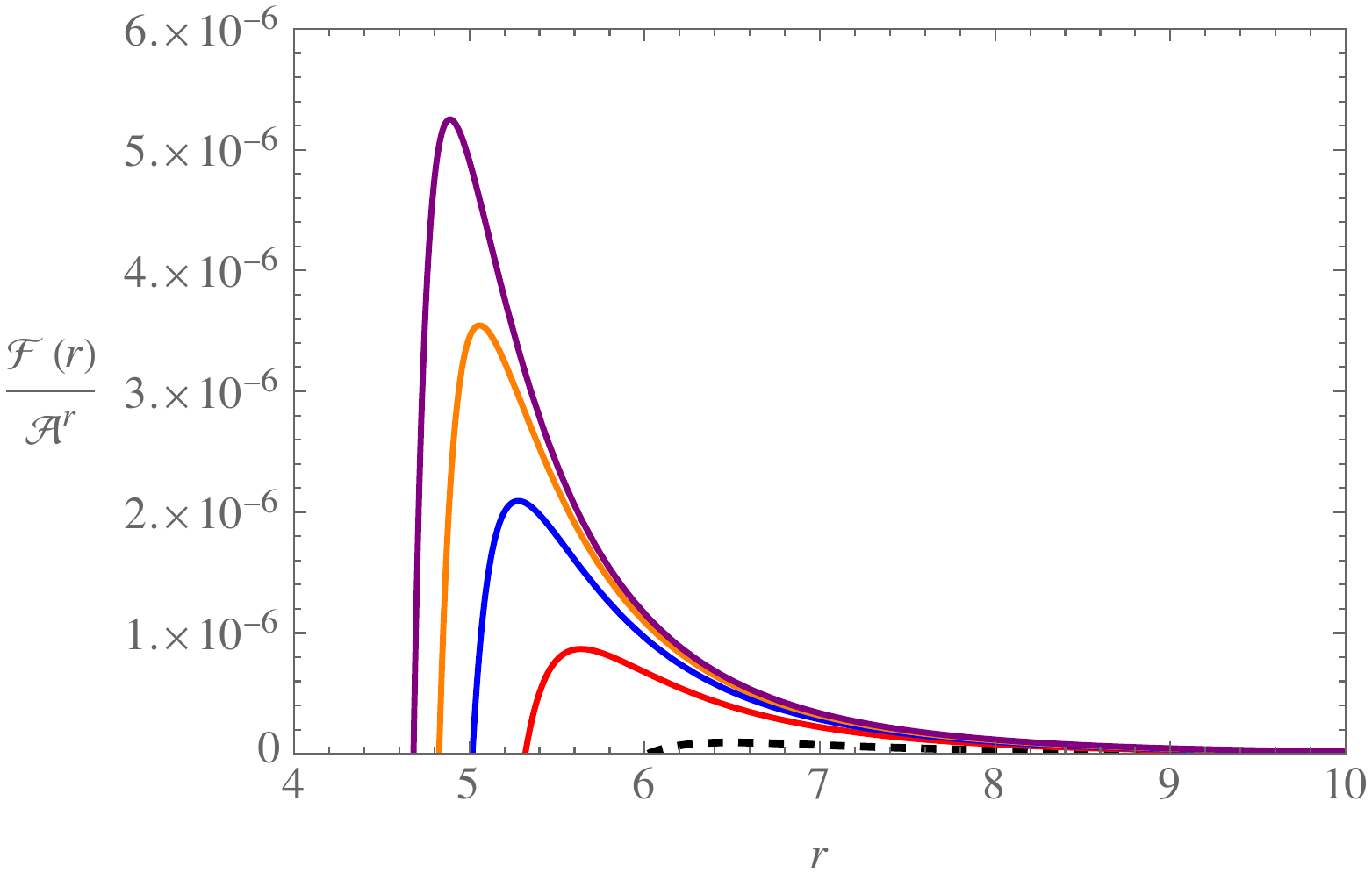}~(a)
     \includegraphics[width=5.495cm]{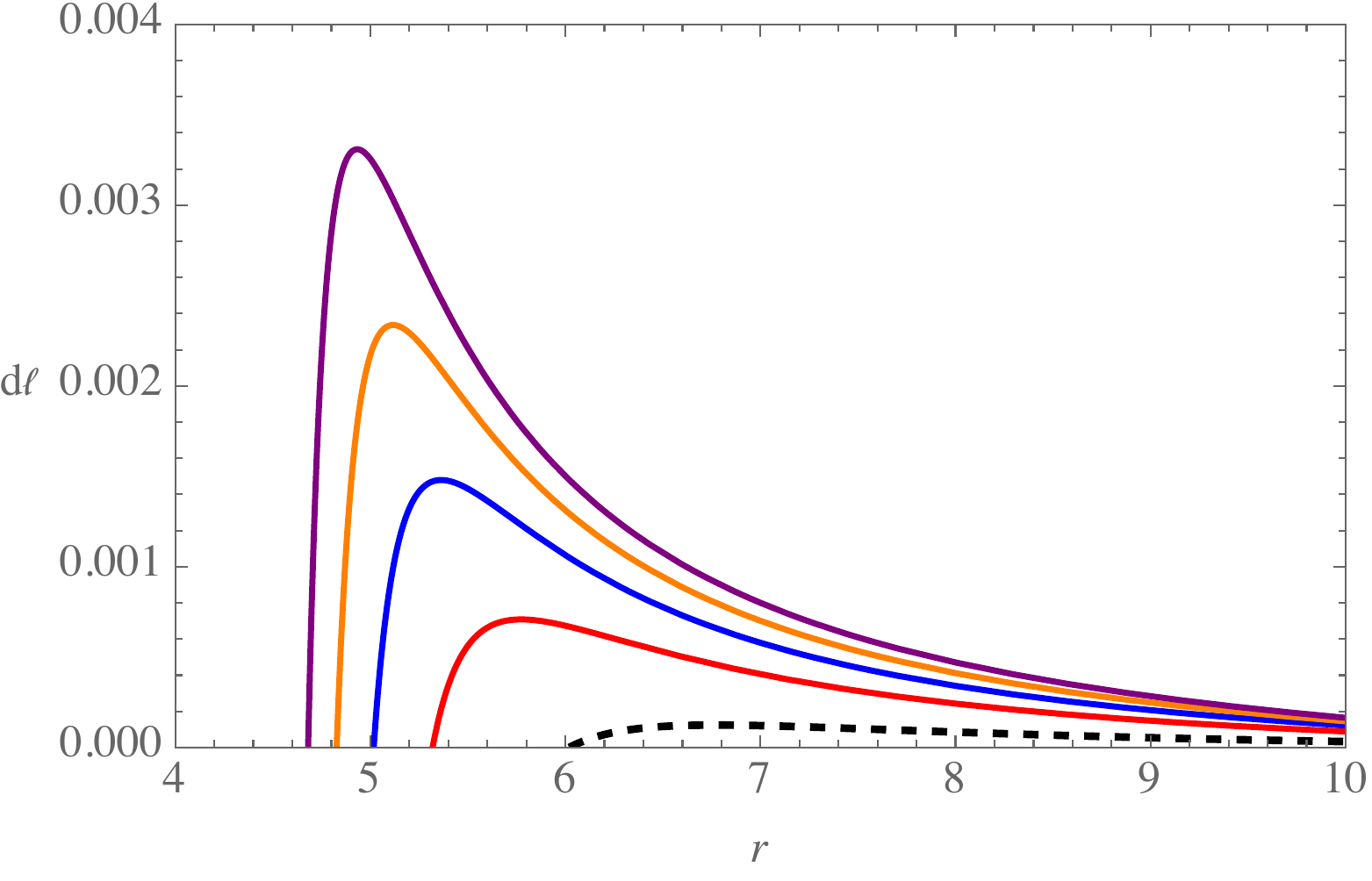}~(b)
     \includegraphics[width=5.495cm]{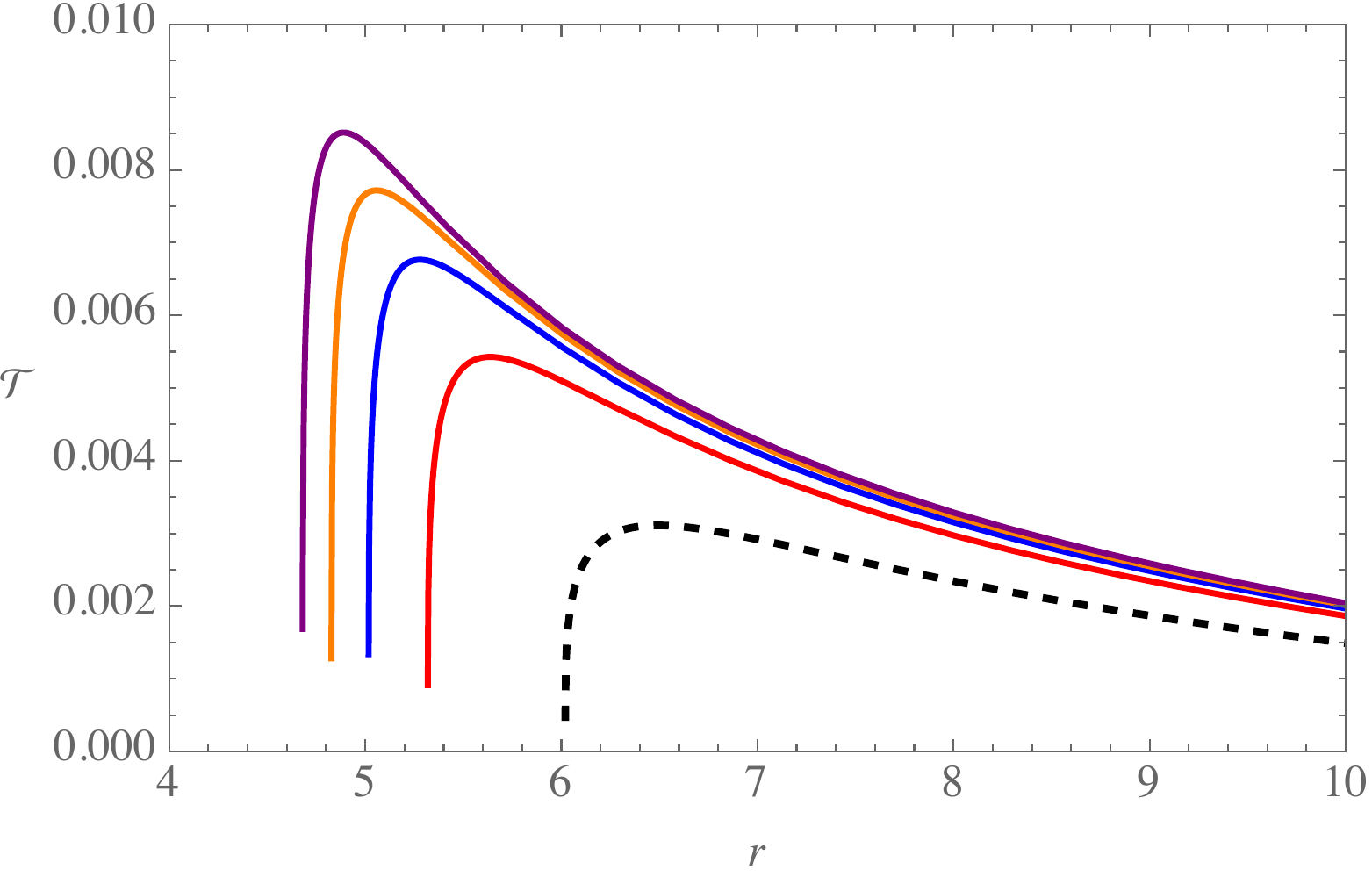}~(c)
    \caption{Radial profiles of (a) flux, (b) differential luminosity, and (c) temperature, for the {same values of the $\beta$-parameter} and  color-coding as in Fig.~\ref{fig:Veff_massive}.}
    \label{fig:FluxTempLum}
\end{figure}

For an accretion disk to be thin, its radius must be large compared to its thickness. In addition, the disk is considered to be in local hydrodynamical equilibrium at each point, which implies low pressure and vertical gradients within the disk. We assume that the cooling process in the disk is fast enough to prevent heat buildup due to particle friction. To ensure the stability of the disk, we assume that the accretion rate along the radial axis, $\mathscr{A}
^{r}$, is constant all the time, in the way that
\begin{equation}
\mathscr{A}^{r} = -2\pi\sqrt{-g}\,\Sigma\, \mathcal{U}^{r}=\mathrm{const.},
    \label{eq:Ar}
\end{equation}
in which $\sqrt{-g}=r^2$, $\Sigma$ is the surface density of the disk, and $\mathcal{U}^{r}=\dot{r}$ is the radial component of the four-velocity of the accreting particles. From the conservation of energy and angular momentum, one can obtain the differential of the luminosity as \cite{page_disk-accretion_1974,Joshi:2014}
\begin{equation}
\frac{\ed\ell}{\ed\ln r}=4\pi r\sqrt{-g}\,\mathcal{E}_c(r)\mathcal{F}(r),
    \label{eq:lum_diff}
\end{equation}
where $\mathcal{F}(r)$ is the flux of the radiated energy from the disk, and is given by
\begin{equation}
\mathcal{F}(r) = -\frac{\mathscr{A}^{r}}{4\pi\sqrt{-g}}\frac{\varpi'_c(r)}{\Big[\mathcal{E}_c(r)-\varpi_c(r)\mathcal{L}_c(r)\Big]^2}\int_{r_c}^{r}\Big[\mathcal{E}_c(r)-\varpi_c(r)\mathcal{L}_c(r)\Big]\mathcal{L}_c'(r)\,\ed r.
    \label{eq:flux_0}
\end{equation}
By considering the fact that $\mathcal{E}_c'(r)=\varpi_c(r)\mathcal{L}_c'(r)$, one can write \cite{page_disk-accretion_1974}
\begin{equation}
\int_{r_c}^{r}\Big[\mathcal{E}_c(r)-\varpi_c(r)\mathcal{L}_c(r)\Big]\mathcal{L}_c'(r)\,\ed r = \mathcal{E}_c(r)\mathcal{L}_c(r)-\mathcal{E}_c(r_c)\mathcal{L}_c(r_c)-2\int_{r_c}^r\mathcal{L}_c(r)\mathcal{E}'_c(r)\,\ed r.
 \label{eq:flux_0_int}
\end{equation}
Now taking the expressions in Eqs.~\eqref{eq:EISCO}, \eqref{eq:LISCO} and \eqref{eq:varpi_c} up to the first-order in $\Lambda$, and employing them in the integrand of the above relation, one can obtain the analytical solution 
\begin{equation}
\mathcal{F}(r) = -\frac{\mathscr{A}^{r}}{4\pi\sqrt{-g}}\frac{\varpi_c'(r)}{\Big[\mathcal{E}_c(r)-\varpi_c(r)\mathcal{L}_c(r)\Big]^2}\Big[
\mathcal{E}_c(r)\mathcal{L}_c(r)-\mathcal{E}_c(r_c)\mathcal{L}_c(r_c)
-2\mathscr{J}(r)+2\mathscr{J}(r_c)
\Big],
    \label{eq:flux_1}
\end{equation}
for the flux, for which $\mathscr{J}(r)$ has been given in appendix \ref{app:A}.  Since the disk is thin, we can assume that the emission follows the radiation of a black body whose temperature profile is given by
\begin{equation}
\mathcal{T}(r)^4=\frac{\mathcal{F}(r)}{\sigma},
    \label{eq:BBR_0}
\end{equation}
in which, $\sigma$ is the Stefan-Boltzmann constant. In Fig. \ref{fig:FluxTempLum}, the above relations have been employed to plot the radial profiles of the flux, temperature, and differential luminosity for different values of the $\beta$-parameter. It can be observed that by increasing $\beta$, all of the above quantities will increase, which implies that the more the black hole differs from the Schwarzschild-de Sitter solution, the more intense the radiation of the accretion disk is, and the disk's temperature becomes higher.  


\subsection{Shadow and rings of the black hole with thin accretion }\label{subsec:EmissionandRings}

To a distant observer, a real black hole appears as a dark shaded region surrounded by an illuminated area. This illuminated area is formed by light rays originating from various parts of the accretion disk, which have the potential to escape from the black hole. As discussed in Subsec. \ref{subsec:massless_motion}, certain photons with specific impact parameters can escape the black hole either through direct deflection or by following critical orbits (OFK and COFK) as shown in diagrams (a) and (c) of Fig.\ref{fig:orbits_0}. Consequently, incoming photons from the accretion disk may undergo different numbers of orbits around the black hole before exiting, giving rise to several light rings that confine the shadow.

\subsubsection{Direct emission, lensing rings and photon rings}\label{subsubsec: DirectLensedPhoton}

\begin{figure}[t]
    \centering
    \includegraphics[width=5.4cm]{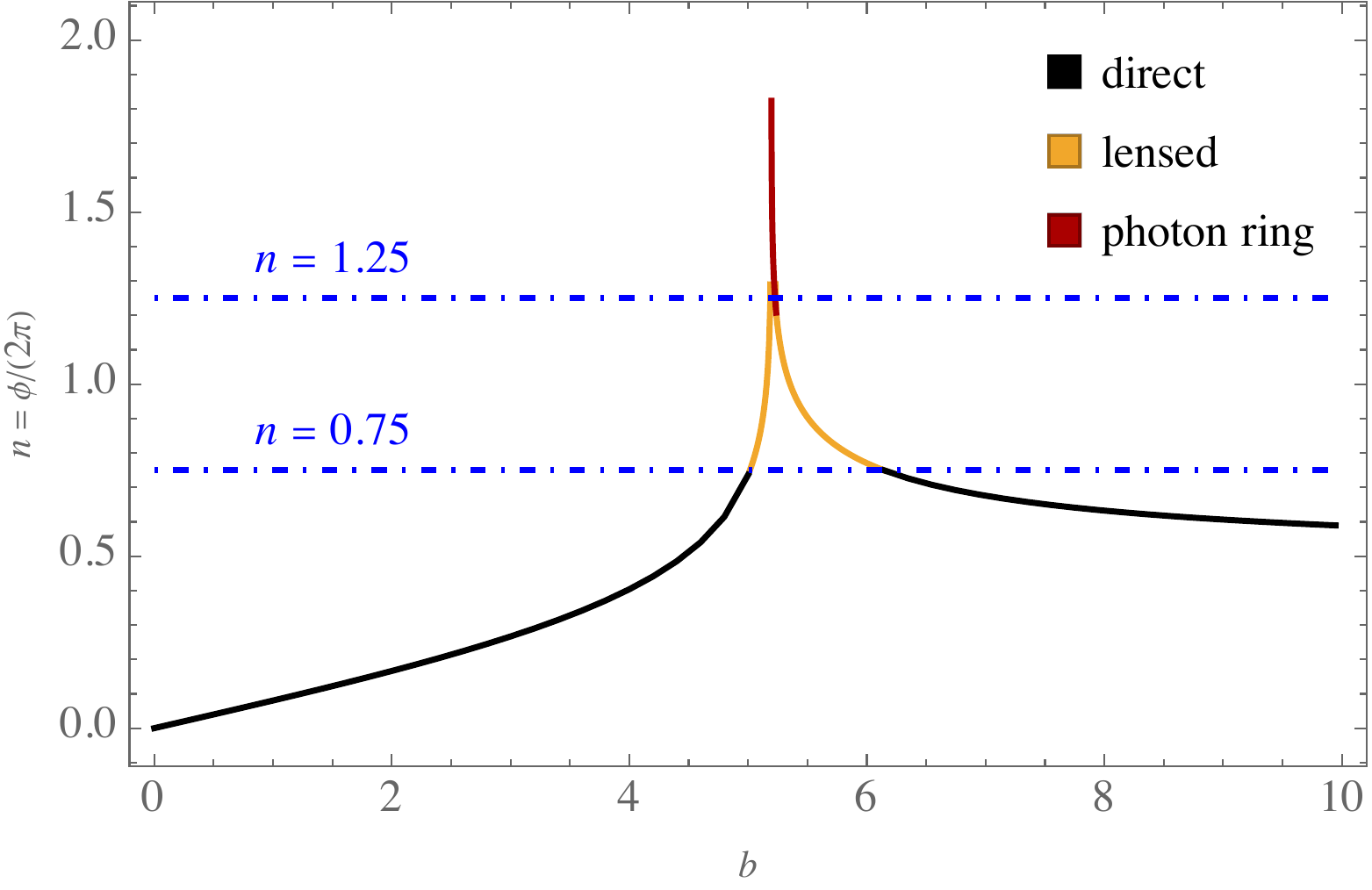}~(a)
    \includegraphics[width=5.4cm]{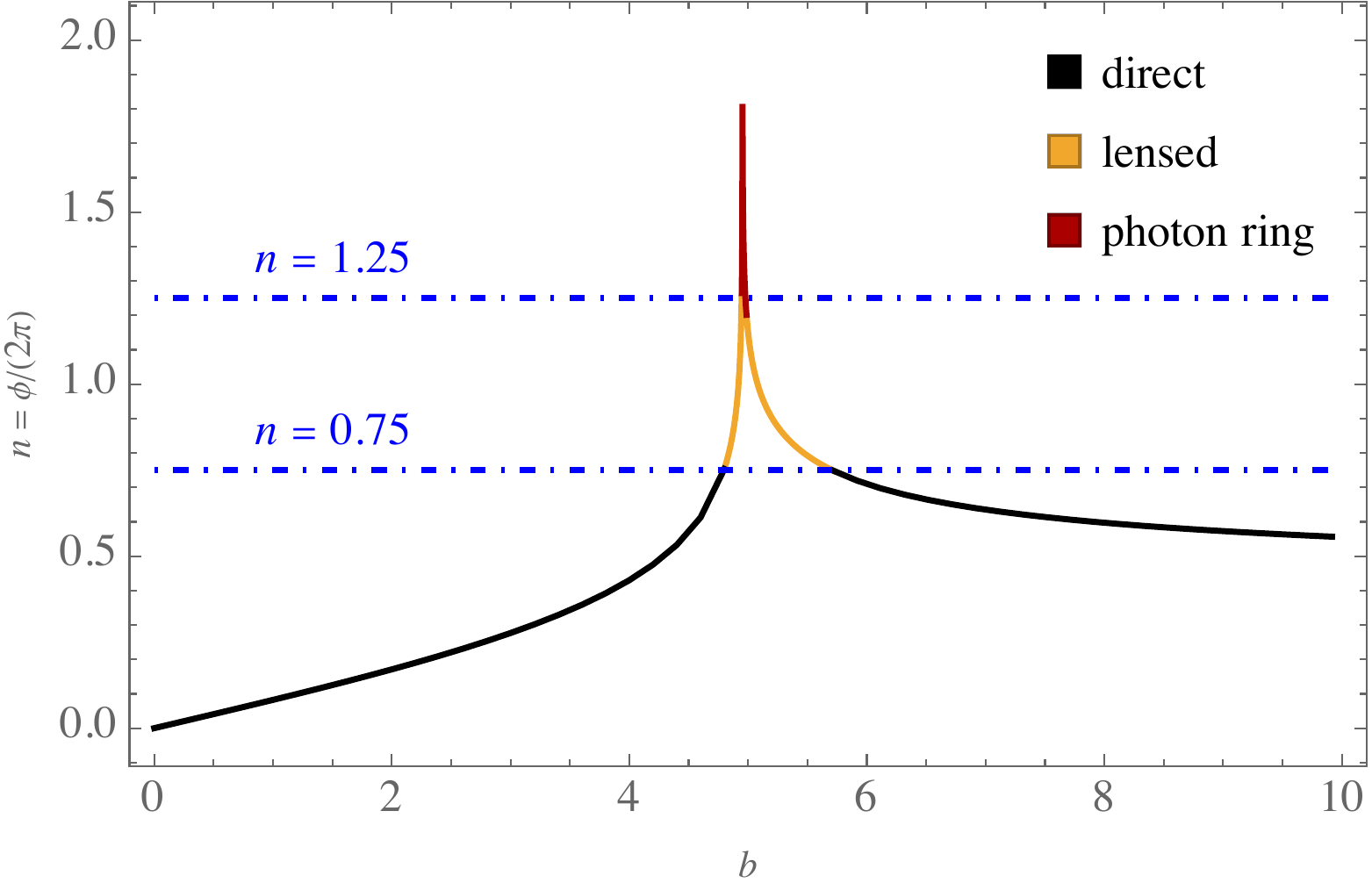}~(b)
    \includegraphics[width=5.4cm]{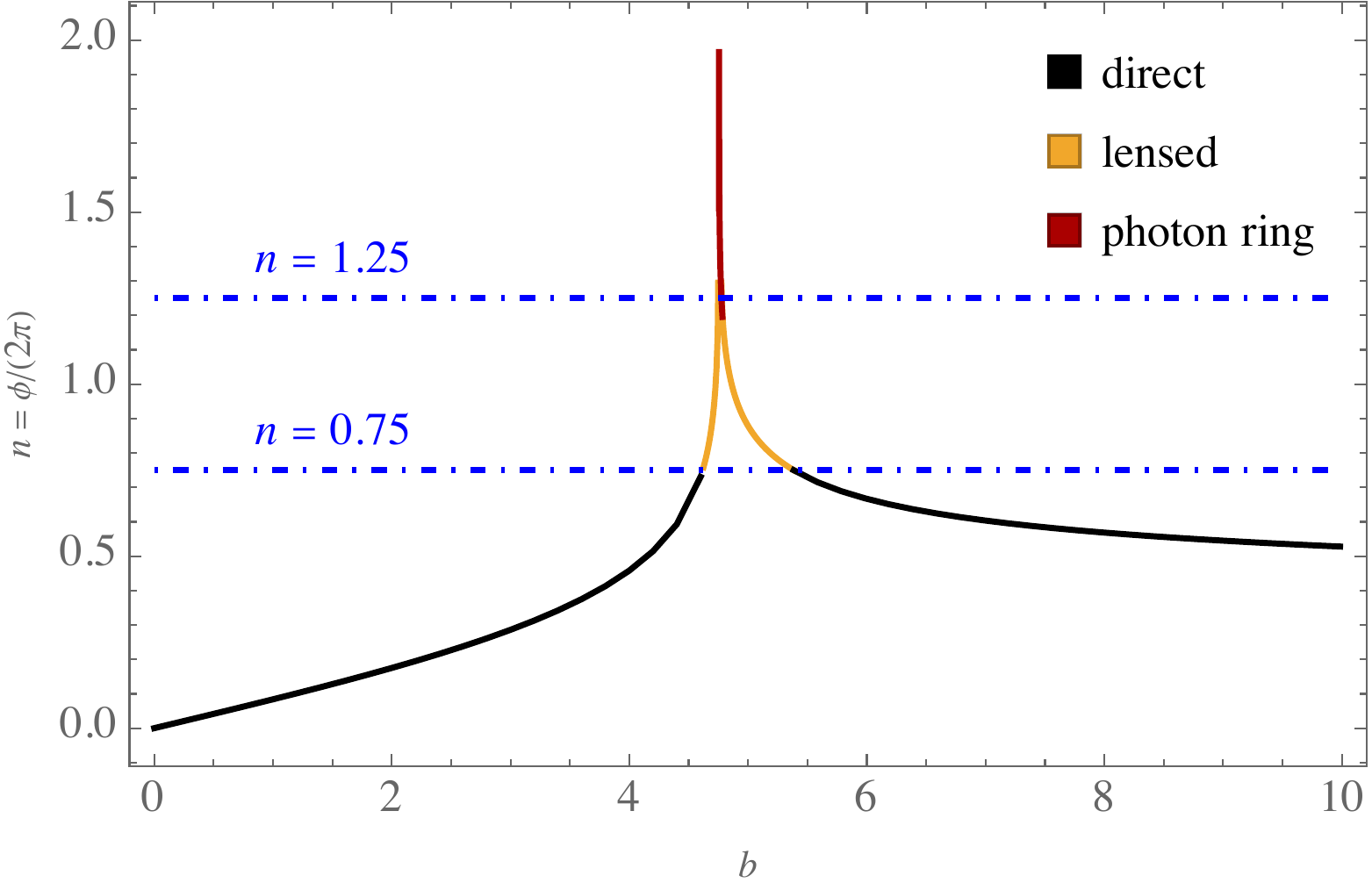}~(c)
    \includegraphics[width=5.4cm]{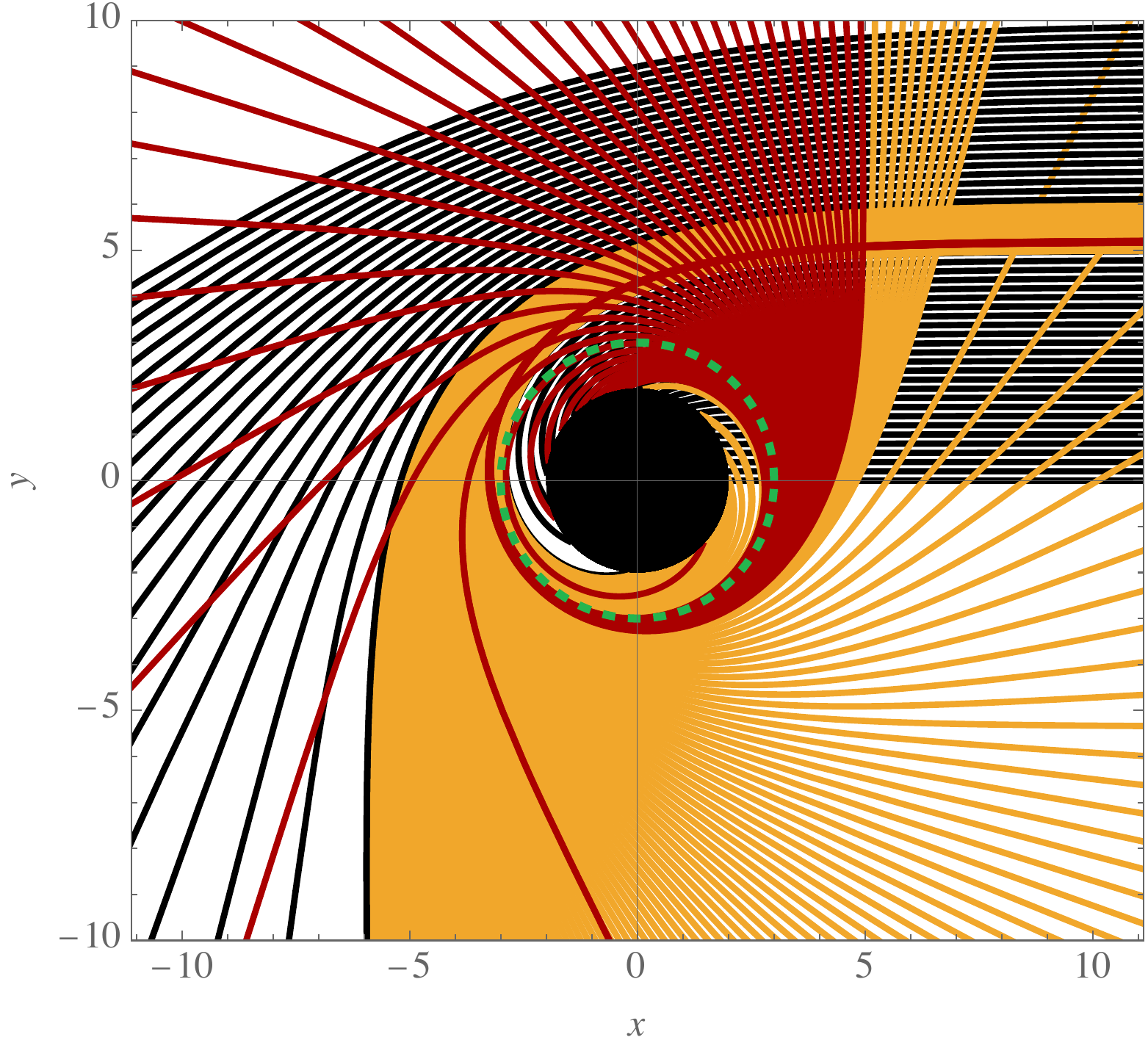}~(d)
    \includegraphics[width=5.4cm]{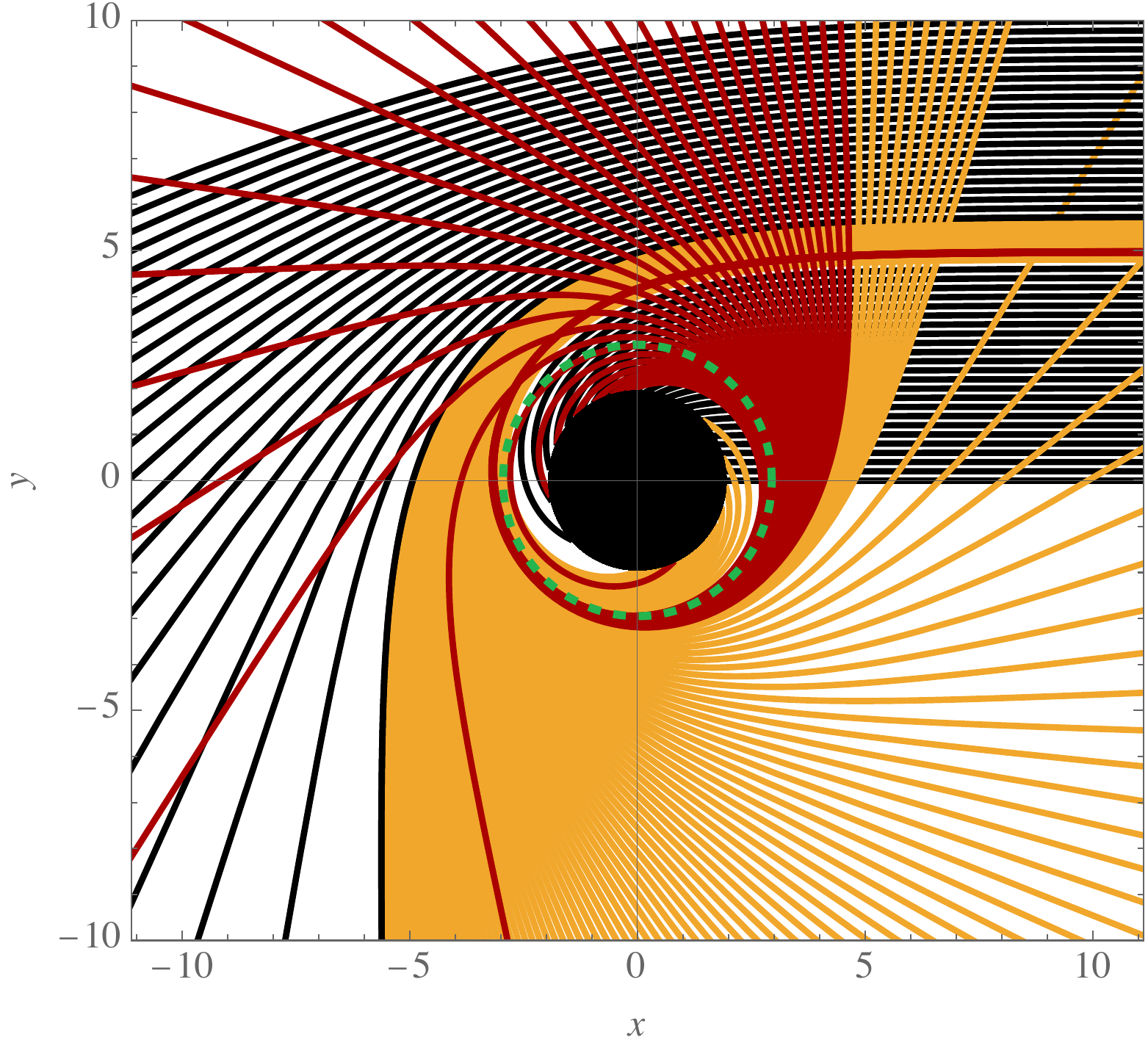}~(e)
    \includegraphics[width=5.4cm]{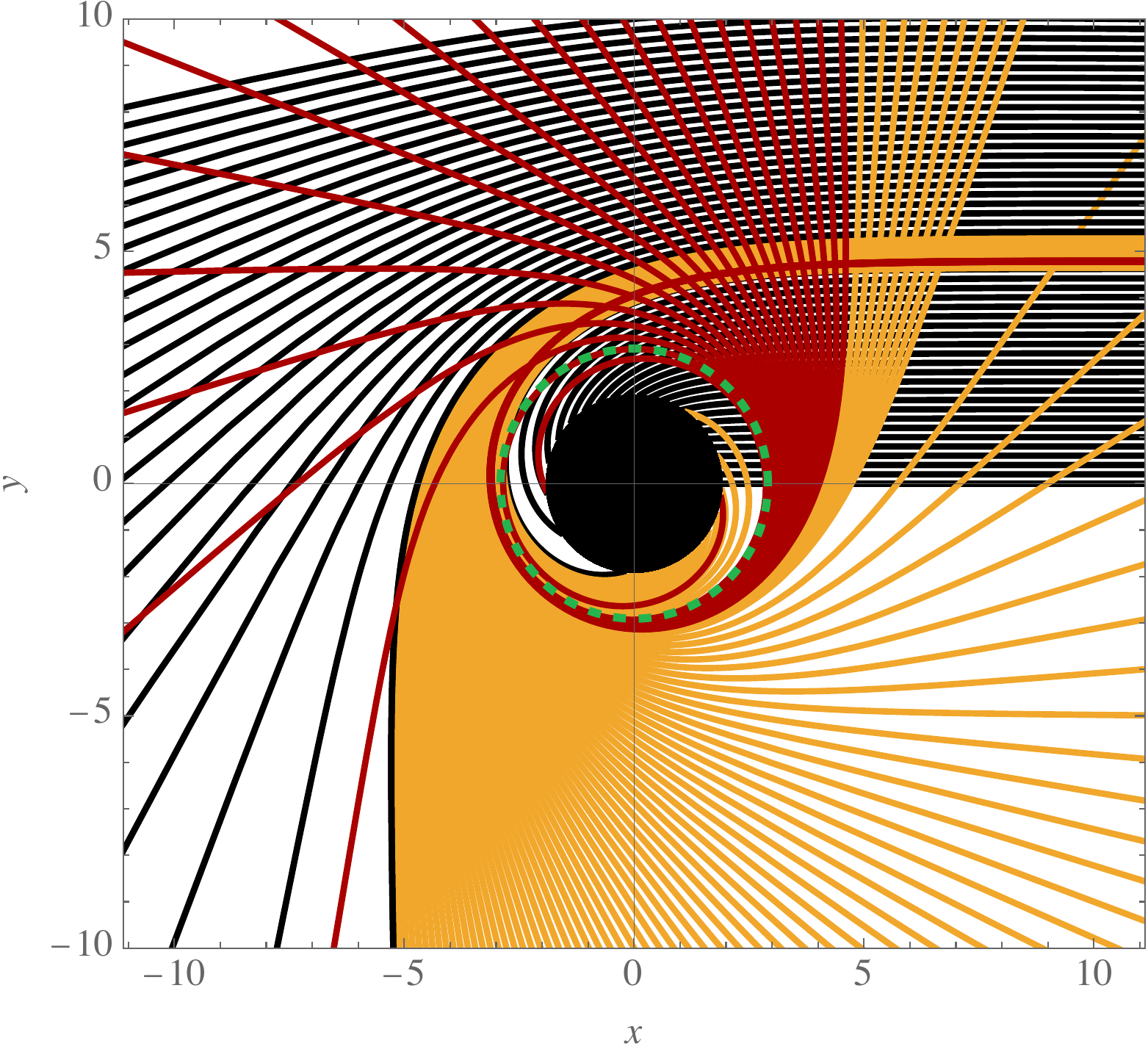}~(f)
    \includegraphics[width=5.4cm]{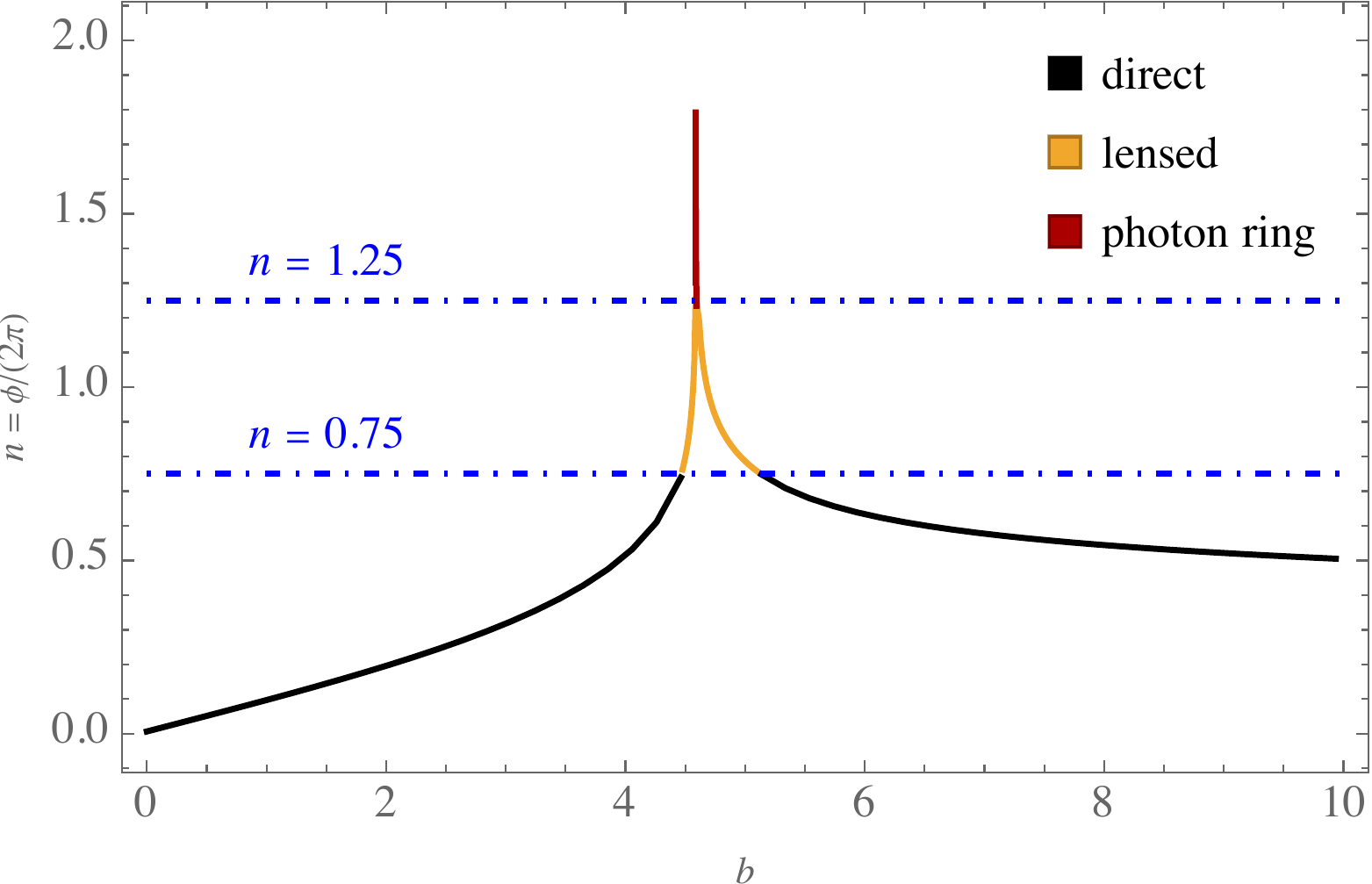}~(g)
     \includegraphics[width=5.4cm]{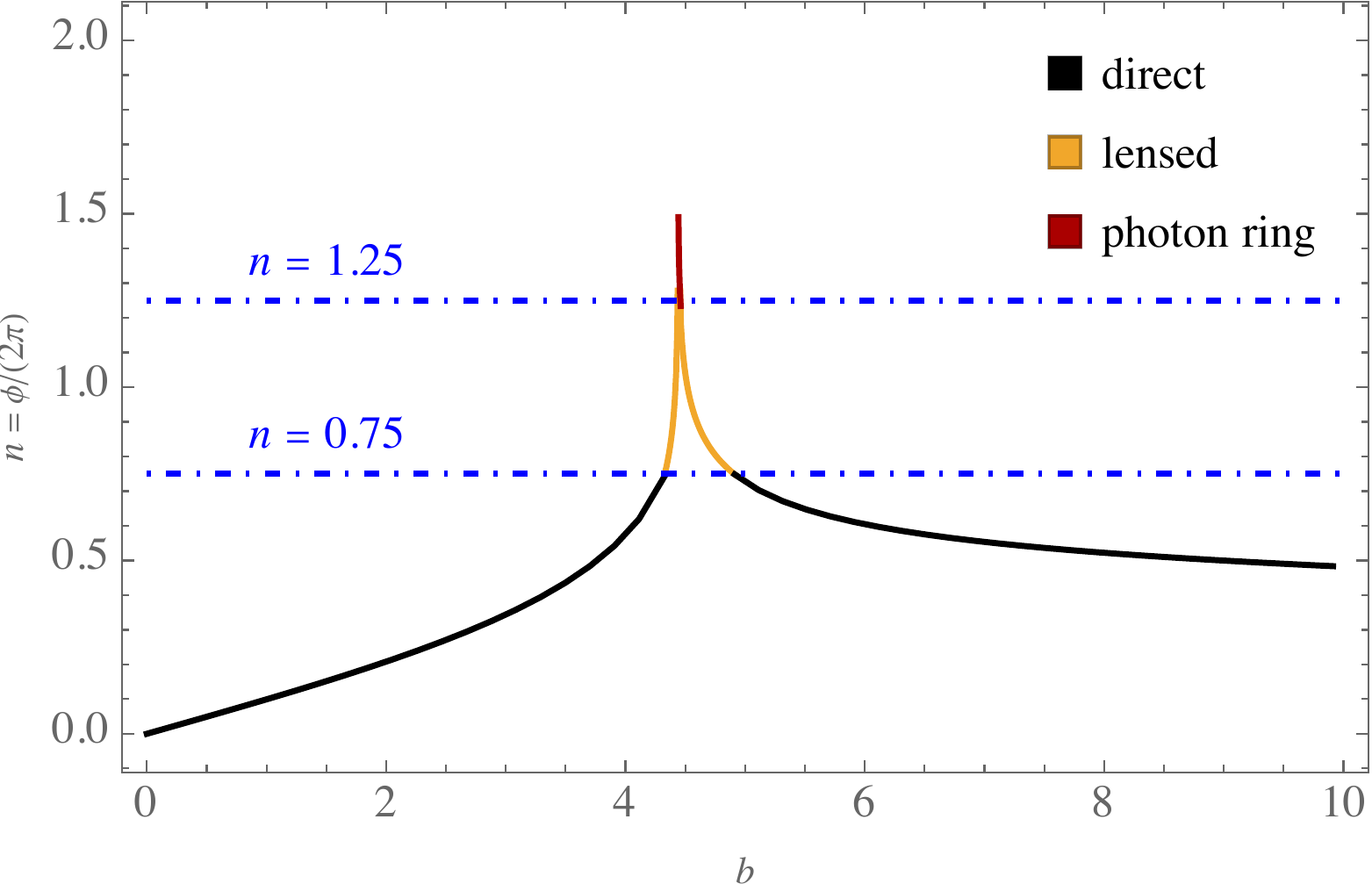}~(h)\qquad\qquad
    \includegraphics[width=5.4cm]{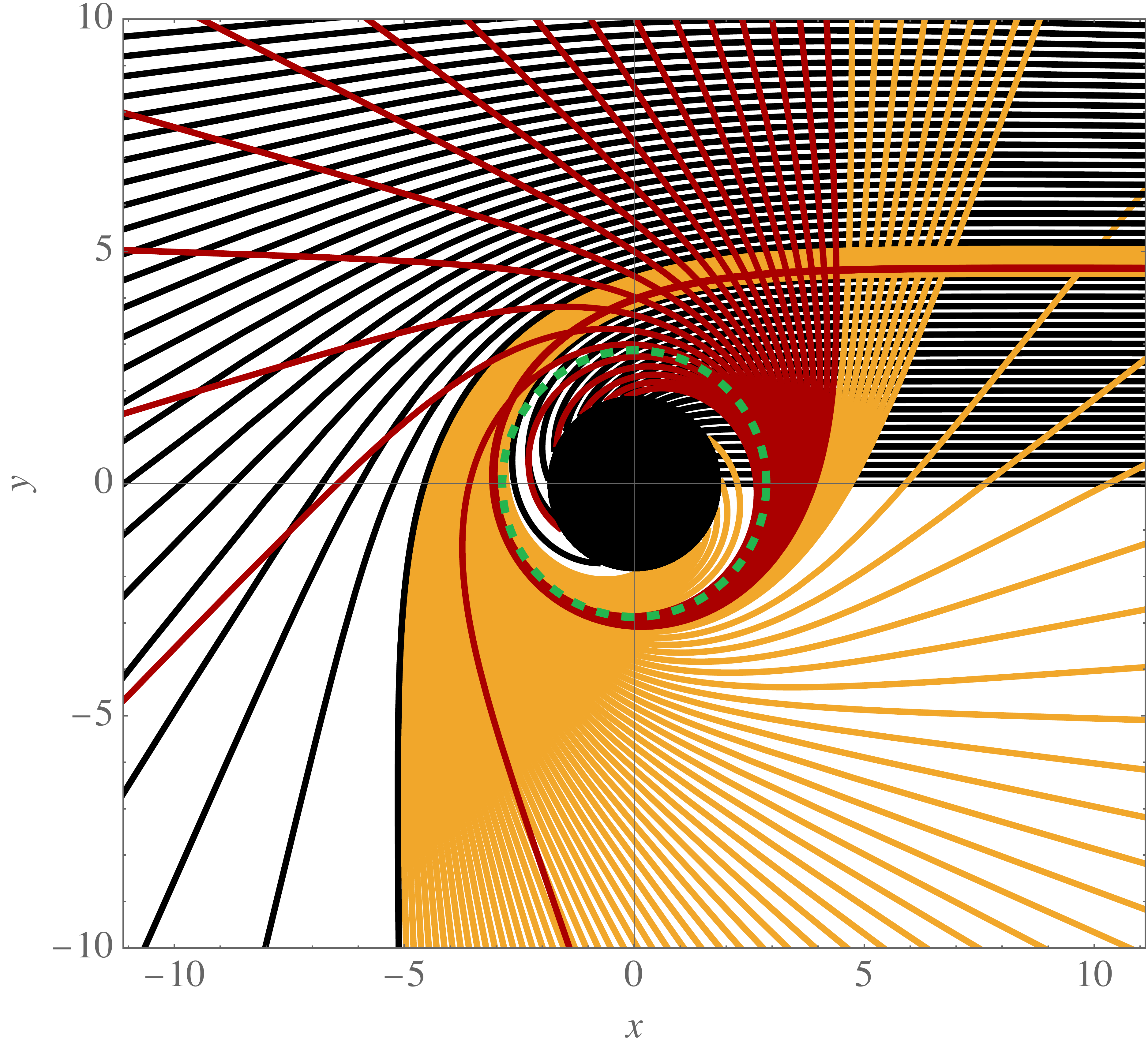}~(i)
       \includegraphics[width=5.4cm]{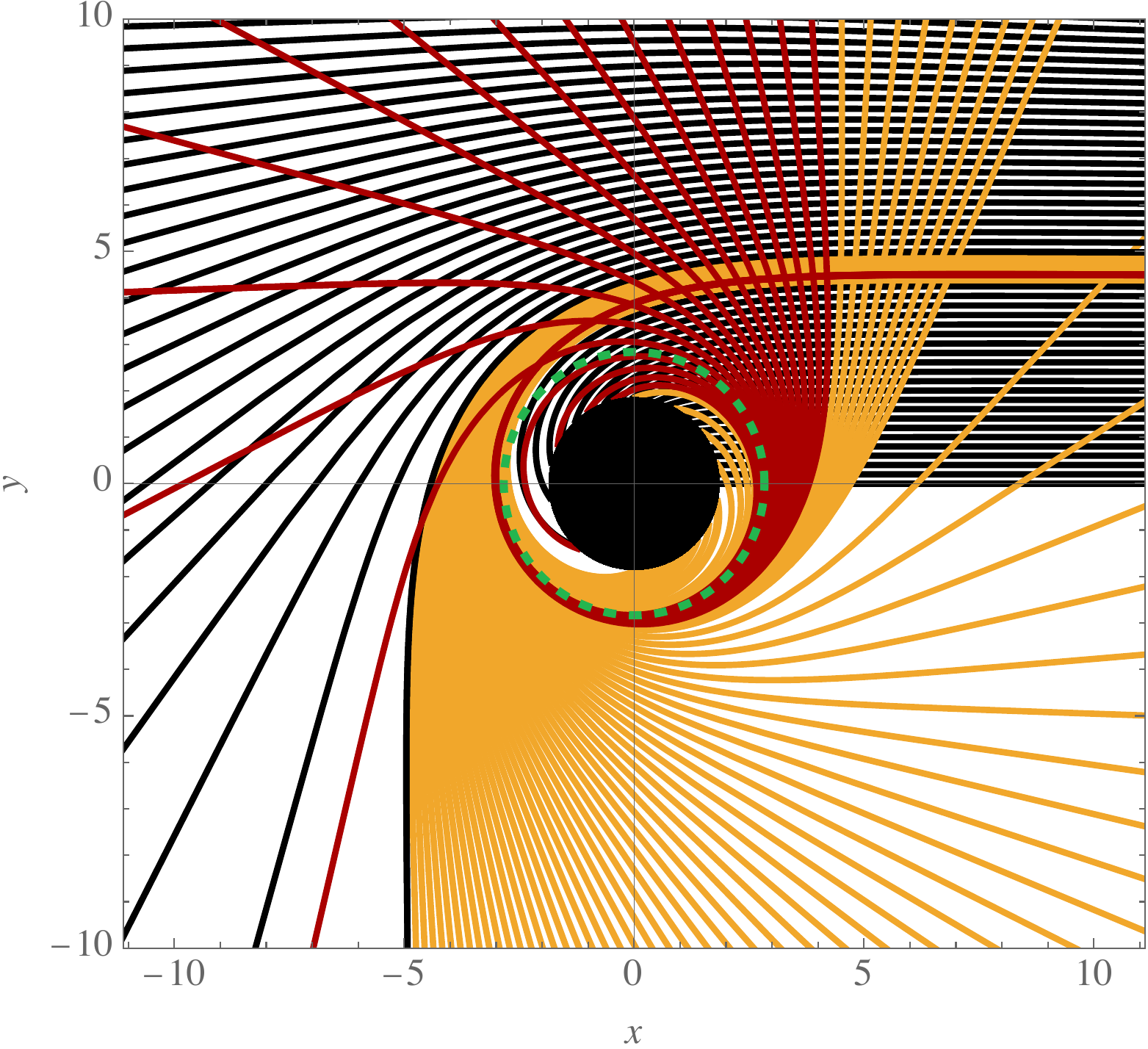}~(j)
    \caption{The $b$-profile of the total number of photon orbits $n$, together with the behavior of the null geodesics in the near-horizon regions. In these diagrams, the direct, lensing ring and photon ring emissions have been color-coded appropriately. The black disk indicates the event horizon of the black hole whereas the green dashed circle denotes the radius of unstable (critical) orbits, $r_p$. The diagrams correspond to the cases of (a,d) $\beta = 0$, (b,e) $\beta = 0.011$, (c,f) $\beta = 0.022$, (g,i) $\beta=0.031$, and (h,j) $\beta = 0.041$.}
    \label{fig:n-b}
\end{figure}

Here, we follow the method introduced in Ref.~\cite{Gralla:2019}, to characterize the light rings in the observer's sky, where the number of orbits is defined as 
\begin{equation}
n=\frac{\phi}{2\pi},
    \label{eq:n_0}
\end{equation}
in which, $\phi$ represents the final azimuth angle of photons just before they escape the black hole. The parameter $n$ corresponds to the number of times that the light ray geodesics cross the plane of the accretion disk. In Ref. \cite{Gralla:2019}, these rays were classified into three cases: $0.25 < n < 0.75$, where the light rays intersect the accretion disk only once, forming the direct emission; $0.75 < n < 1.25$, in which the rays cross the accretion disk twice, generating the lensed (lensing) ring; and $n > 1.25$, corresponding to the formation of the photon ring, where the rays intersect the accretion disk more than twice. Fig. \ref{fig:n-b} illustrates the behavior of $n$ with respect to the impact parameter $b$ for different values of the $\beta$-parameter, with distinct colors indicating the domains of direct emissions, lensing rings, and photon rings. 
%
\begin{table}[h]
\centering
 \begin{tabular}{c  c c c} 
 \hline
 
 $\beta$  & Direct emission $(0.25>n>0.75)$ & Lensing ring $(0.75<n<1.25)$ & \qquad Photon ring $(n>1.25)$ \\ [0.5ex] 
 \hline\hline
 
 0.0 & $b<5.01685$; $b>6.14685$ & $5.01685<b< b_p$; $5.23685<b<6.14685$ & \qquad$b_p<b< 5.23685$ \\ 
 0.011 & $b<4.80681$; $b>5.71683$ & $4.80681<b<b_p$; $4.98682<b<5.71683$ & \qquad$ b_p<b<4.98682$ \\
 0.022 & $b<4.62449$; $b>5.38449$ & $4.62449<b<b_p$; $4.78449<b<5.38449$ & \qquad$b_p<b<4.78449$ \\
 0.031 & $b<4.47621$; $b>5.13621$ & $4.47621<b<b_p$; $4.61621<b<5.13621$ & \qquad$b_p<b<4.61621$ \\
 0.041 & $b<4.33281$; $b>4.91281$ & $4.33281<b<b_p$; $4.46281<b<4.91281$ & \qquad $b_p<b<4.46281$ \\ [1ex] 
\hline
 \end{tabular}
 \caption{The impact parameter domains corresponding to the direct emission, lensing rings and photon rings of the black hole given for different values of the $\beta$-parameter.}
 \label{table:Table2}
\end{table}
In this diagram, a large number of geodesics have been simulated for each case, including the OFK, OSK, COFK, and COSK. Notably, for the remainder of this section, we assume the value of $\Lambda = 10^{-8}$, which does not significantly alter the photon orbit properties or the characteristic distances of the spacetime but is essential for proper functioning of our ray-tracing codes. As seen from the diagrams, increasing $b$ leads to an increase in the total number of orbits within the domain $b < b_p$ until it reaches a narrow peak, after which it decreases within the domain $b > b_p$. On the other hand, for larger values of the $\beta$-parameter, the width of the lensing rings and photon rings shrinks. In Table~\ref{table:Table2}, this has also been shown numerically by writing down the range of $b$ for the direct, lensing ring, and photon ring emissions, for different values of $\beta$. According to the data presented in this table, it can be checked that by an increase in the $\beta$-parameter, the range of $b$ for all emission types is shrunk. Therefore, the thickness of the photon and lensing rings is decreased in this sense. Accordingly, the angular size of the shadow is also decreased for larger $\beta$, and hence, the contribution to the brightness of the rings is reduced. 

We continue by studying the observed emission intensity from the thin accretion disk in the framework of the $f(R)$ black hole model.

\subsubsection{Transfer functions and the observed intensities}\label{subsubsec:TransferIobs}

\begin{figure}[t]
    \centering
    \includegraphics[width=5.4cm]{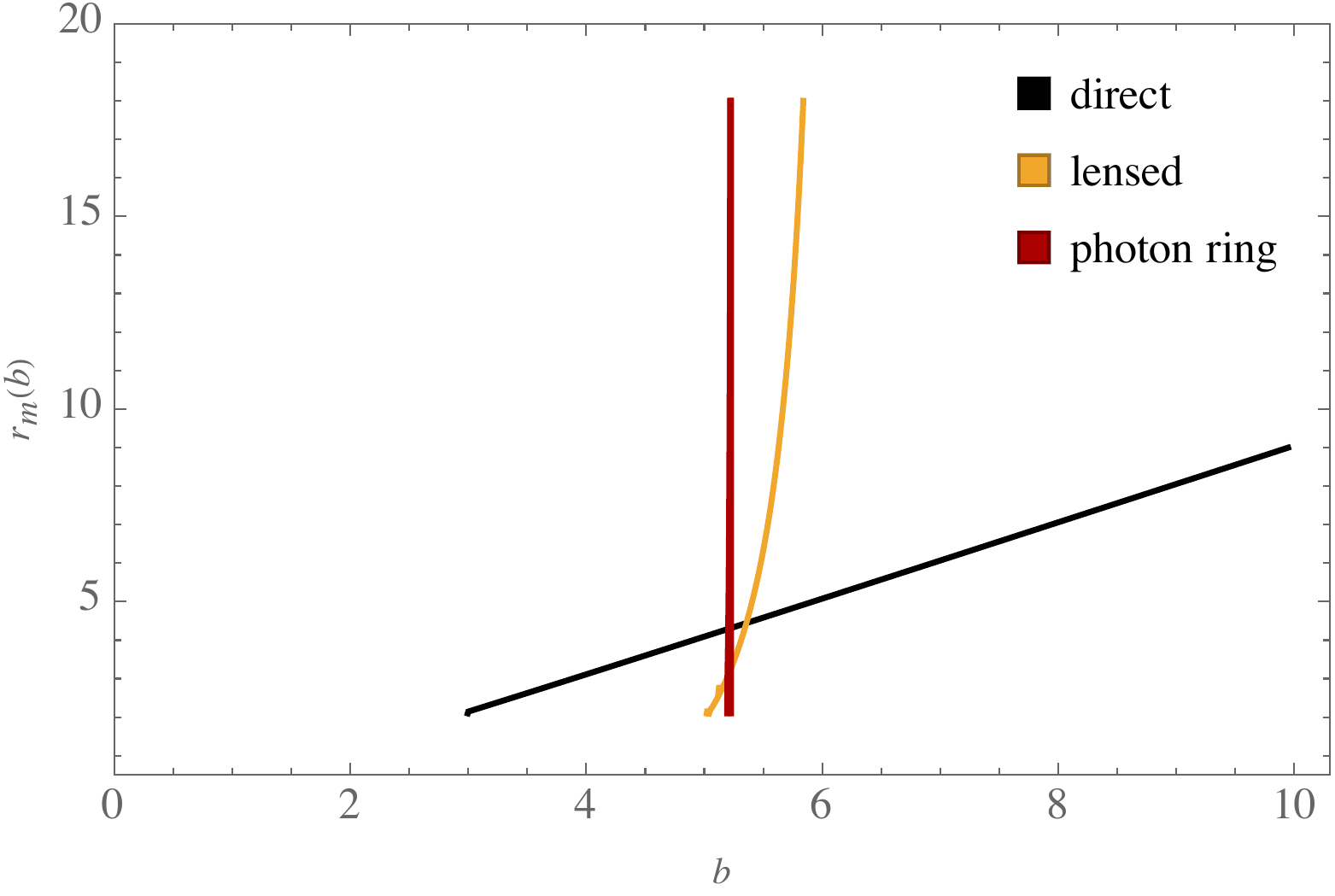}~(a)
      \includegraphics[width=5.4cm]{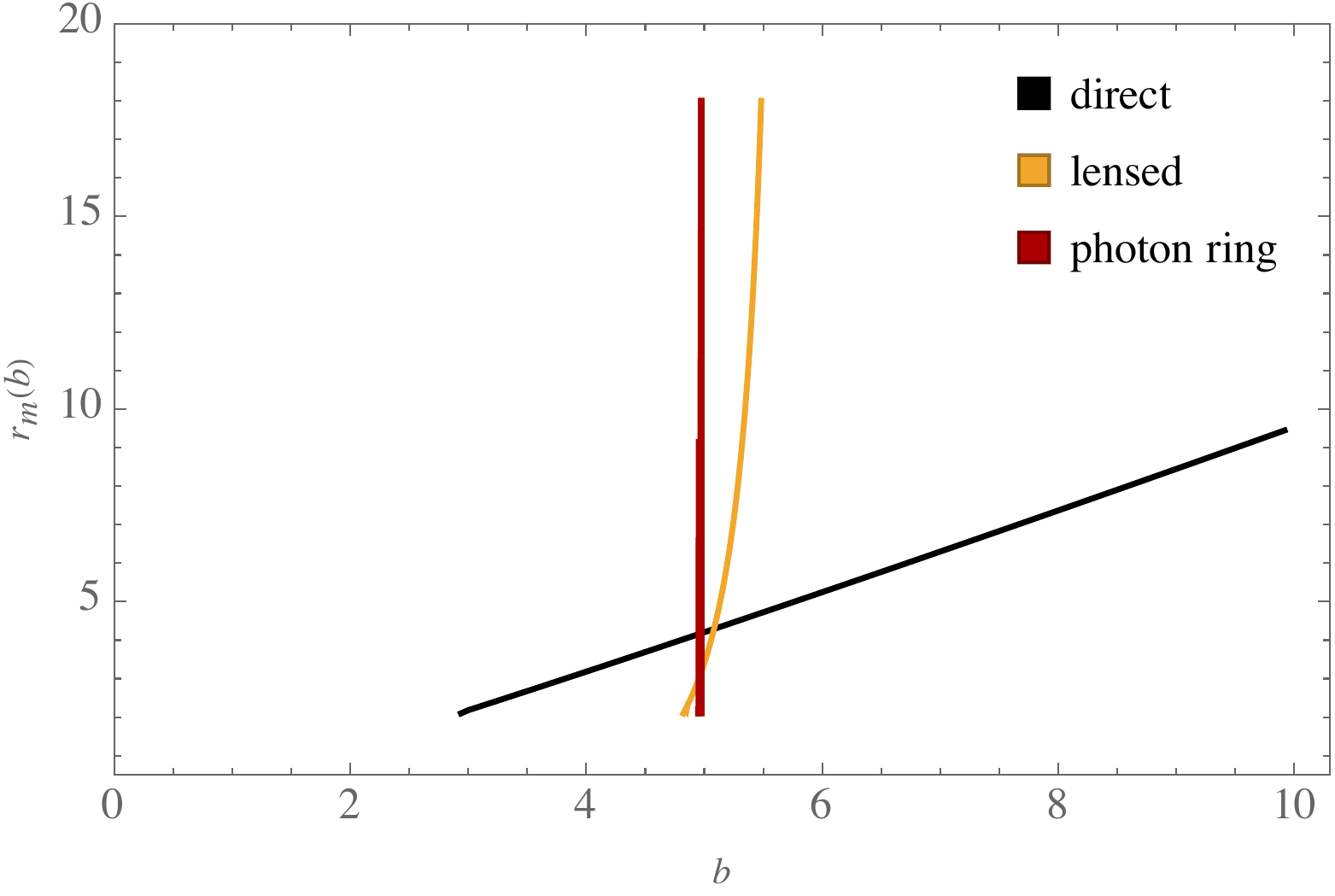}~(b)
        \includegraphics[width=5.4cm]{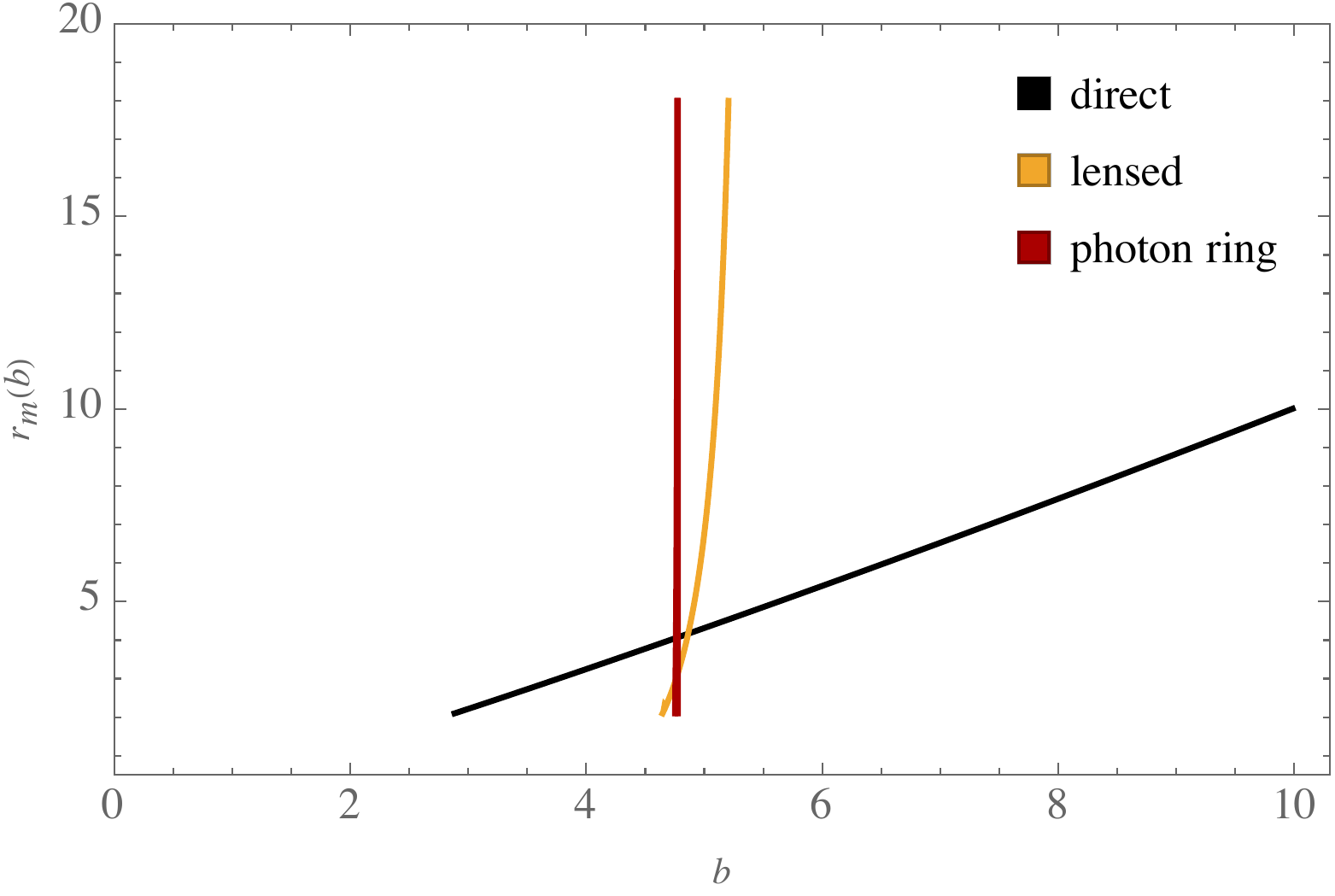}~(c)
          \includegraphics[width=5.4cm]{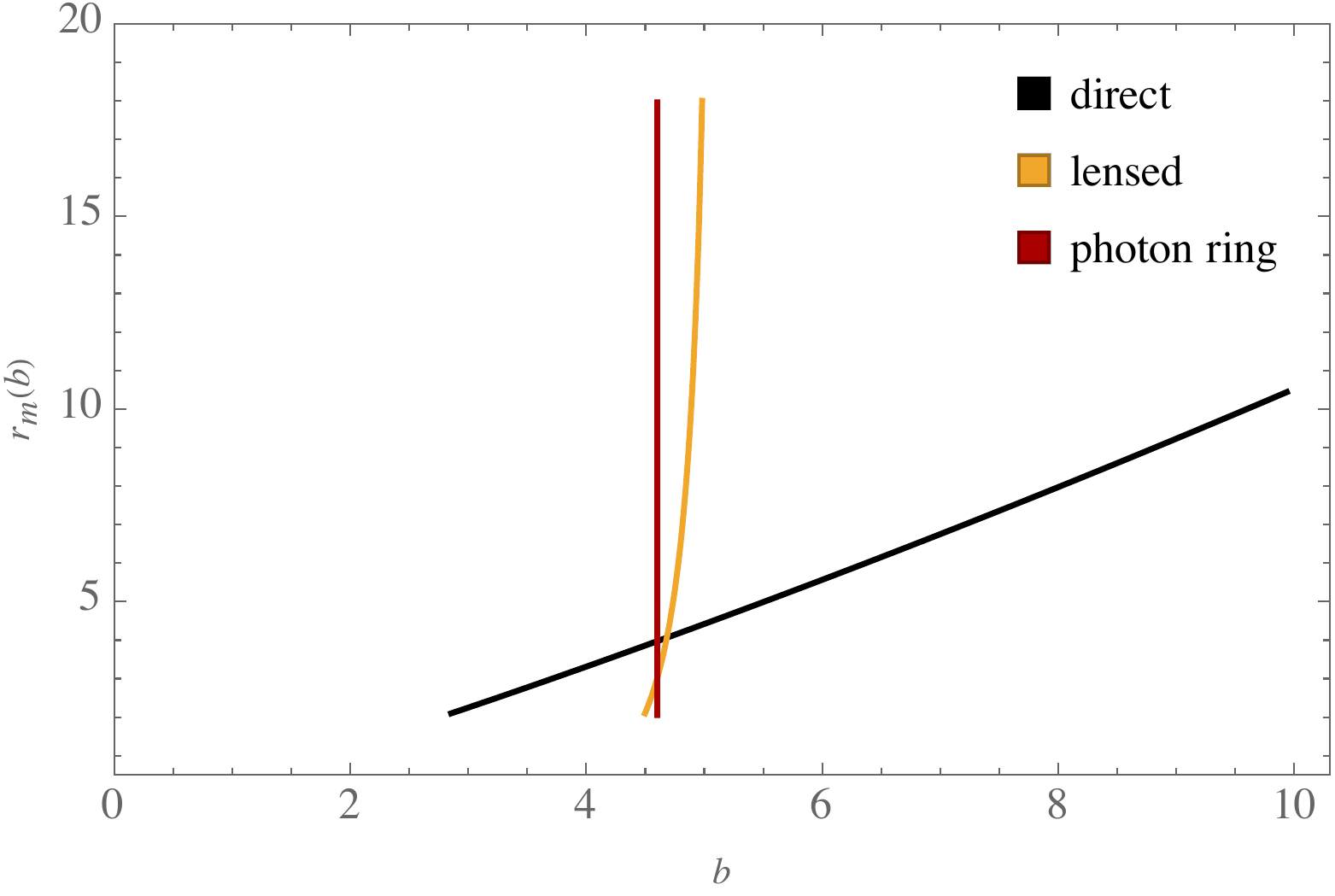}~(d)
            \includegraphics[width=5.4cm]{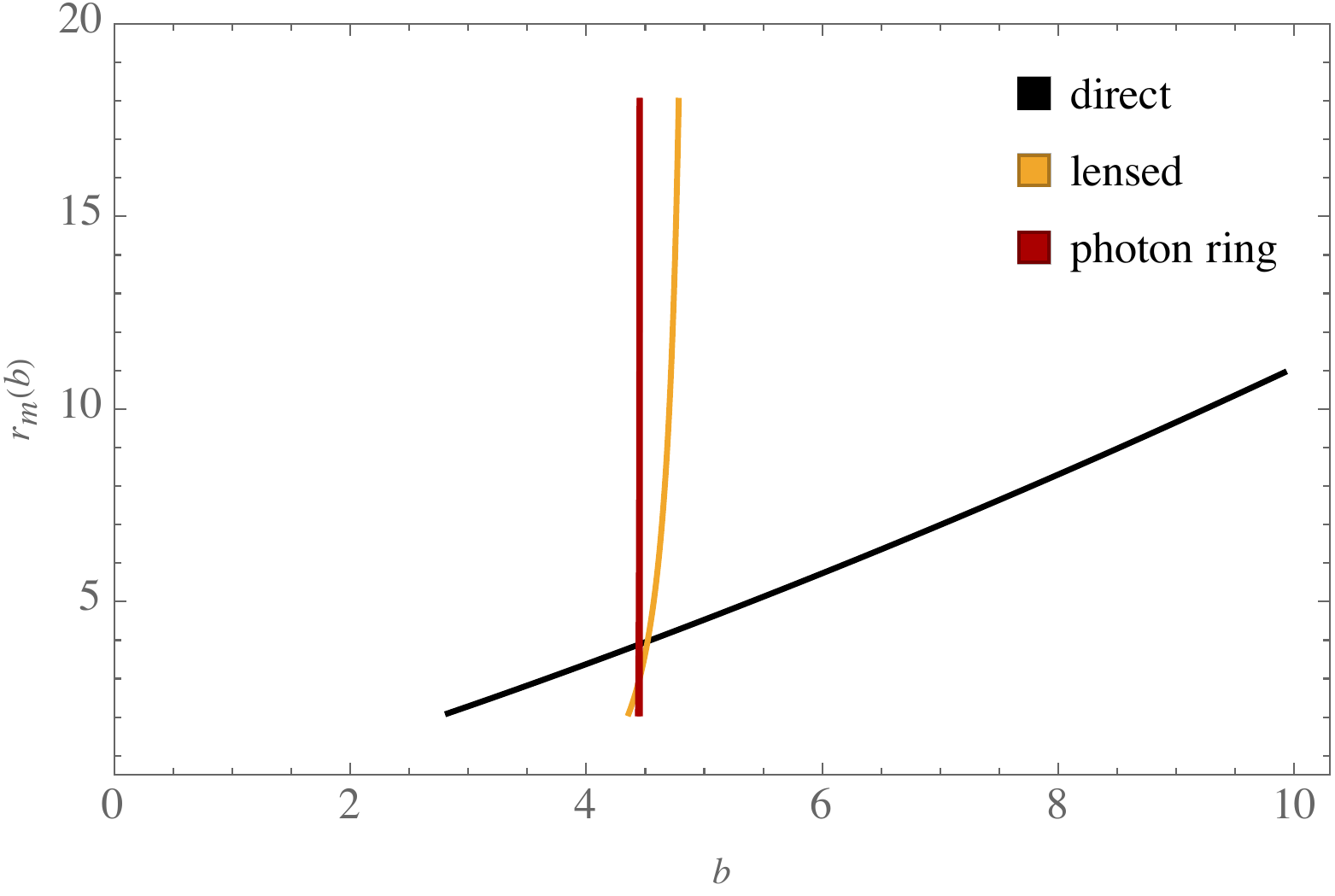}~(e)
    \caption{The transfer function $r_m(b)$ plotted for different values of $\beta$. The panels (a)--(e) correspond respectively to the cases of $\beta=0.011, 0.022, 0.031$ and $0.041$. The color coding is the same as that in Fig.~\ref{fig:n-b}.}
    \label{fig:transfer}
\end{figure}

The radiation of the accretion disk is supposed to be isotropic in its rest frame. By $I_\e(r)$, we denote the specific intensity of an emitted radiation of frequency $\nu_\e$ from the disk. From Liouville's theorem, we know that the quantity ${I_\e(r)}/{\nu^3_\e}$ is conserved along the entire path of light propagation. Hence, the observed intensity $I_\oo$ of frequency $\nu_\oo$ are related in terms of the relation ${I_\e(r)}/{\nu^3_\e}={I_\oo(r)}/{\nu^3_\oo}$ \cite{bromley_line_1997}. Accordingly, we have
\begin{equation}
I_\oo(r)= \mathfrak{h}^3 I_\e(r),
    \label{eq:Iobs_0}
\end{equation}
which in our model $\mathfrak{h}=\sqrt{B(r)}$. Now by integrating over the range of all the observed frequencies, the total observed specific intensity is obtained as 
\begin{equation}
I_\oo^t(r)=\int I_\oo(r)\,\ed\nu_\oo=\mathfrak{h}^4 I_\emit(r),
    \label{eq:Iobs_1}
\end{equation}
in which, the total emission intensity is given by $I_\emit(r) = \int I_\e(r) \ed\nu_\e$. Note that, since each intersection of the light rays with the accretion disk generates an additional brightness, the reliable total observed intensity of the direct emission and the rings is given by 
\begin{equation}
I_\obs(r) = \sum_m I_\oo^t(r)|_{r=r_m(b)},
    \label{eq:Iobs_2}
\end{equation}
where $r_m(b)$ is the transfer function that relates the impact parameter of the light ray trajectories with the radial coordinate of the $m$th intersection of light rays with the accretion disk\footnote{The number of intersections $m$ and the number of orbits $n$, are related as $n={m}/{2}-{1}/{4}$ \cite{hu_observational_2022}.}. Hence, the slope of the transfer function indicates its (de)magnification scale  \cite{Gralla:2019,zeng_influence_2020}. Therefore, this slope is called the (de)magnification factor. In Fig.~\ref{fig:transfer}, we have demonstrated the $b$-profile of the transfer function for different values of $\beta$.  
In these diagrams, the black line with an approximately constant slope corresponds to the case of $m=1$ and represents direct emission. The nearly unit slope indicates a redshifted source. For the lensing ring with $m=2$, the impact parameter $b_p$ is approached, but the slope increases significantly from one, indicating demagnification of the back side image of the accretion disk. In the case of $m=3$ and the formation of the photon ring, the slope tends to infinity, demonstrating an extreme demagnification of the front side image of the accretion disk. Based on these observations, one can infer that the contribution of the lensing ring and photon ring in the observed intensity is negligible compared to direct emission. It is noteworthy that higher-order rings with $n\geq 4$ (black hole subrings) do not possess significant observational features, although they have been found to produce certain interferometric signatures \cite{johnson_universal_2020}.

\subsubsection{Observational signatures of emissions from the accretion disk}\label{subsubsec:ObservationalSignatures}

In this part of the paper, we employ a ray-tracing procedure to generate the black hole shadow along with its accretion disk image. We adopt a face-on view, which provides greater generality and informative insights into the silhouette imaging of black holes.

\begin{figure}
    \centering
    \includegraphics[width=5.4cm,height=4.4cm]{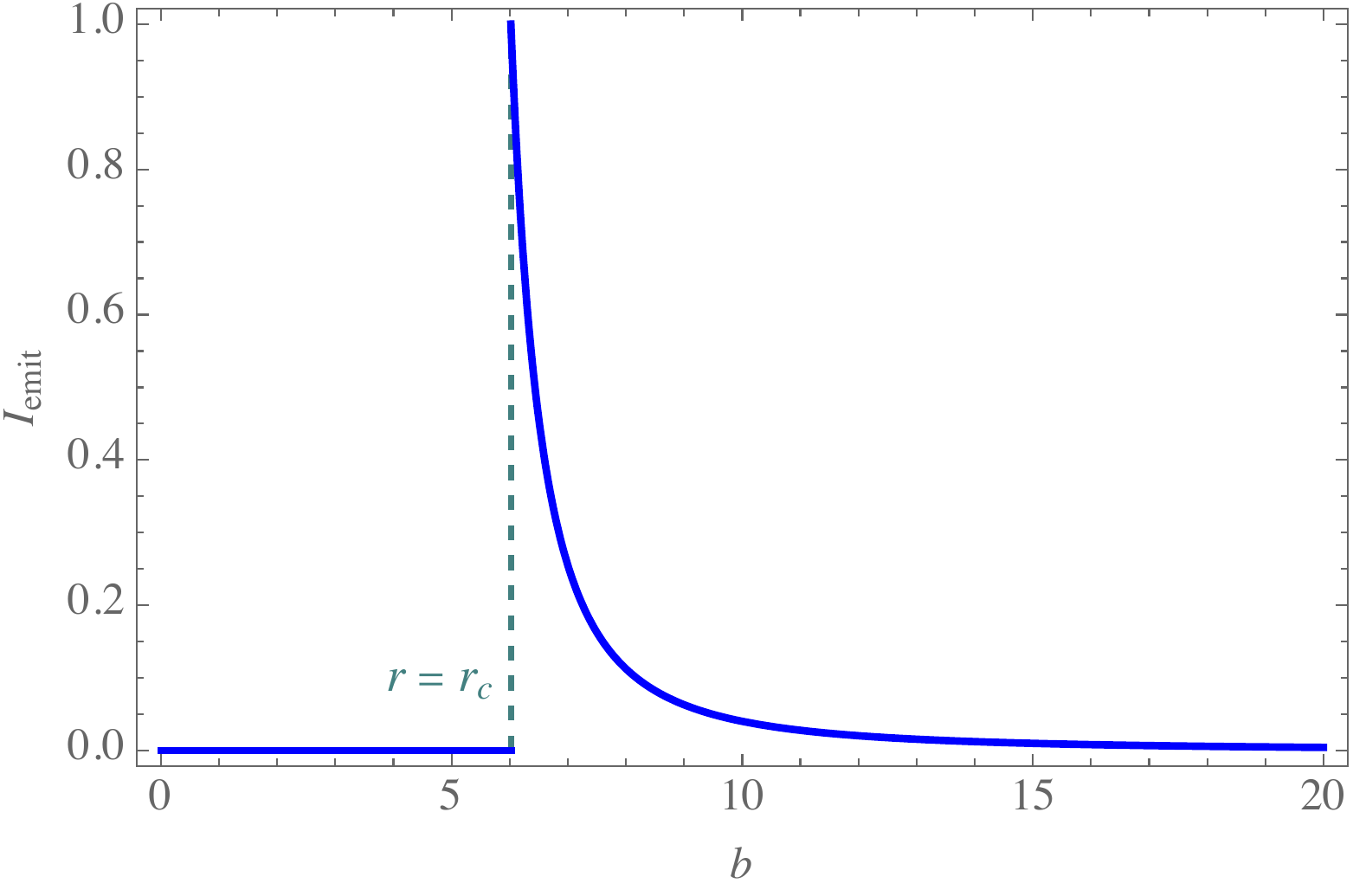}
    \includegraphics[width=5.4cm,height=4.4cm]{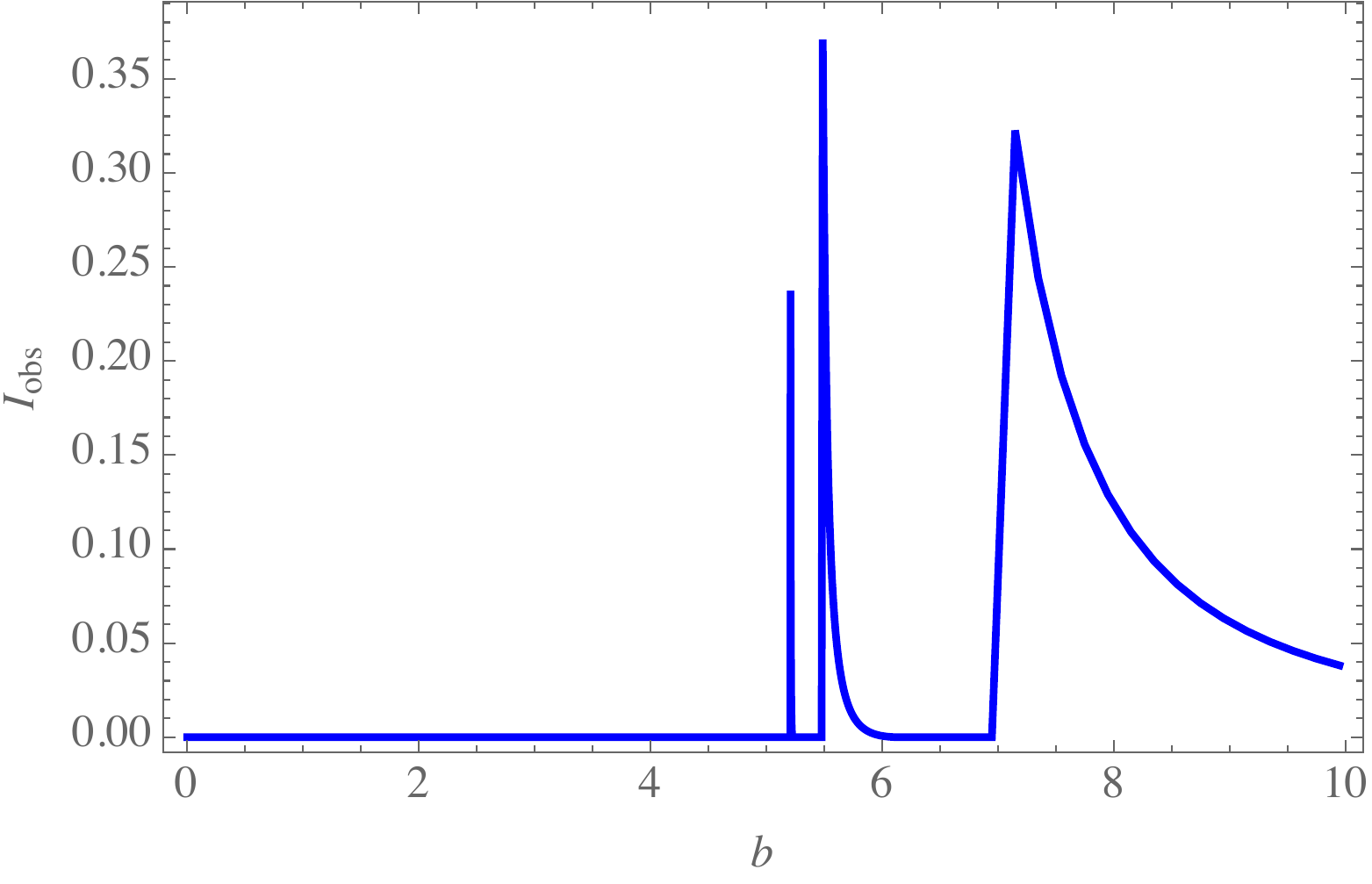}
    \includegraphics[width=5.4cm,height=4.5cm]{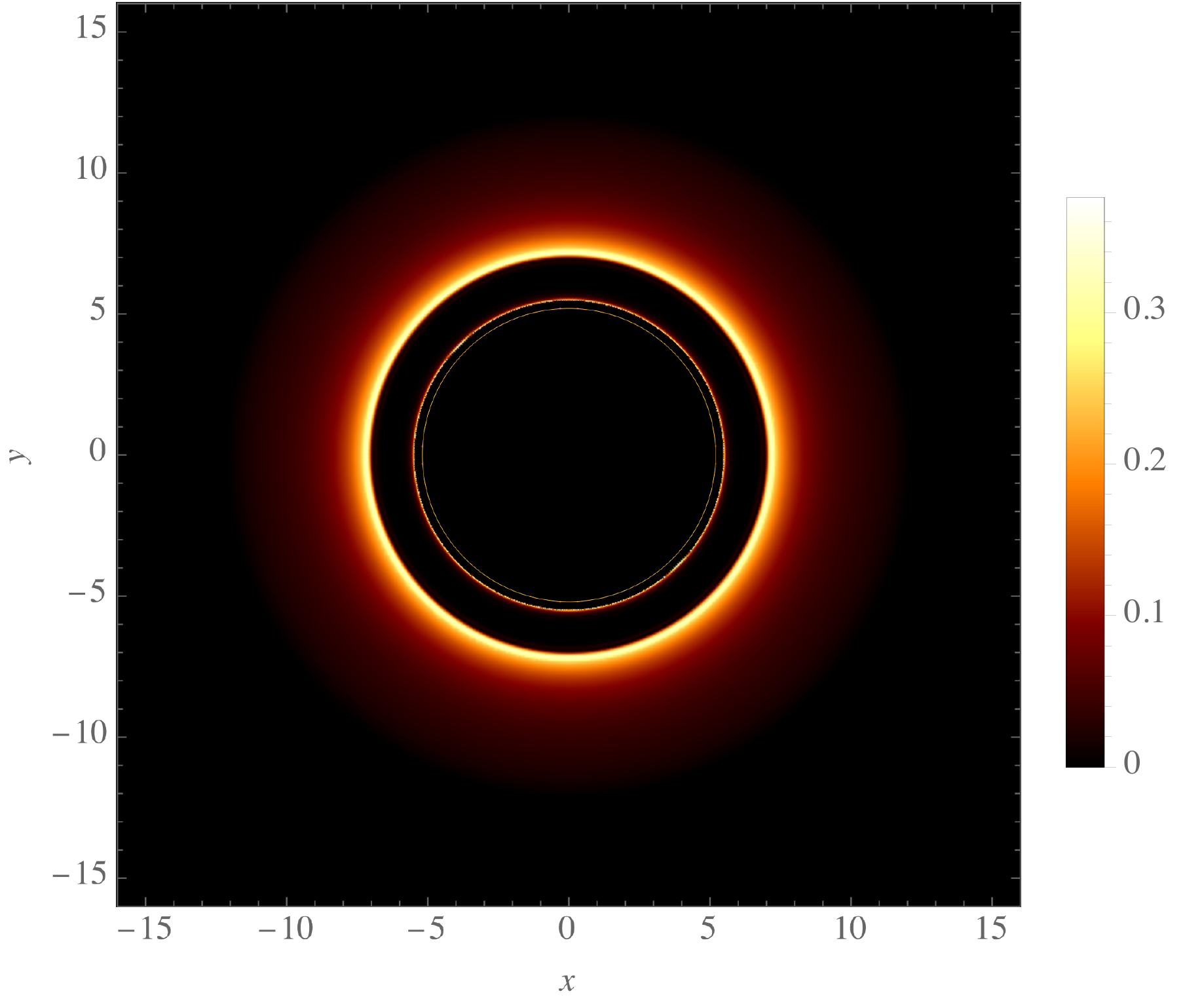}~(a)
    \includegraphics[width=5.4cm,height=4.4cm]{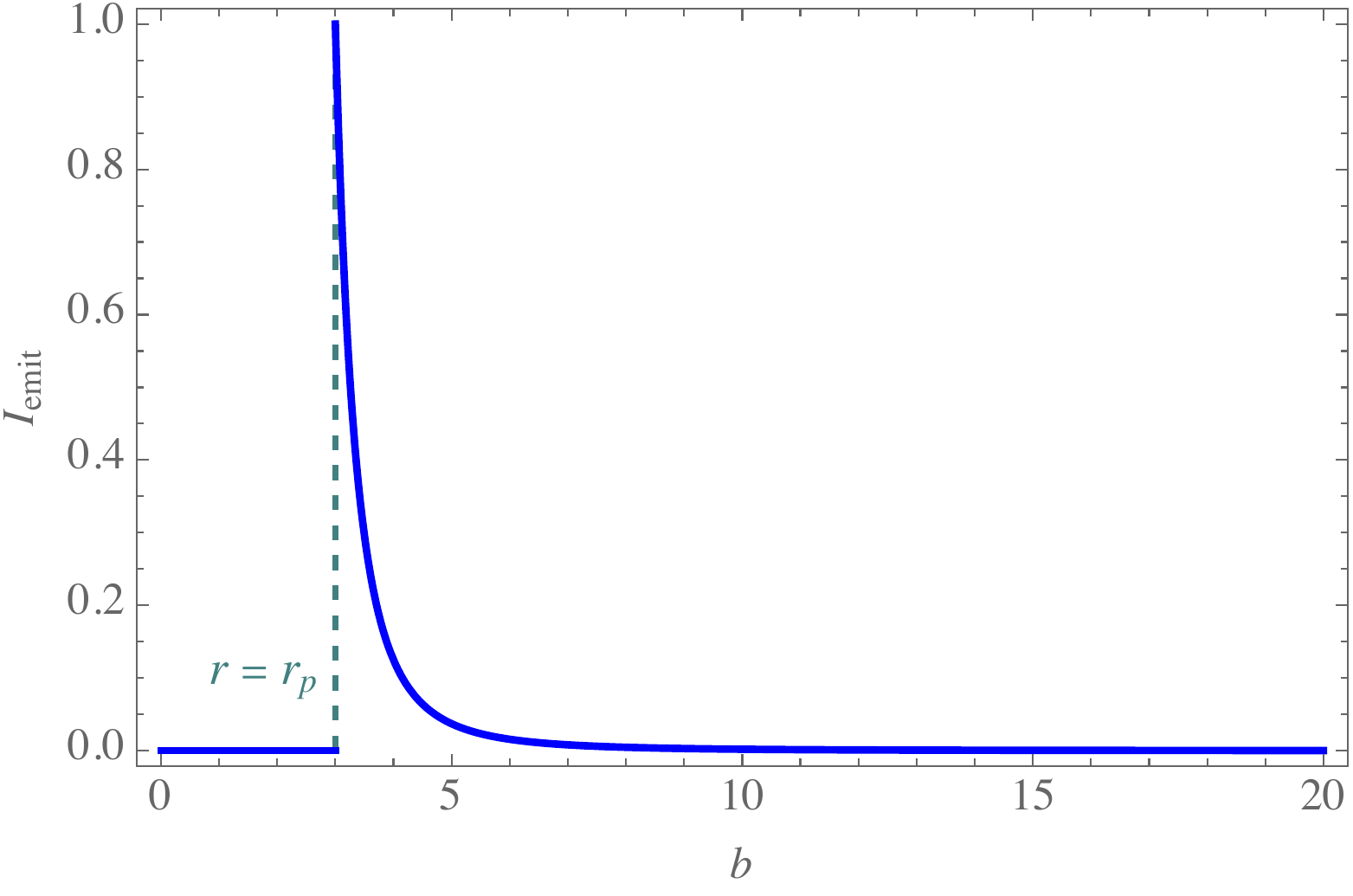}
    \includegraphics[width=5.4cm,height=4.4cm]{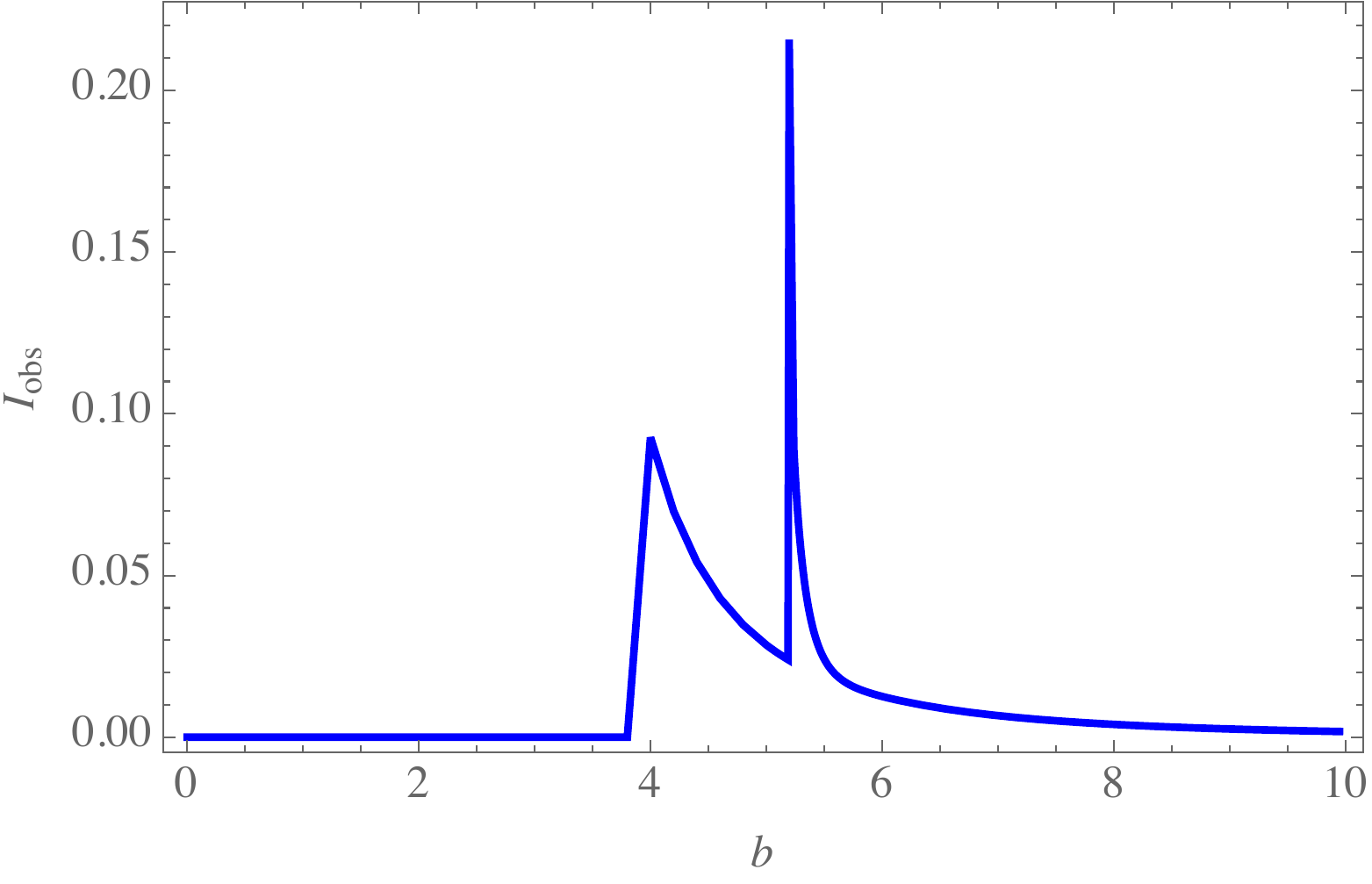}
    \includegraphics[width=5.4cm,height=4.5cm]{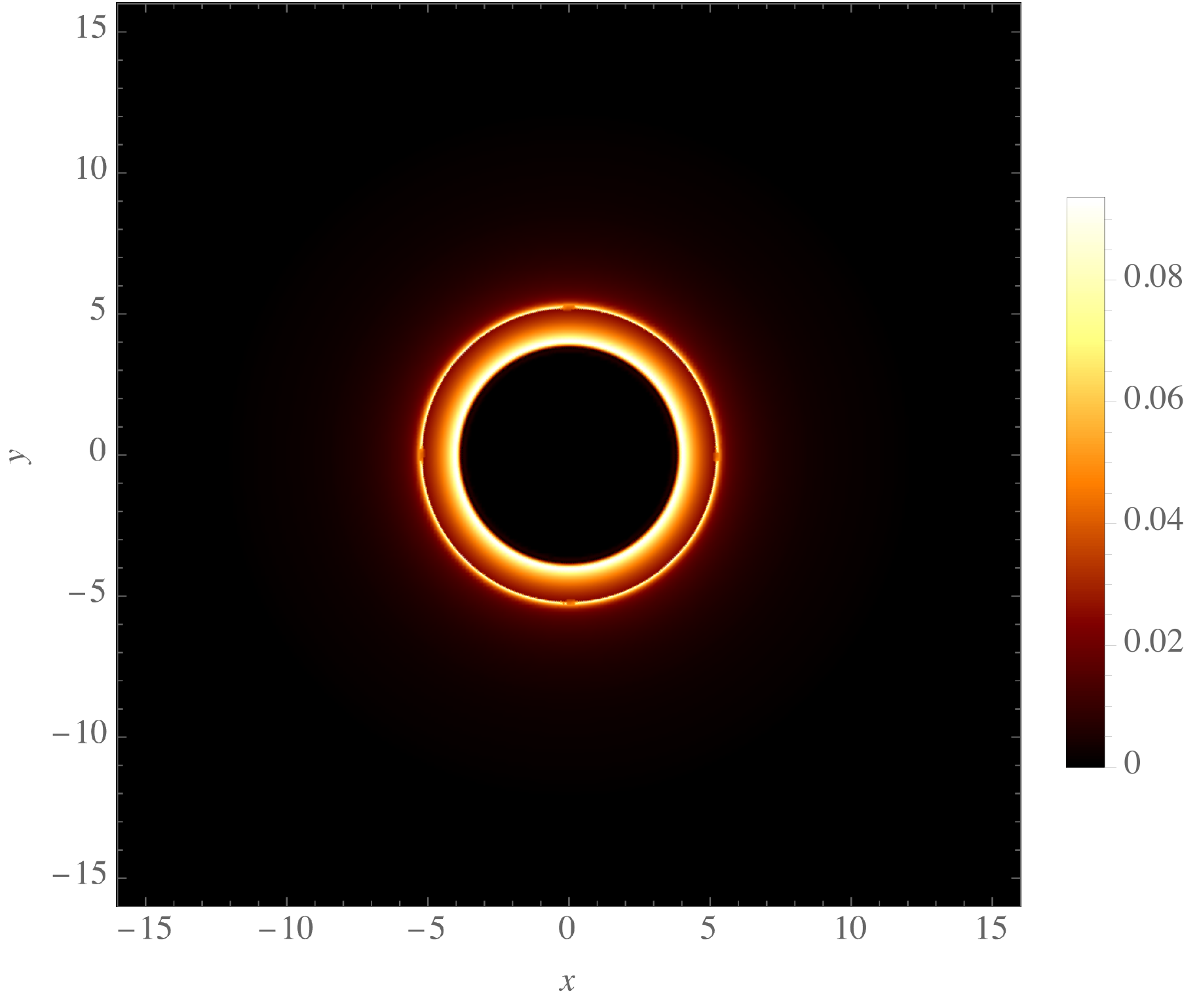}~(b)
    \includegraphics[width=5.4cm,height=4.4cm]{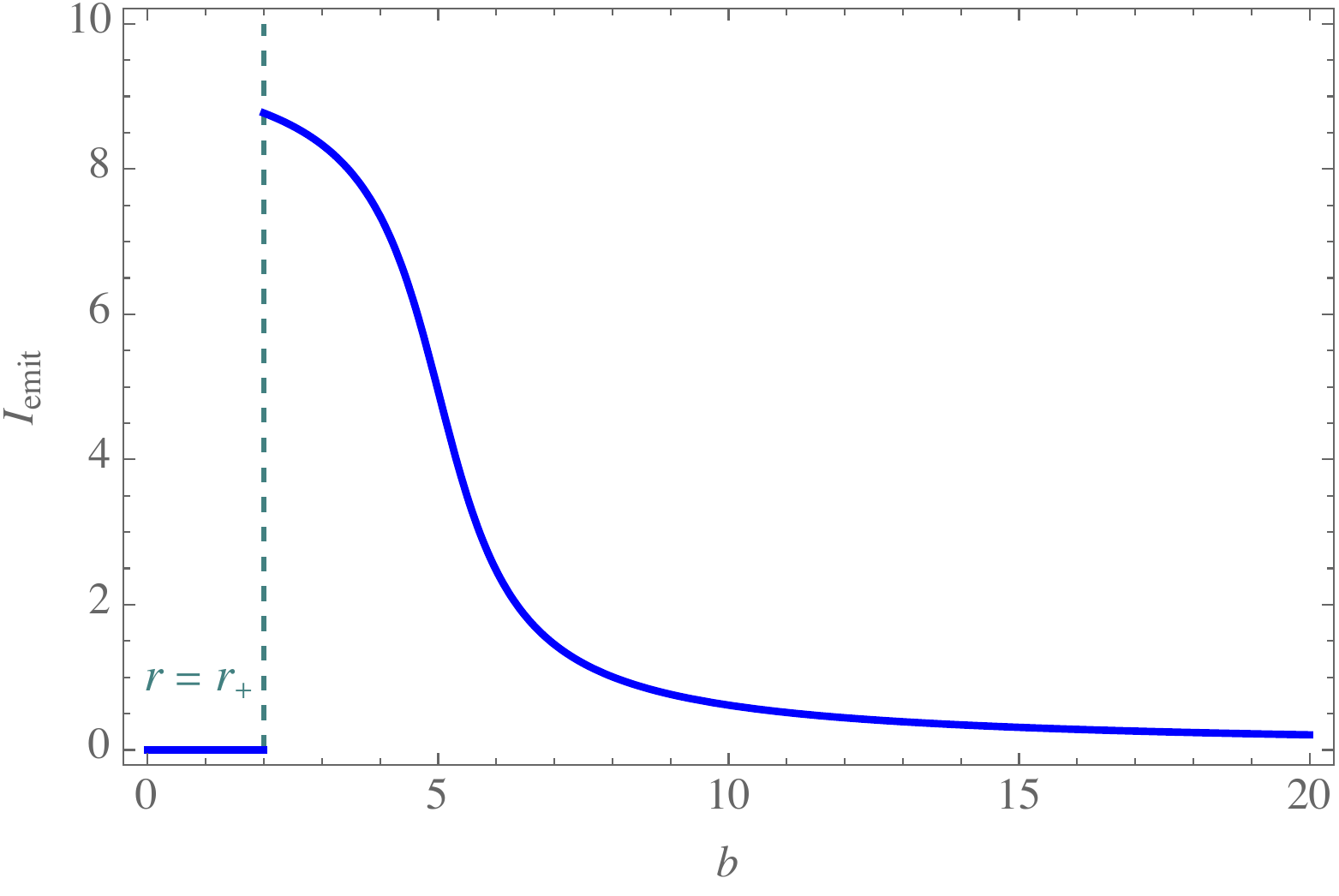}
    \includegraphics[width=5.4cm,height=4.4cm]{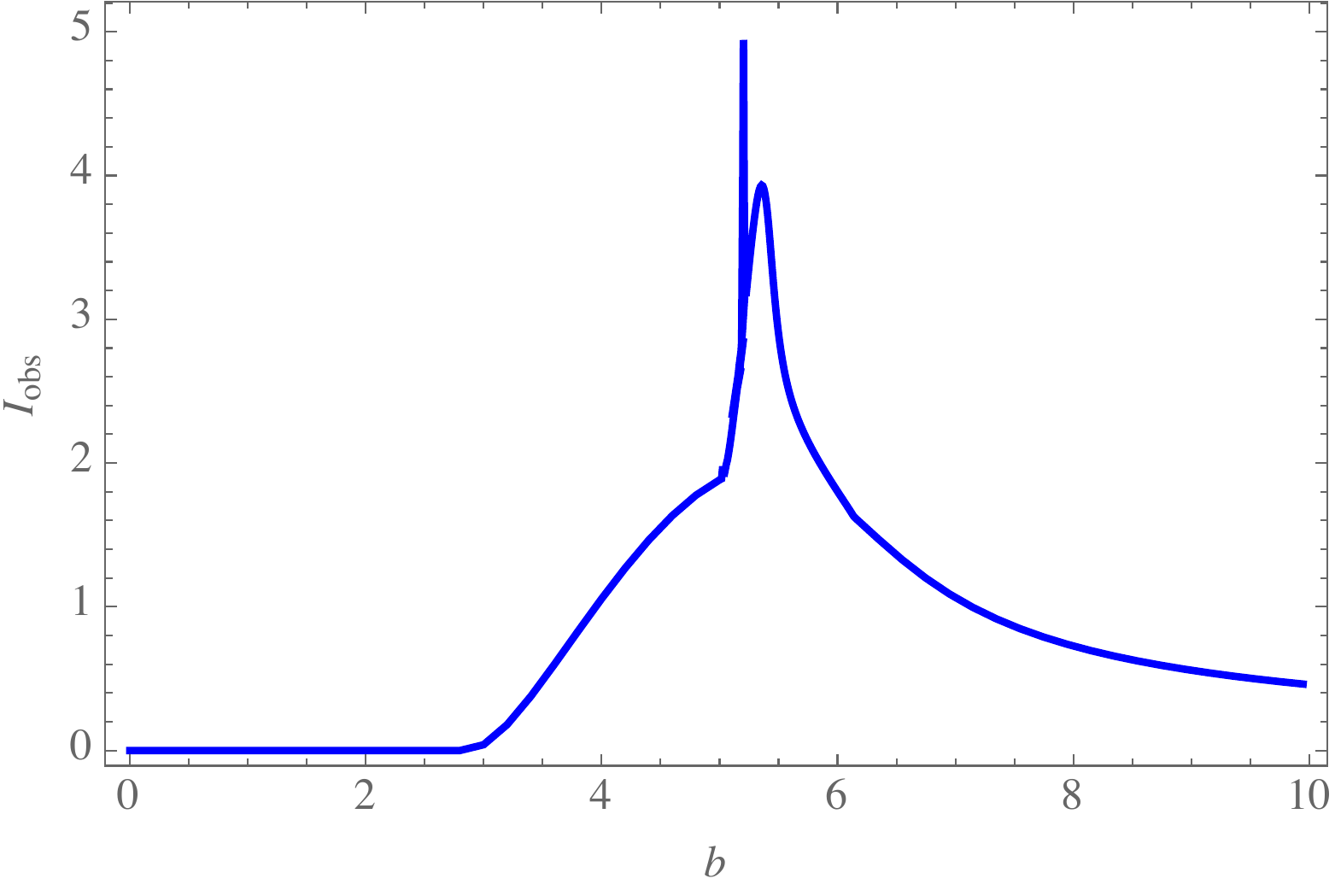}
    \includegraphics[width=5.4cm,height=4.5cm]{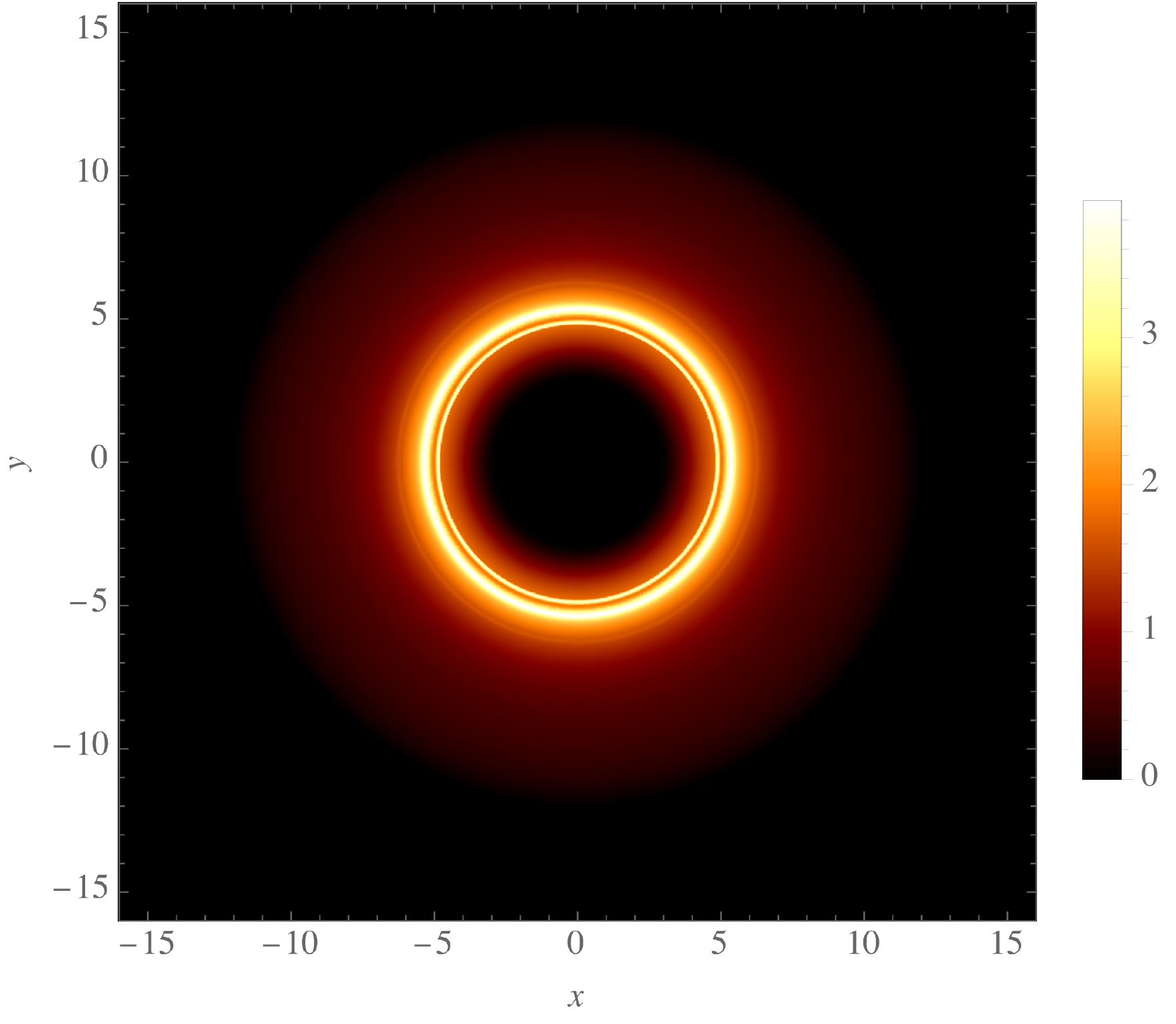}~(c)
    \caption{The observational characteristics of the accretion disk around the $f(R)$ black hole, depicted for the case of $\beta = 0$ (the Schwarzschild-de Sitter black hole). The panels, from top to bottom, correspond to (a) model 1, (b) model 2, and (c) model 3 emission profiles. The left and middle panels in each row display the $b$-profiles of the emitted and observed intensities, respectively. The right panels present two-dimensional face-on ray-traced shadow images for each of the models.}
    \label{fig:ring-beta0}
\end{figure}

\begin{figure}
    \centering
    \includegraphics[width=5.4cm,height=4.4cm]{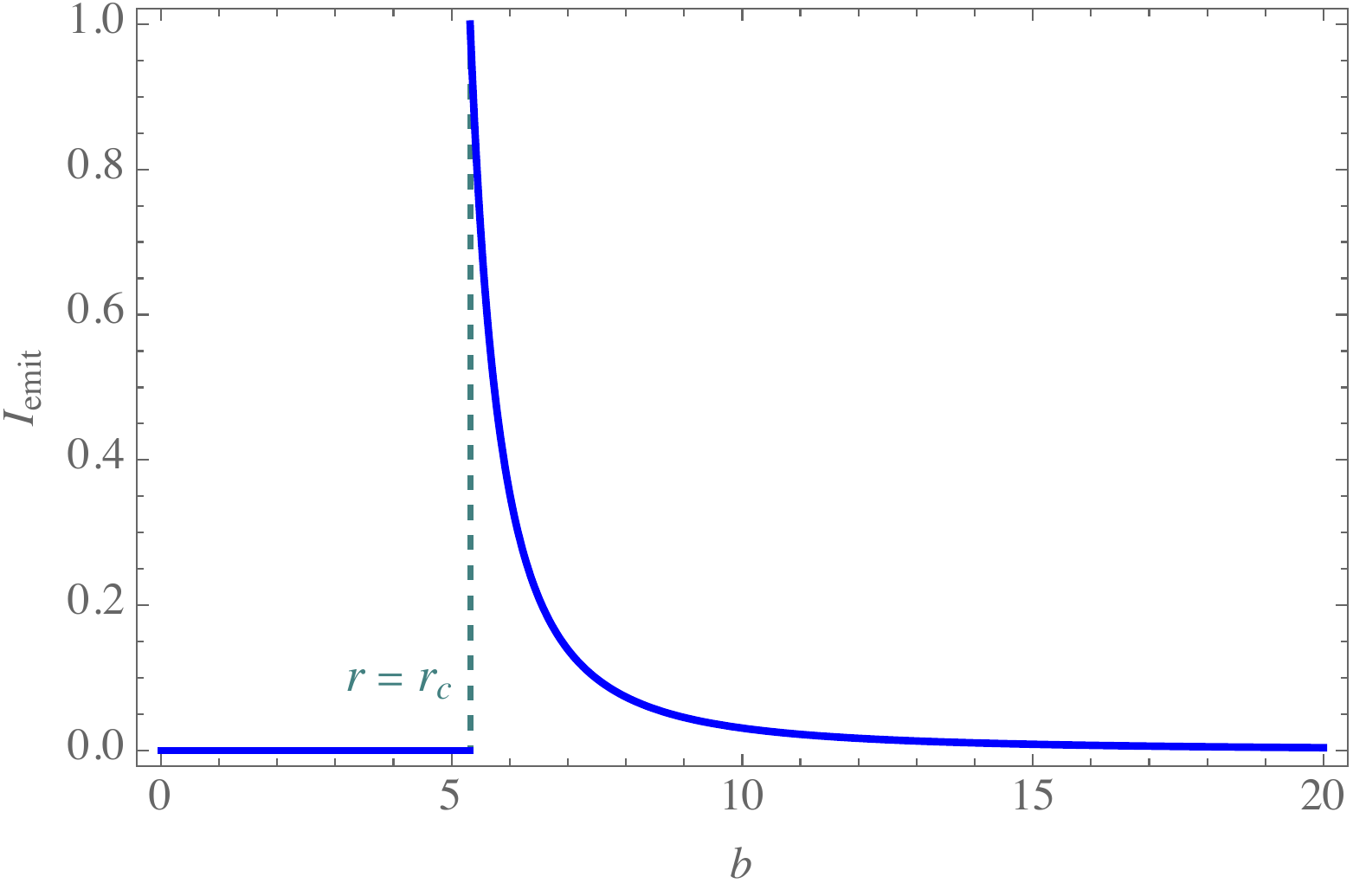}
    \includegraphics[width=5.4cm,height=4.4cm]{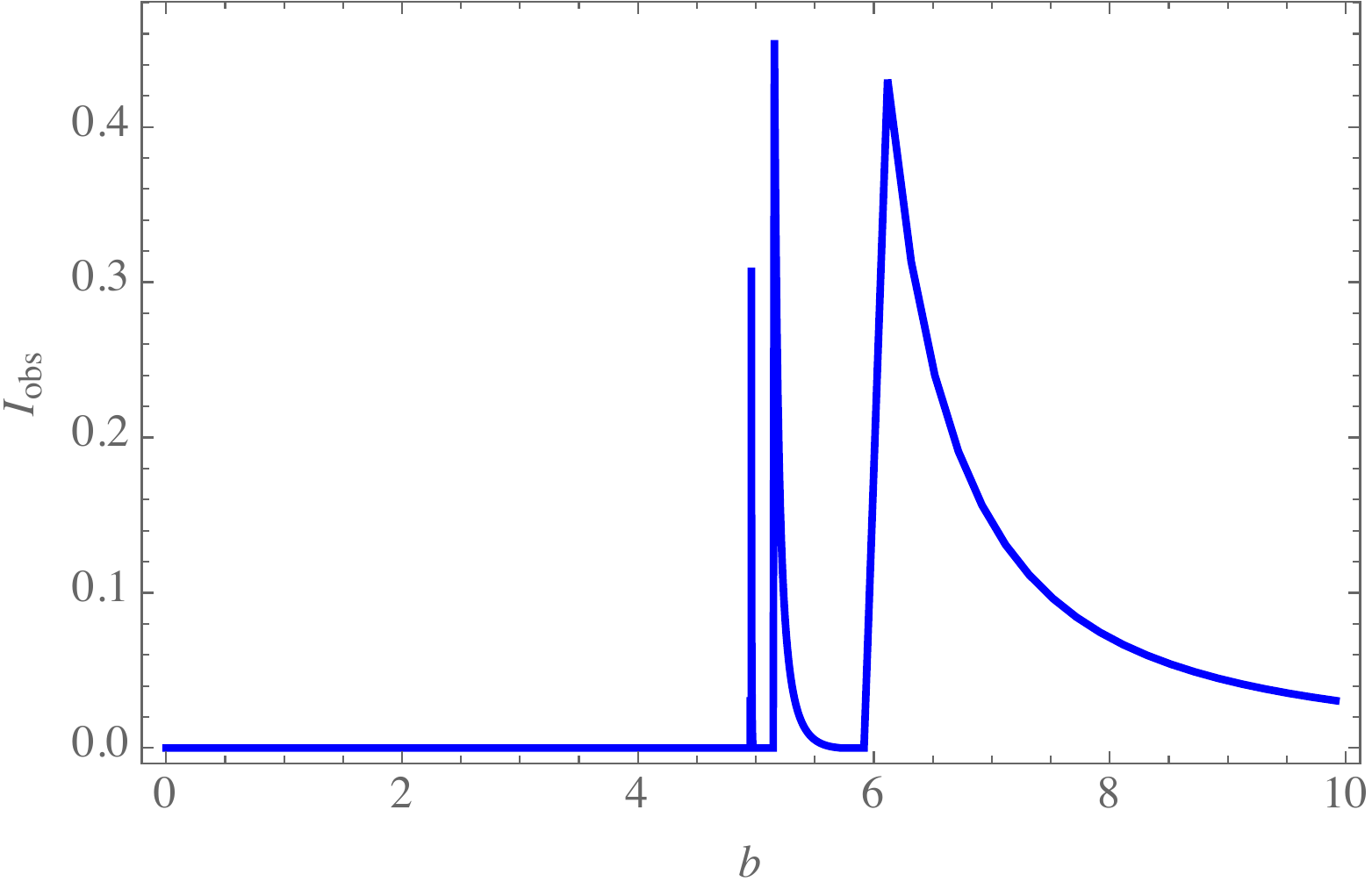}
    \includegraphics[width=5.4cm,height=4.5cm]{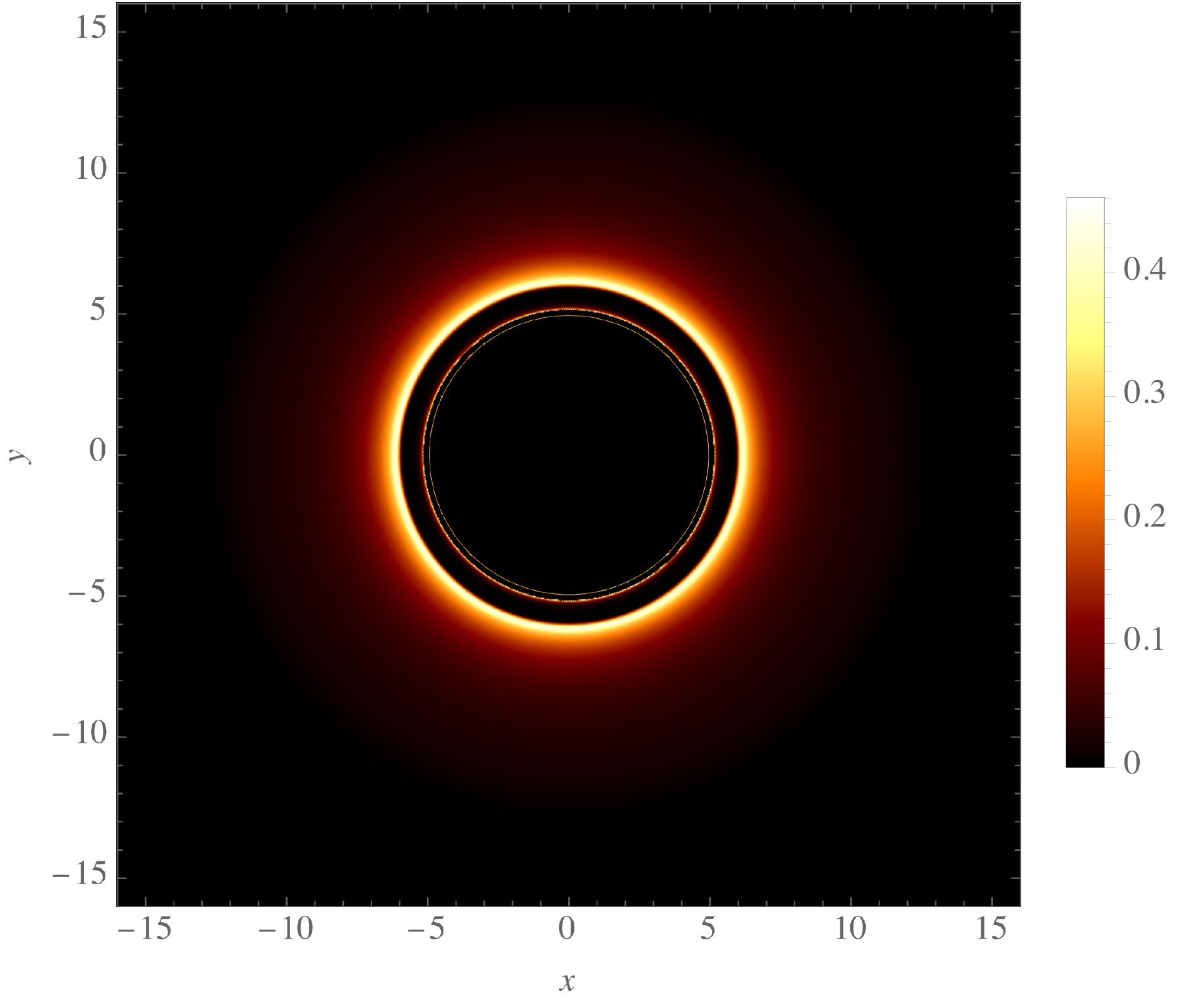}~(a)
    \includegraphics[width=5.4cm,height=4.4cm]{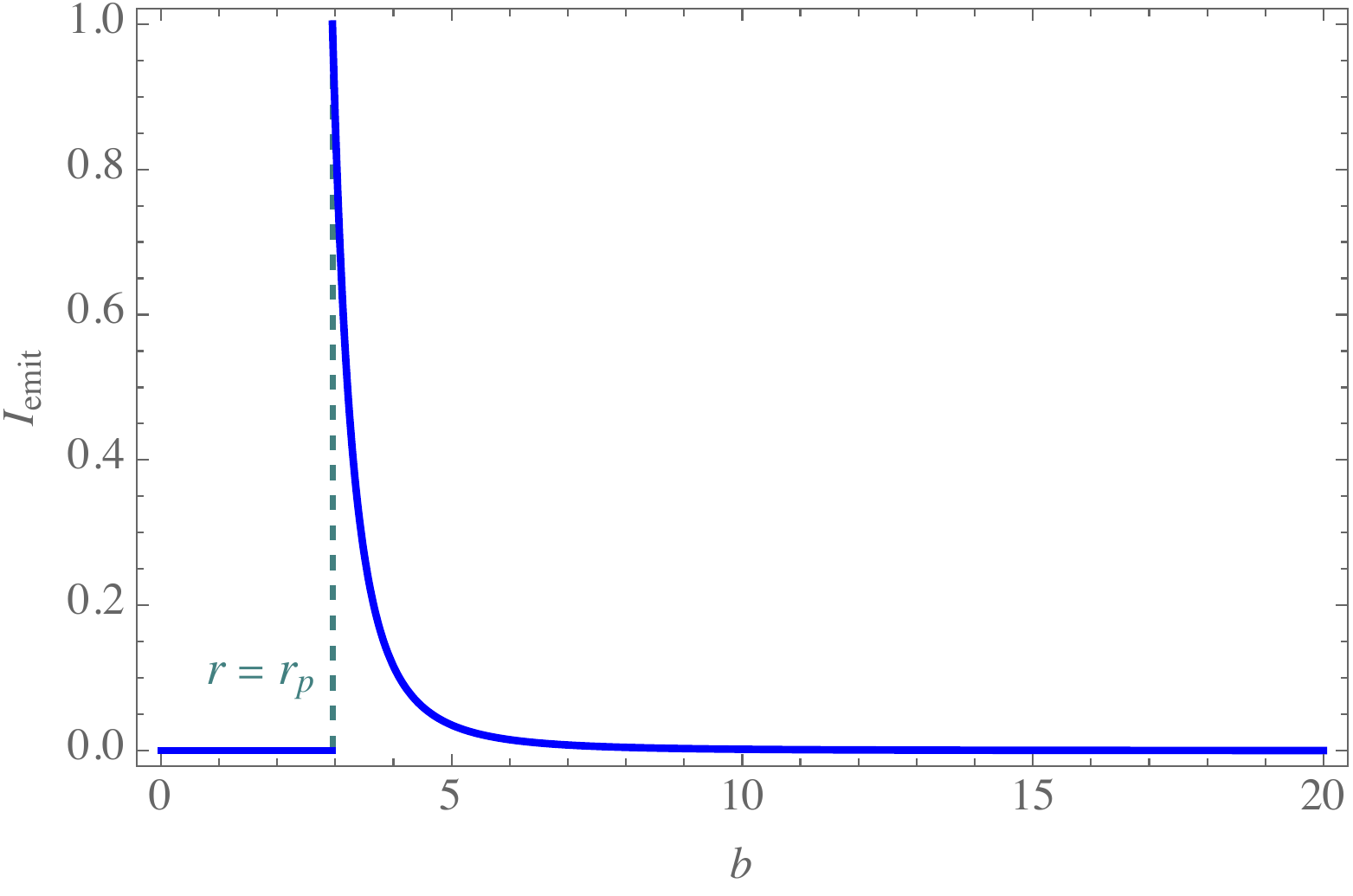}
    \includegraphics[width=5.4cm,height=4.4cm]{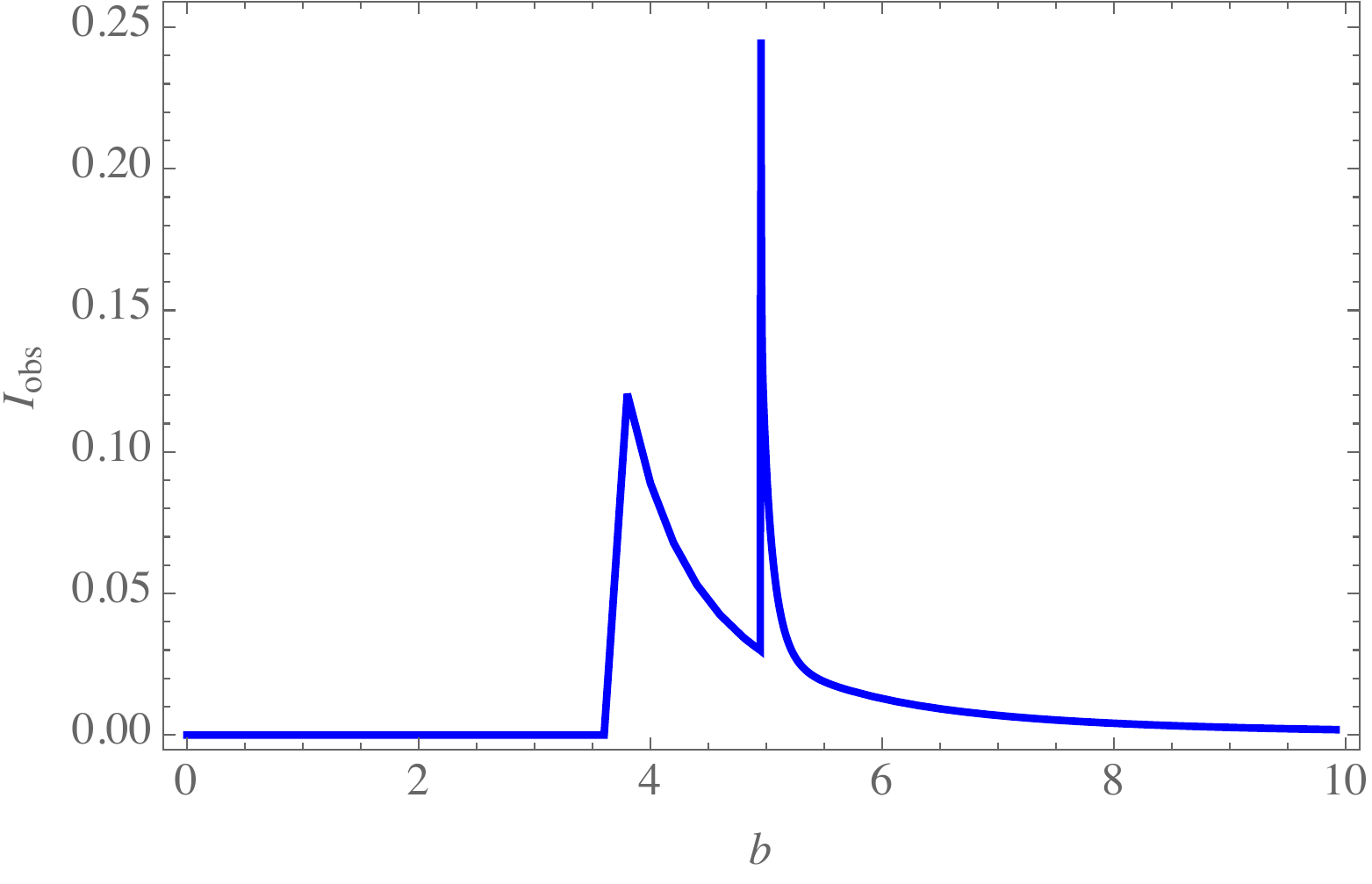}
    \includegraphics[width=5.4cm,height=4.5cm]{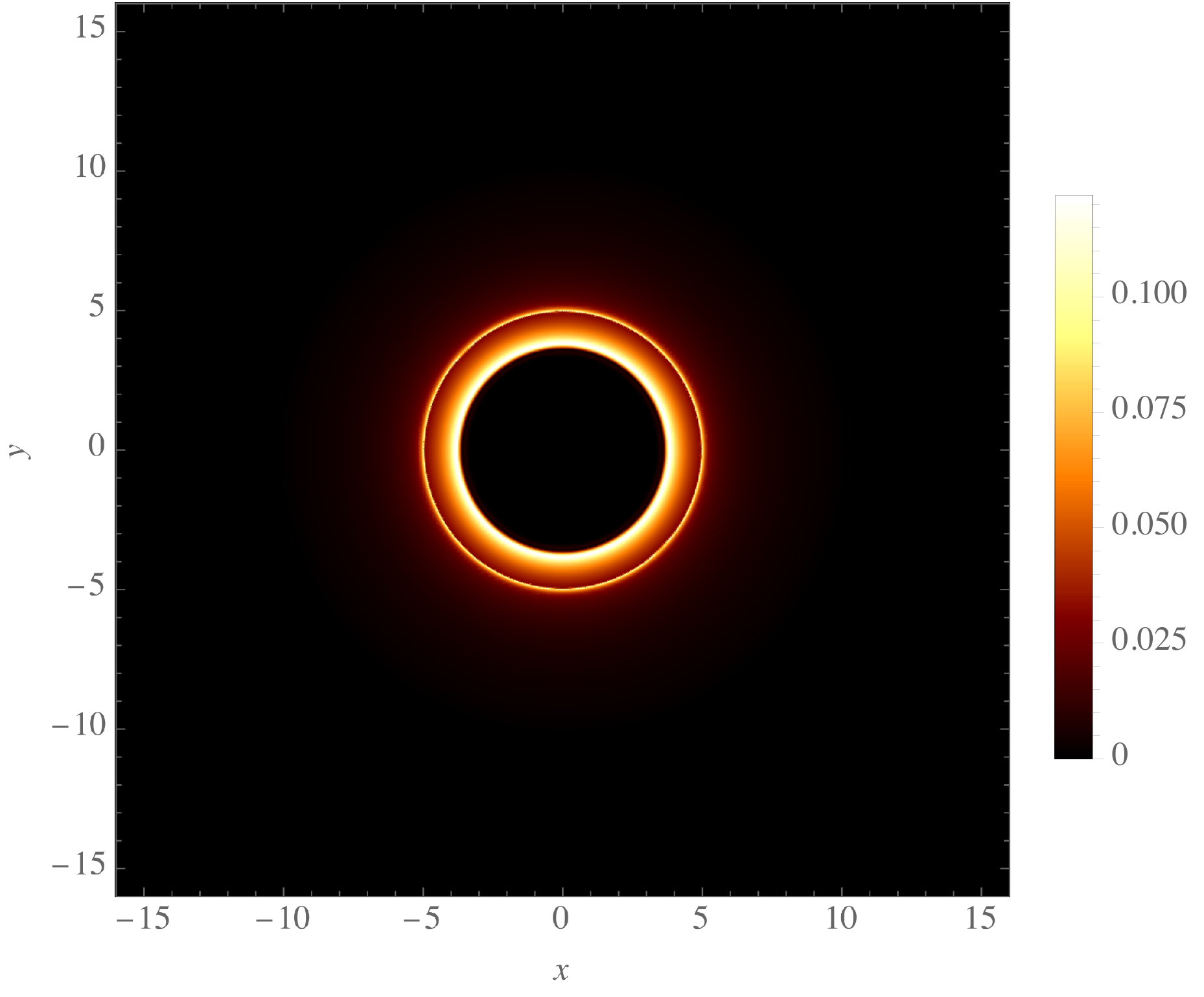}~(b)
    \includegraphics[width=5.4cm,height=4.4cm]{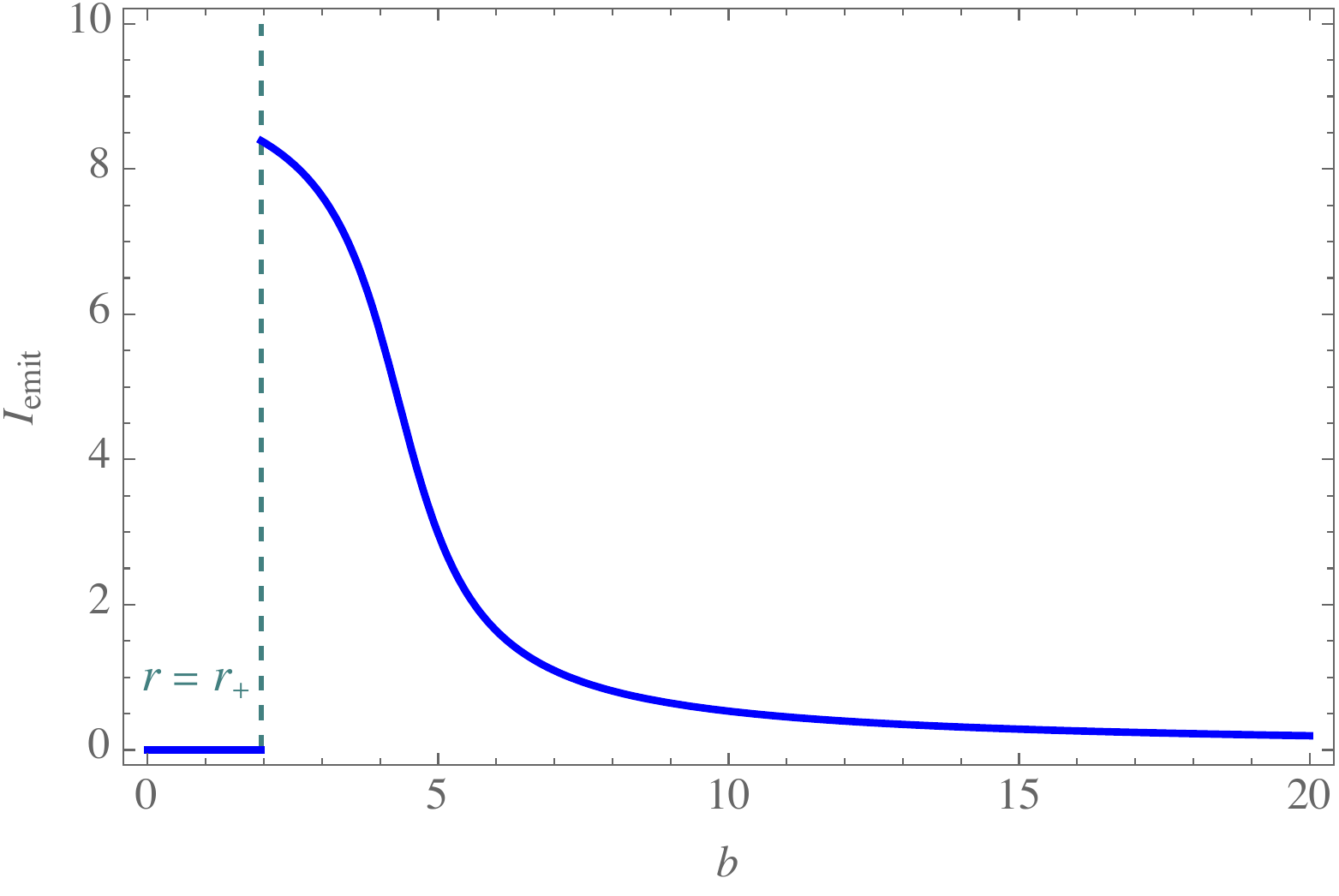}
    \includegraphics[width=5.4cm,height=4.4cm]{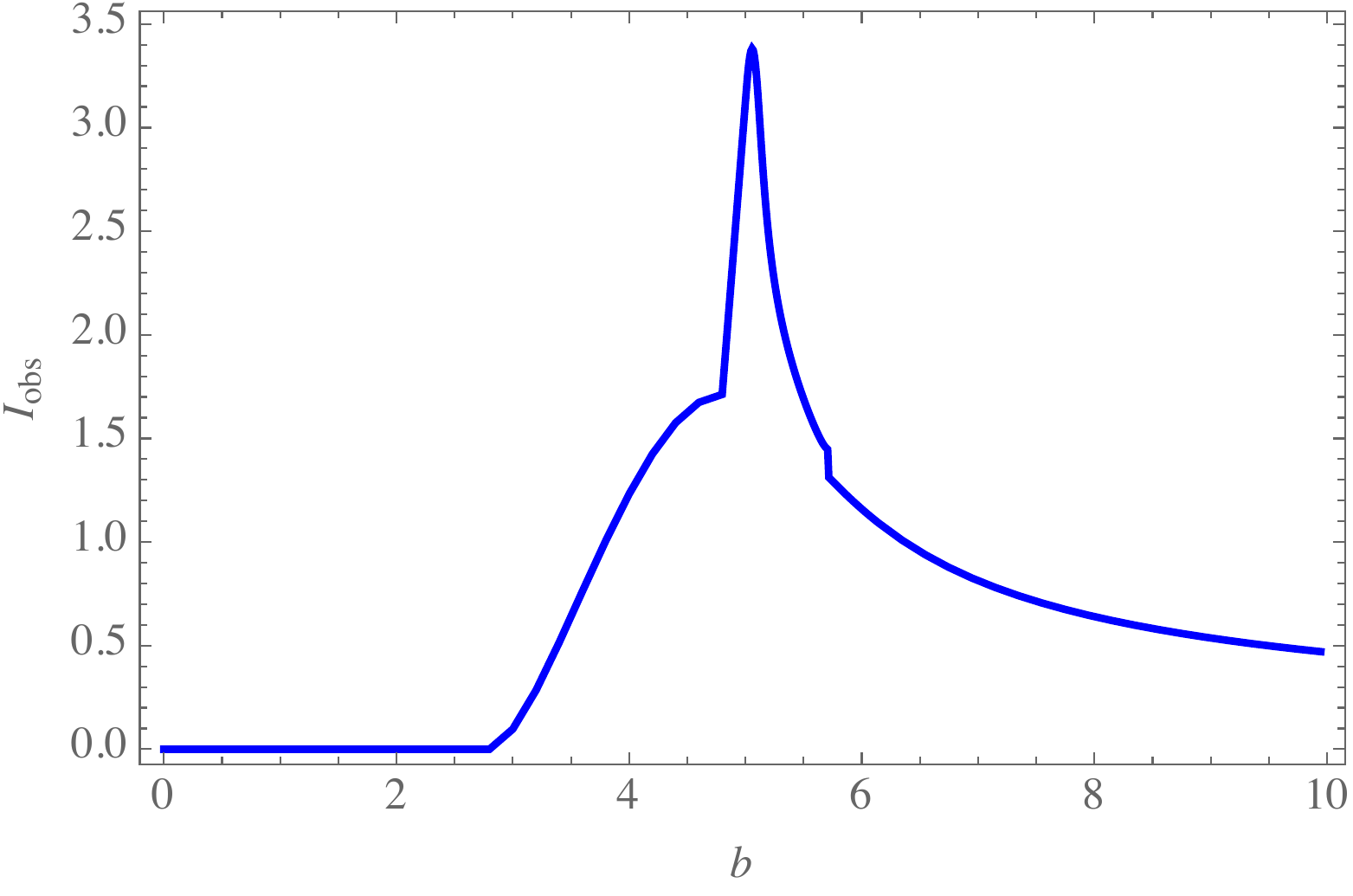}
    \includegraphics[width=5.4cm,height=4.5cm]{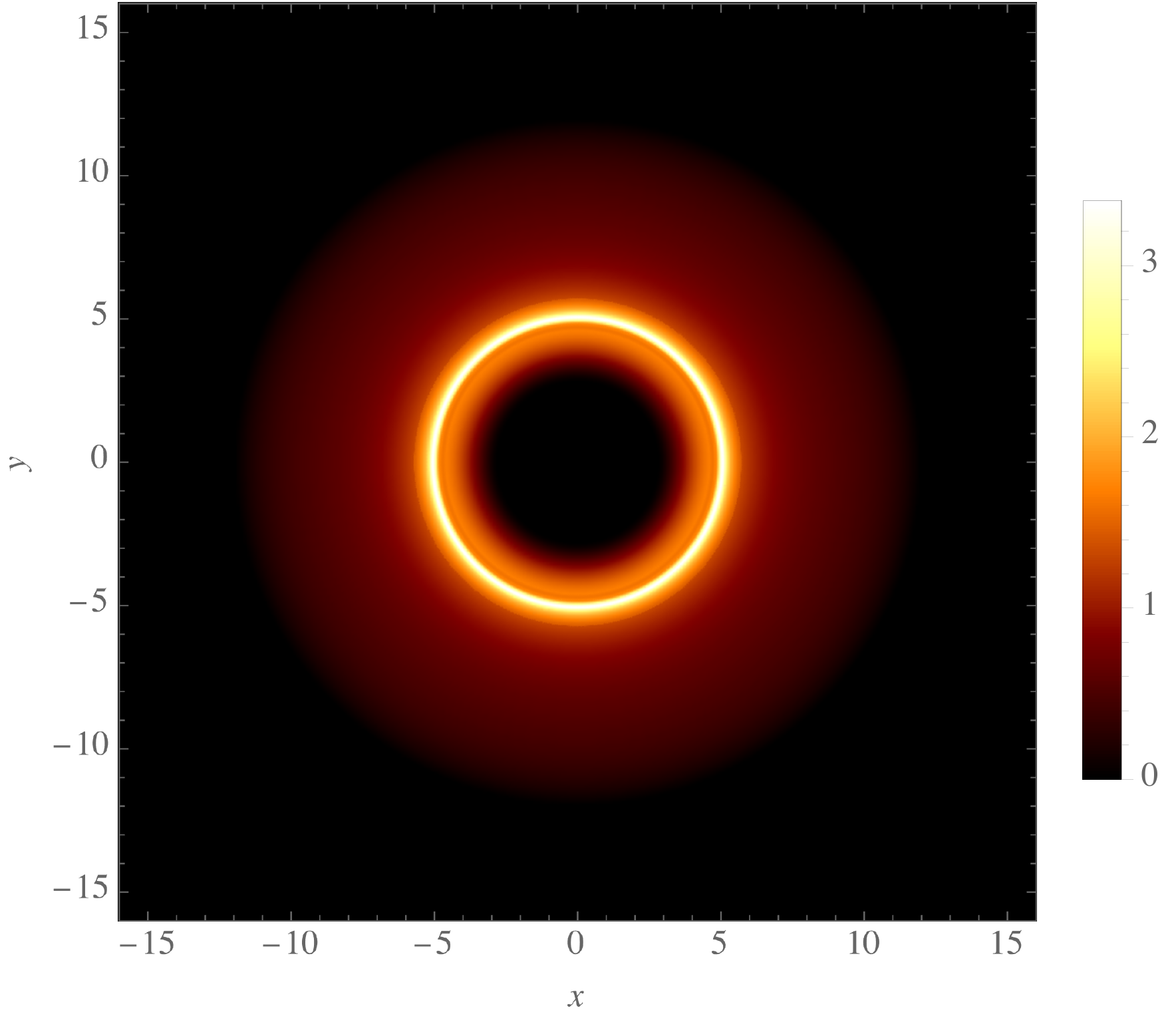}~(c)
    \caption{The case of $\beta=0.011$.}
    \label{fig:ring-beta1}
\end{figure}

\begin{figure}
    \centering
    \includegraphics[width=5.4cm,height=4.4cm]{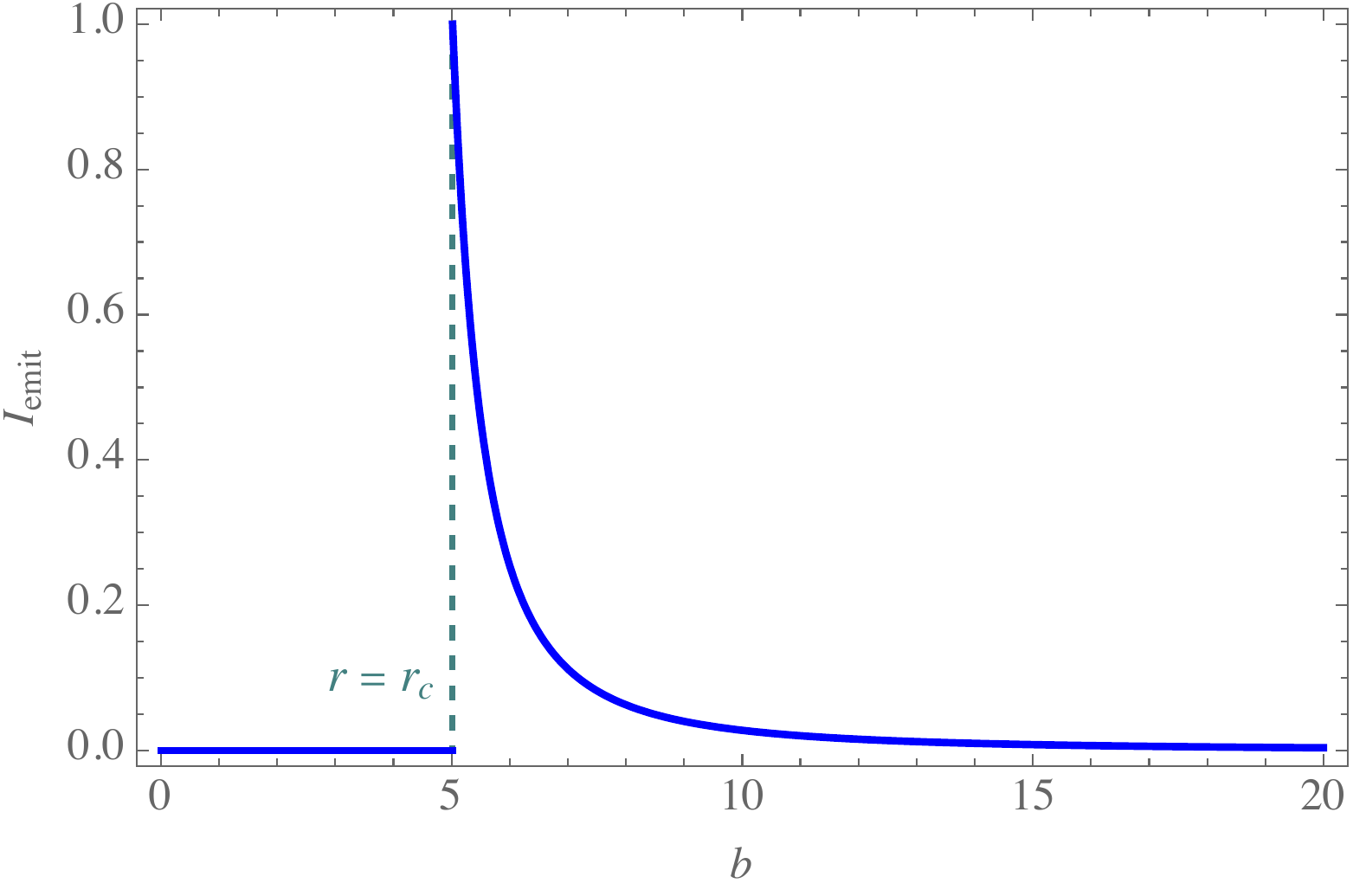}
    \includegraphics[width=5.4cm,height=4.4cm]{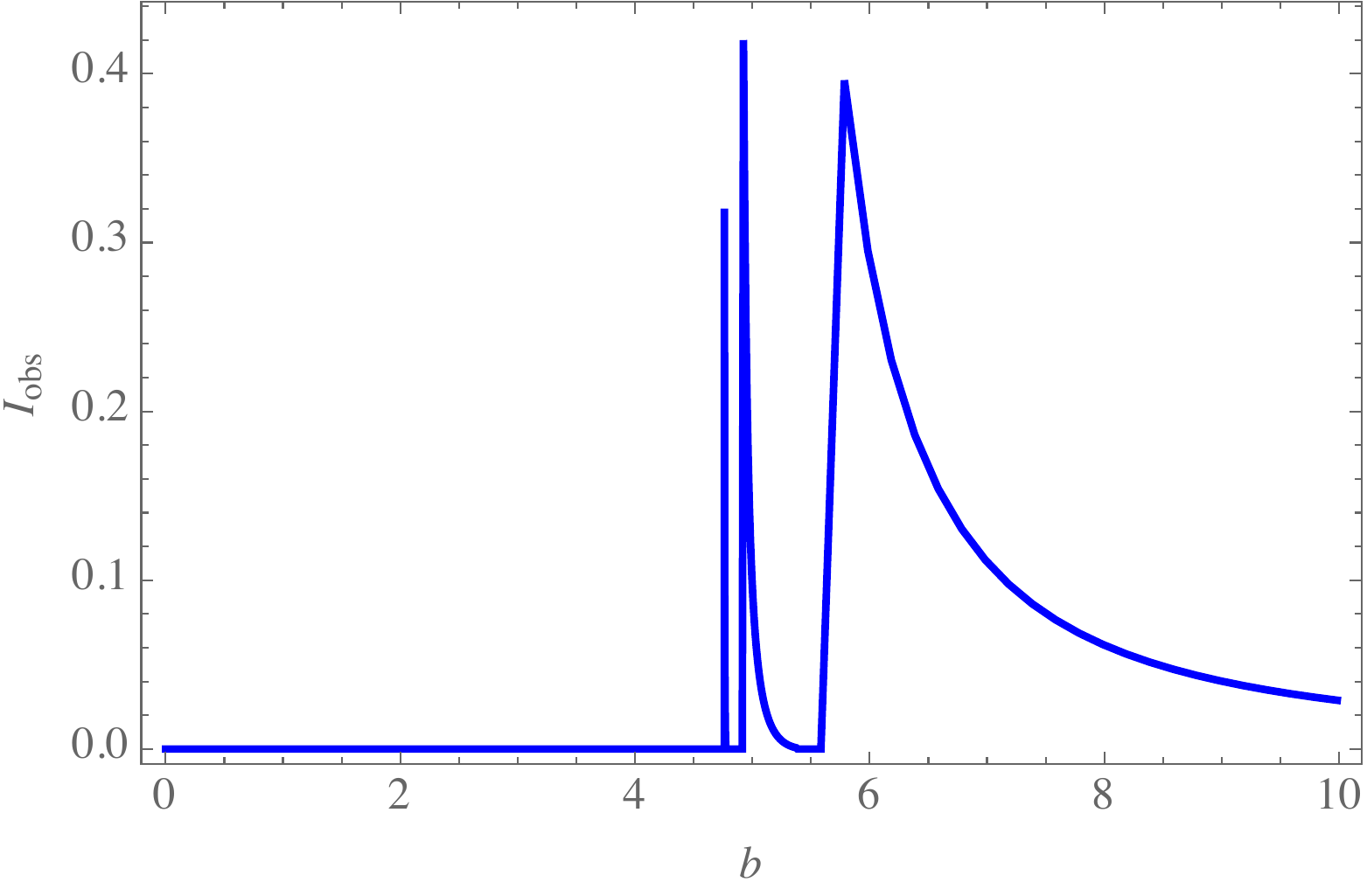}
    \includegraphics[width=5.4cm,height=4.5cm]{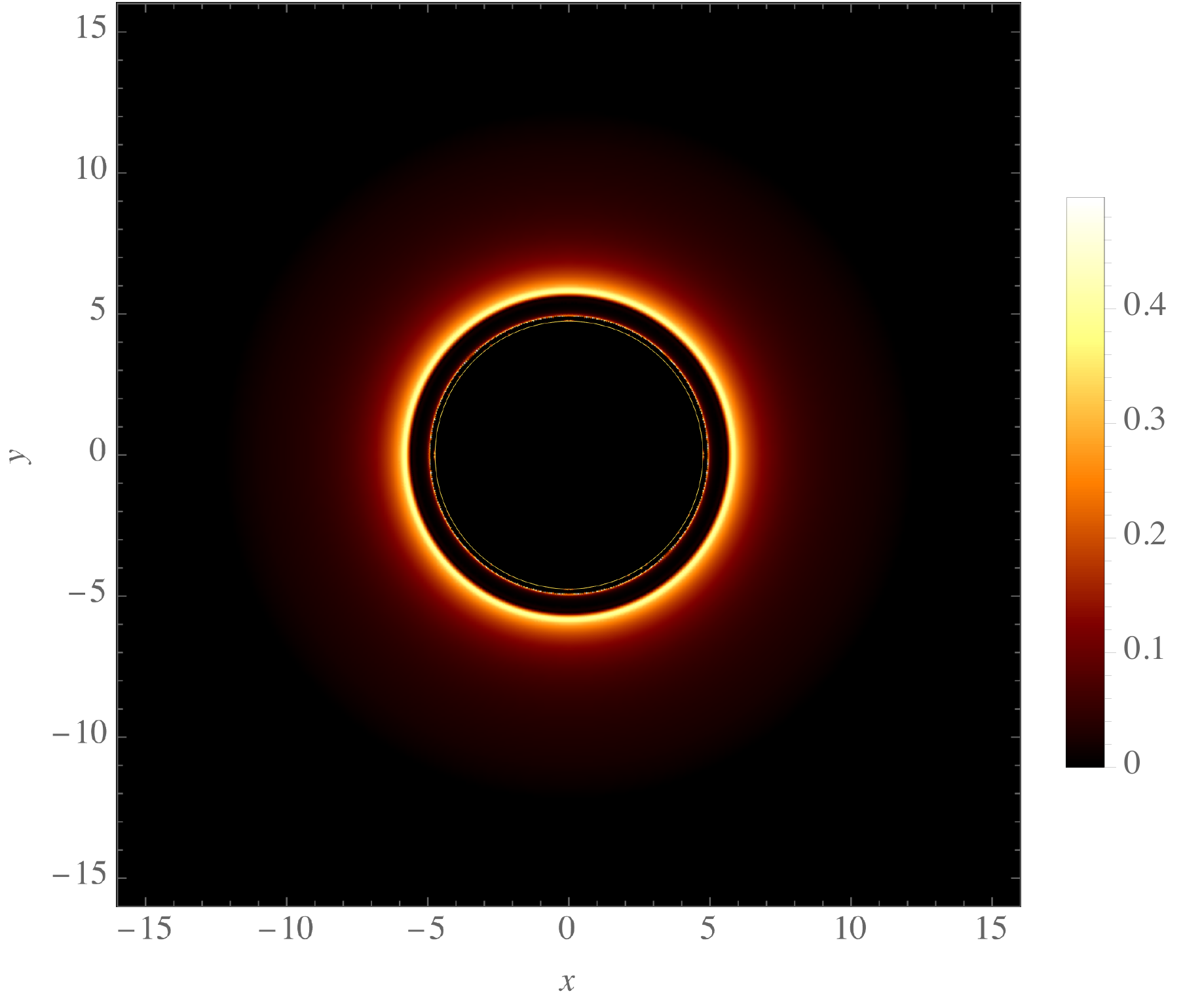}~(a)
    \includegraphics[width=5.4cm,height=4.4cm]{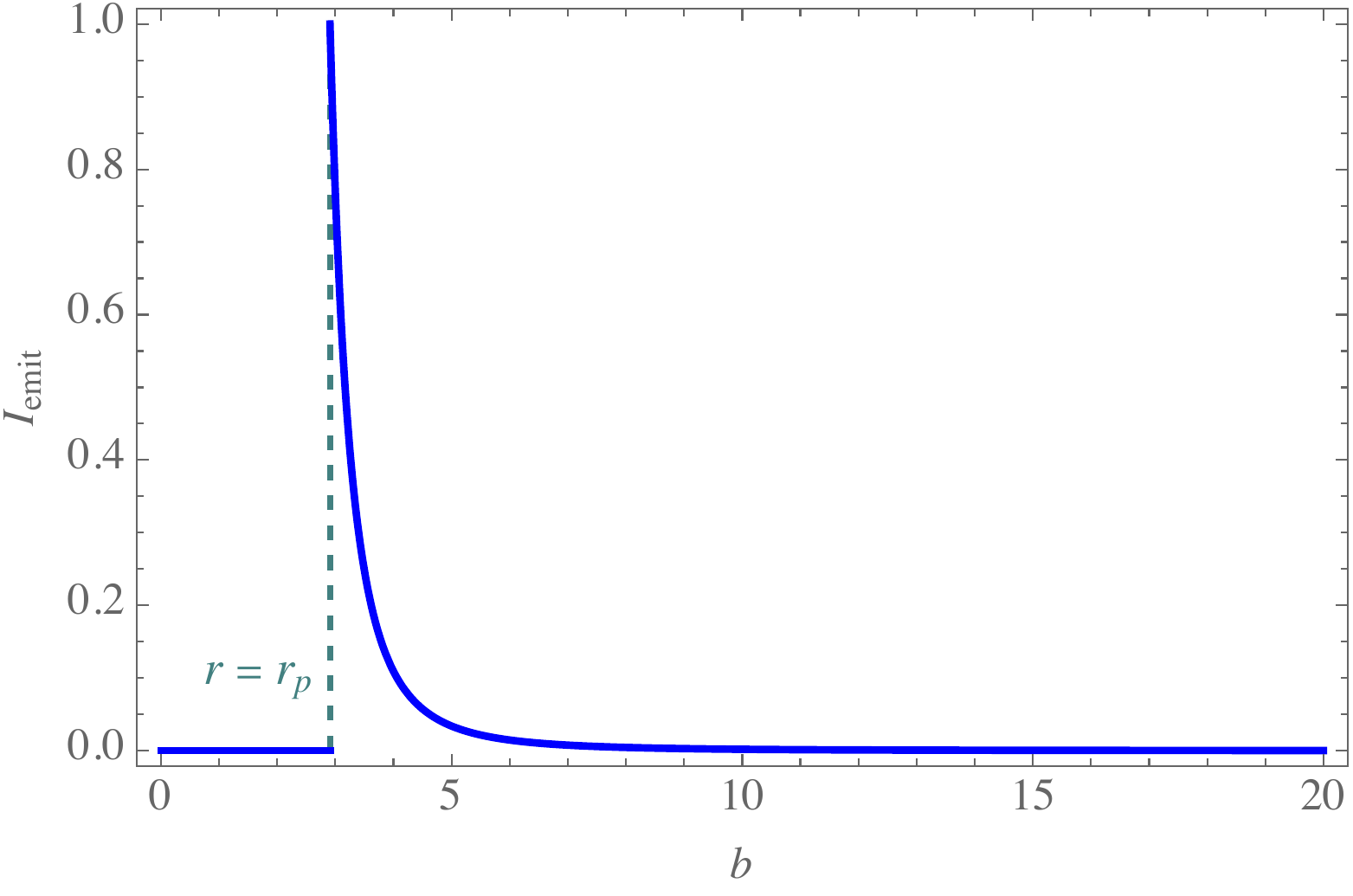}
    \includegraphics[width=5.4cm,height=4.4cm]{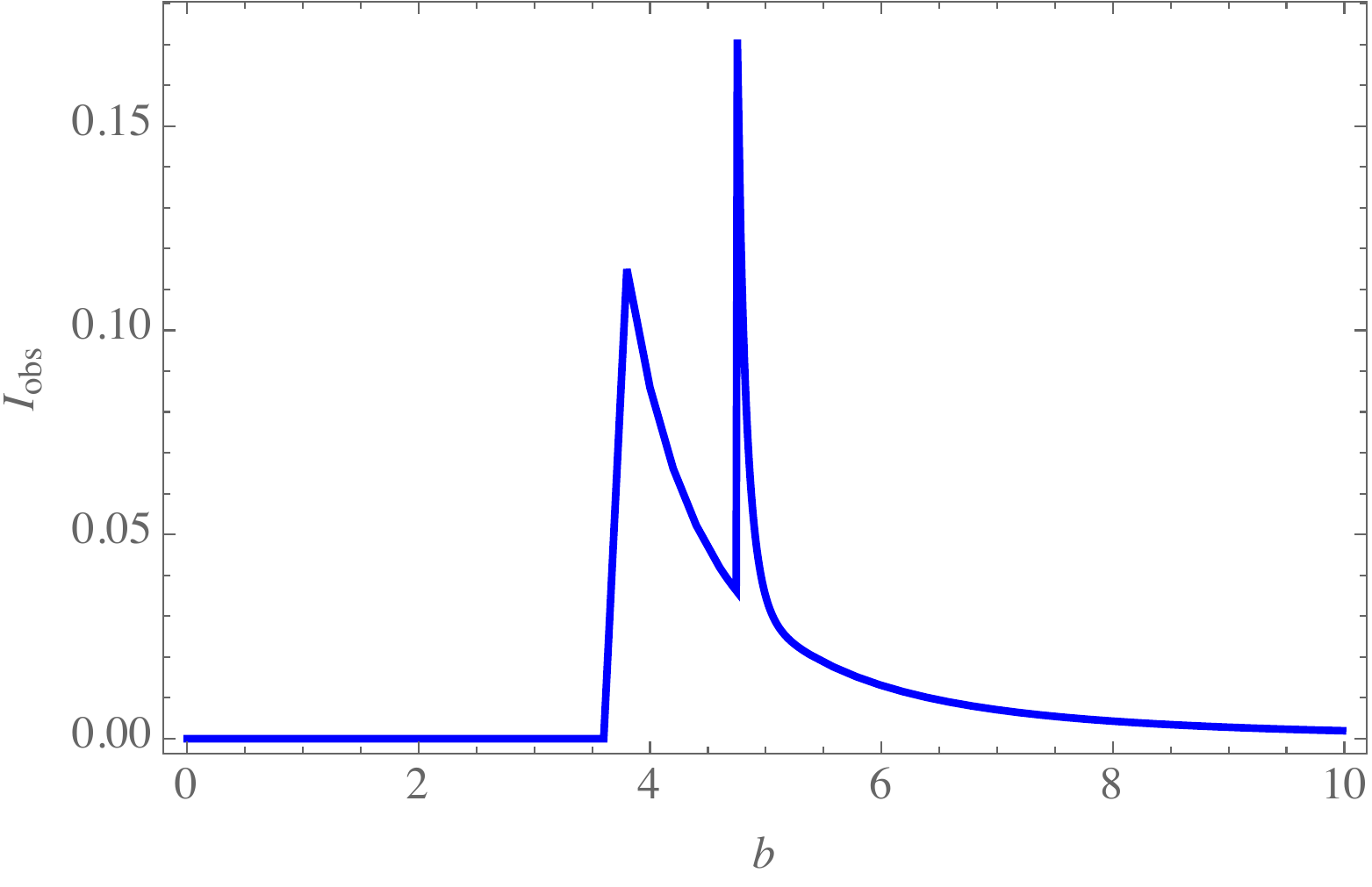}
    \includegraphics[width=5.4cm,height=4.5cm]{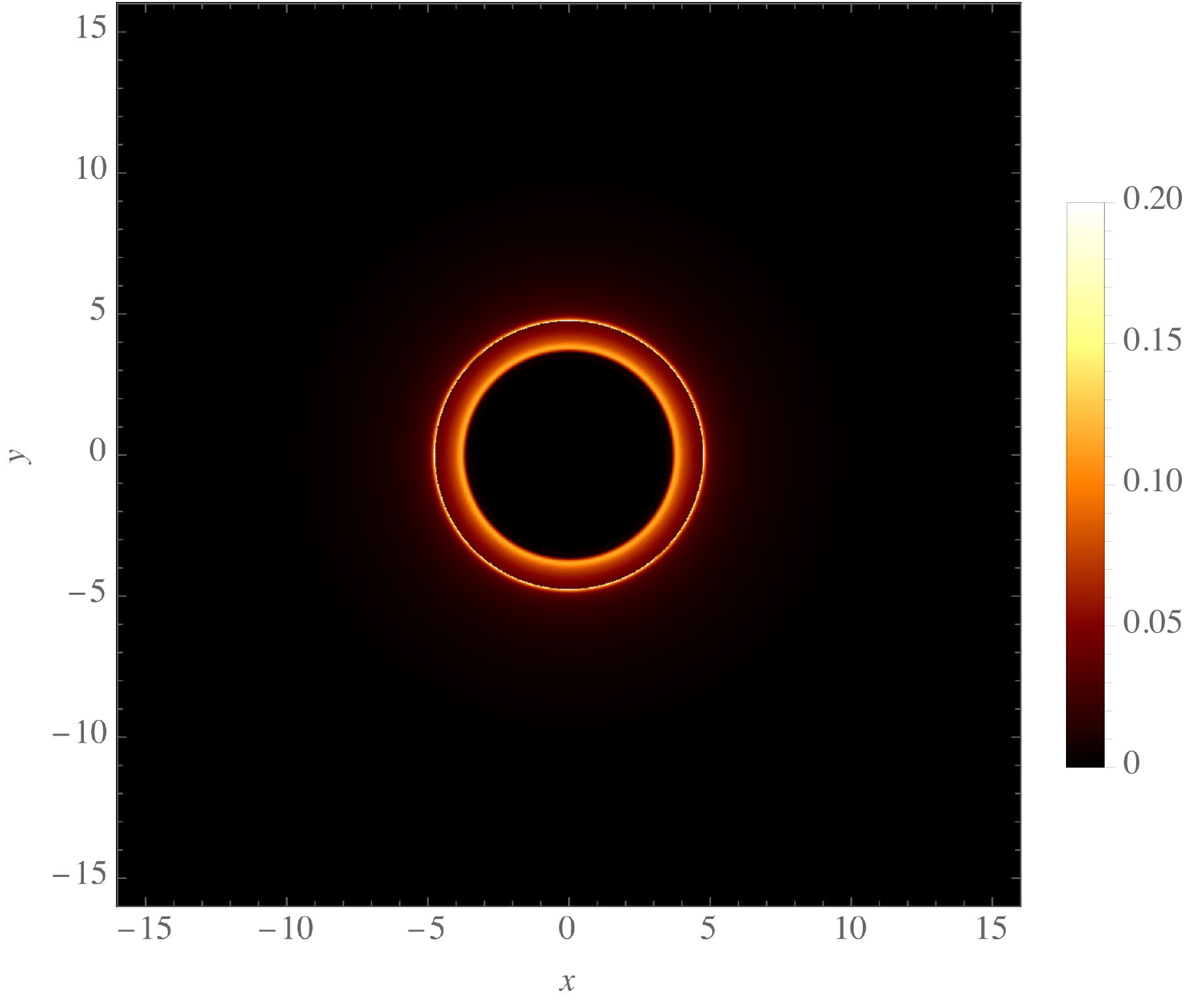}~(b)
    \includegraphics[width=5.4cm,height=4.4cm]{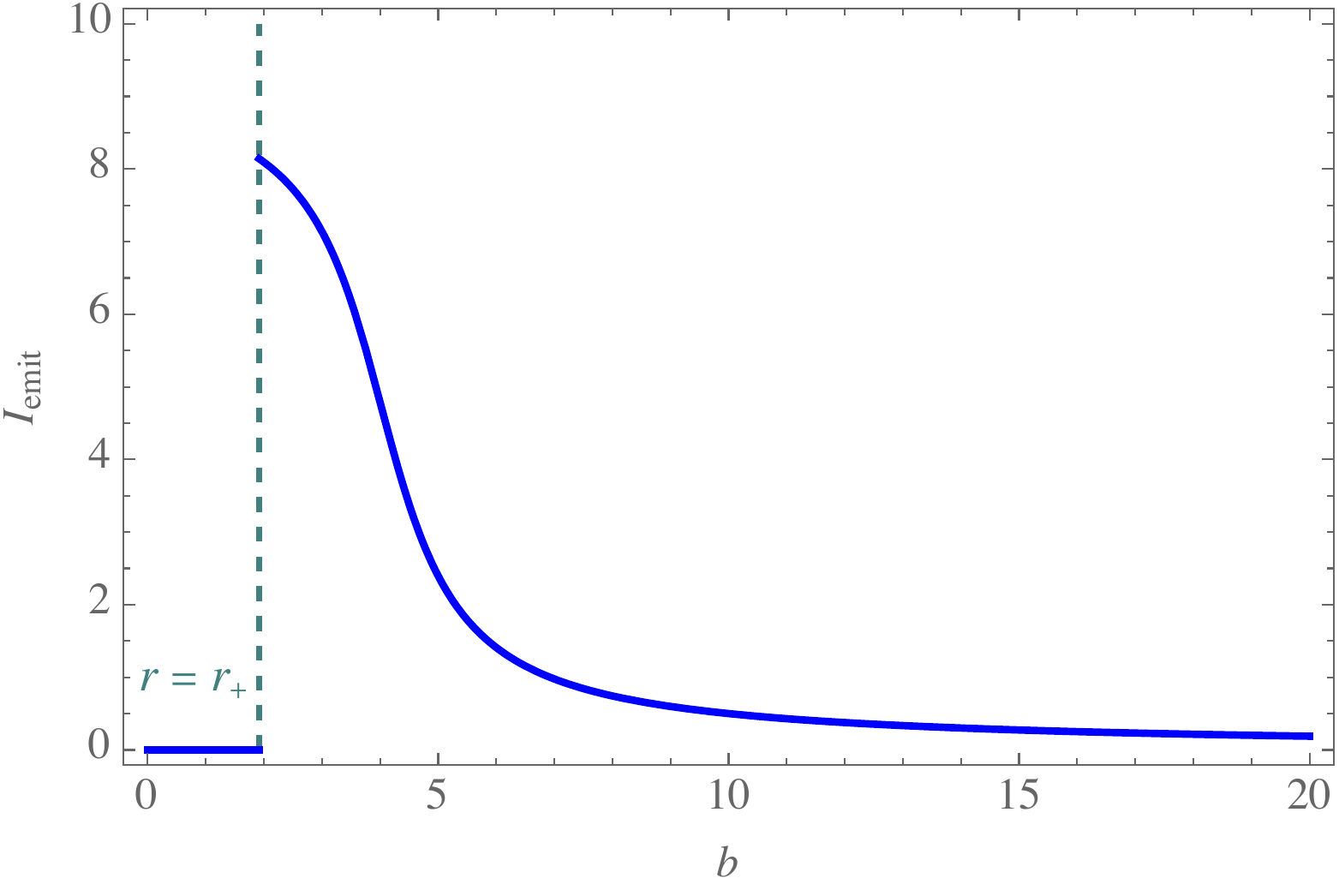}
    \includegraphics[width=5.4cm,height=4.4cm]{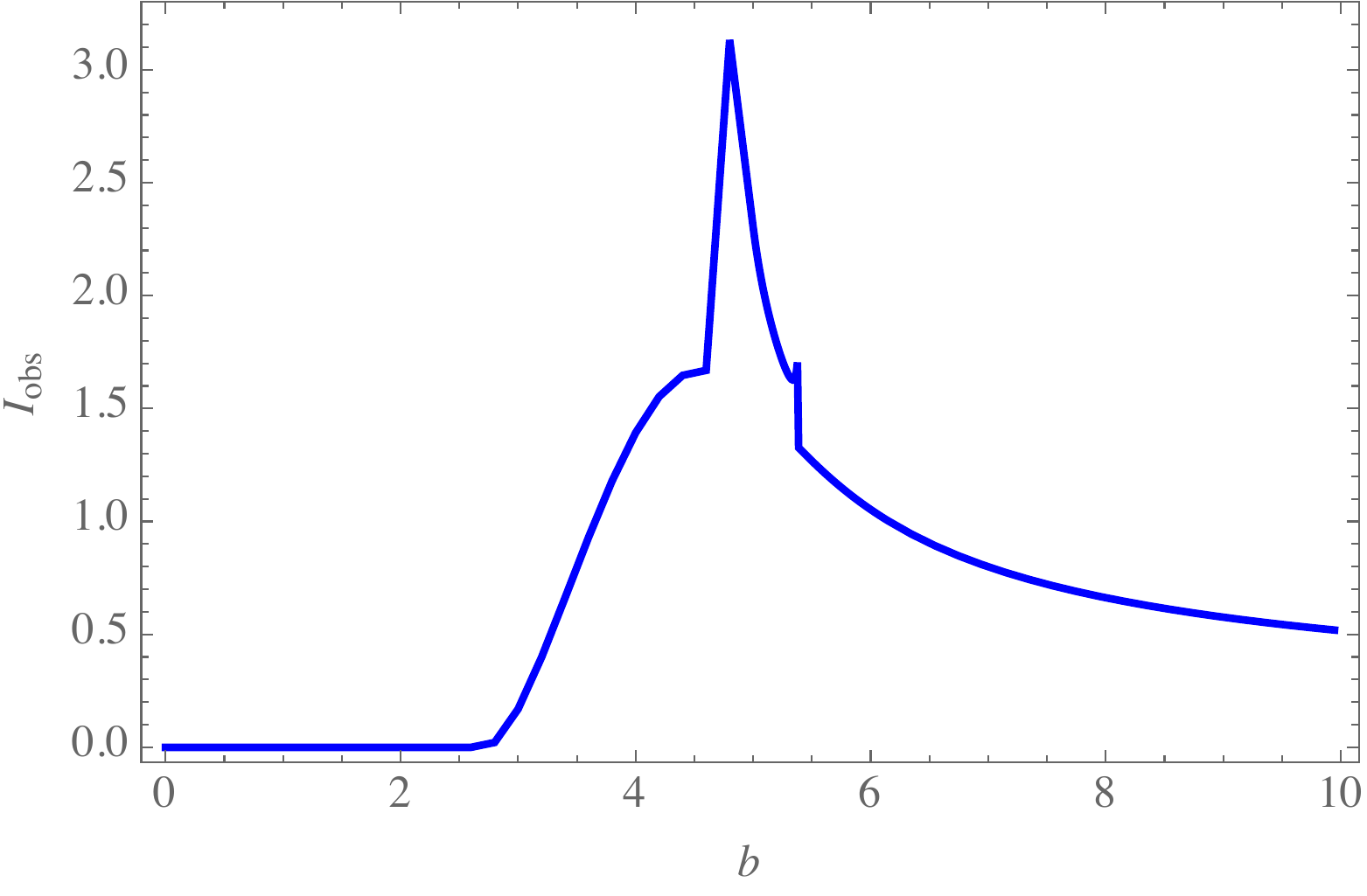}
    \includegraphics[width=5.4cm,height=4.5cm]{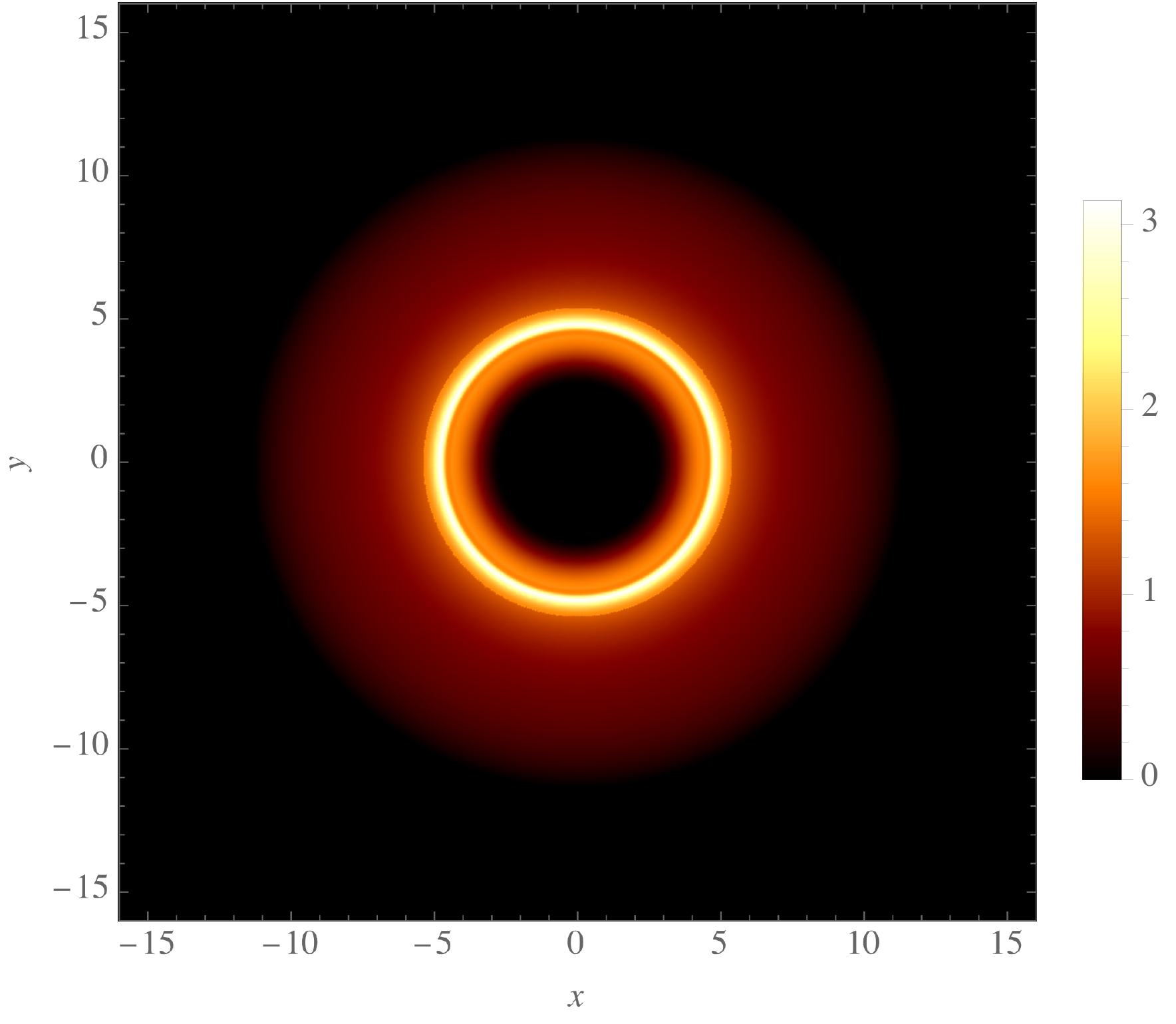}~(c)
    \caption{The case of $\beta=0.022$.}
    \label{fig:ring-beta2}
\end{figure}

\begin{figure}
    \centering
    \includegraphics[width=5.4cm,height=4.4cm]{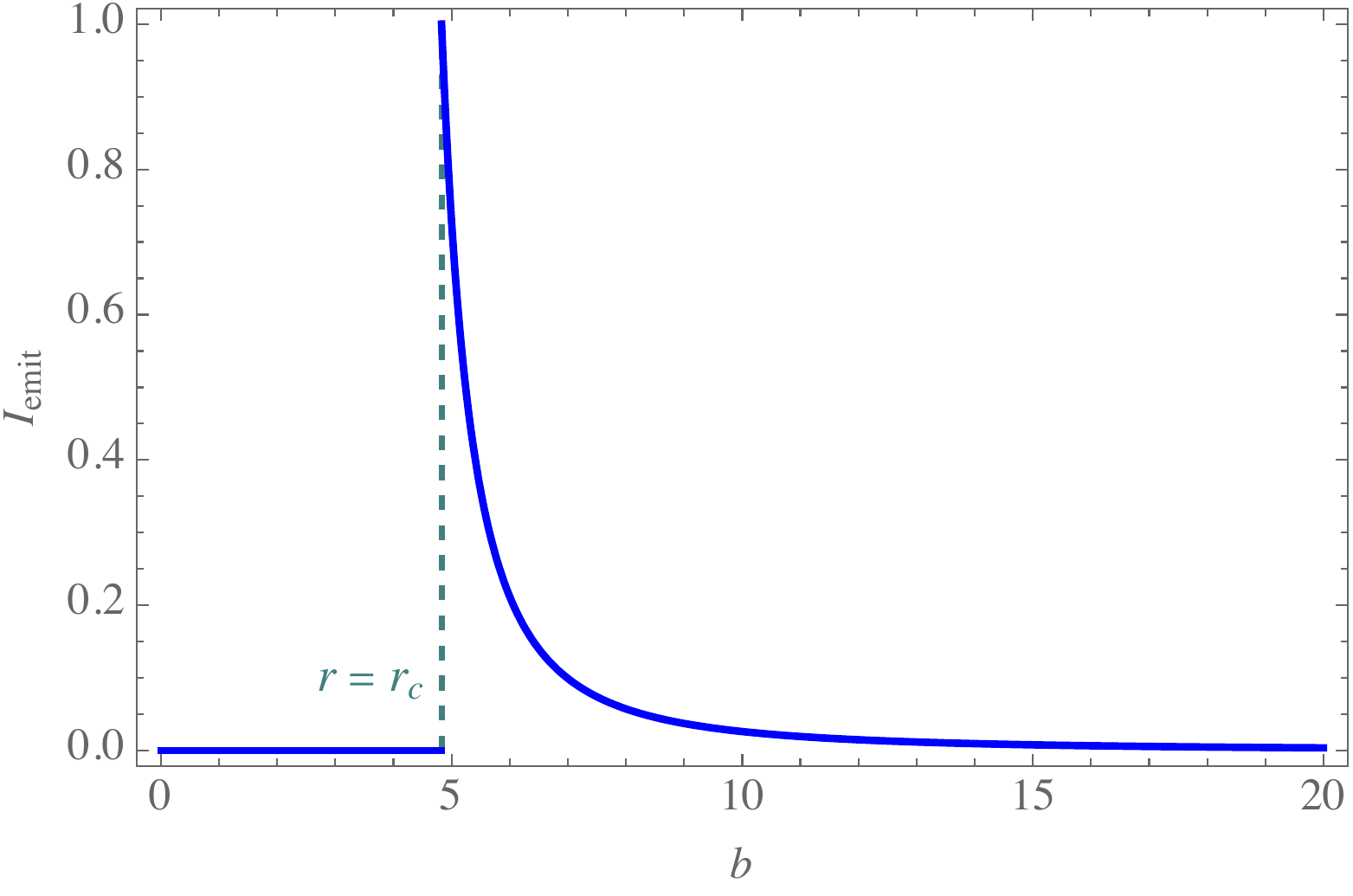}
    \includegraphics[width=5.4cm,height=4.4cm]{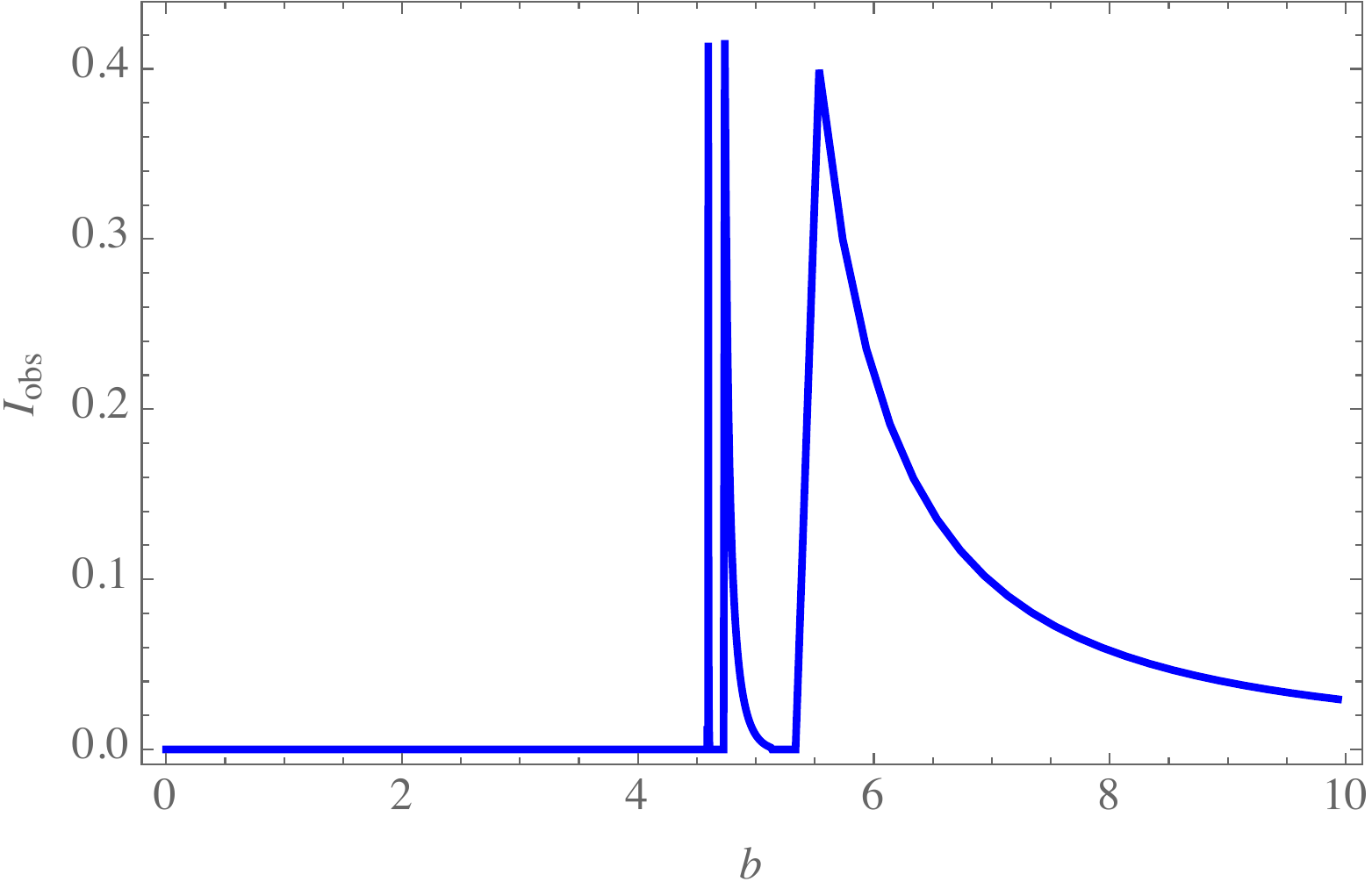}
    \includegraphics[width=5.4cm,height=4.5cm]{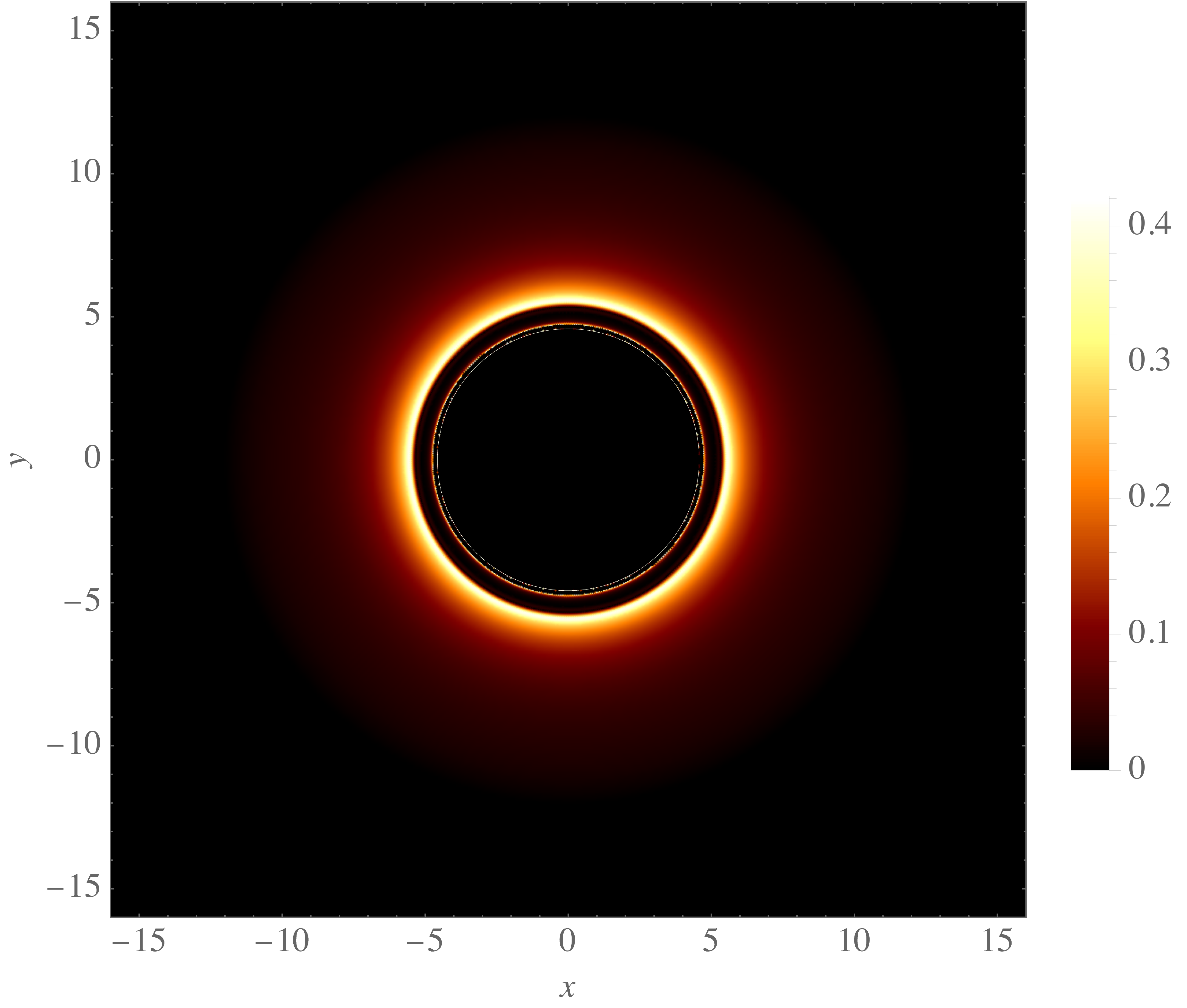}~(a)
    \includegraphics[width=5.4cm,height=4.4cm]{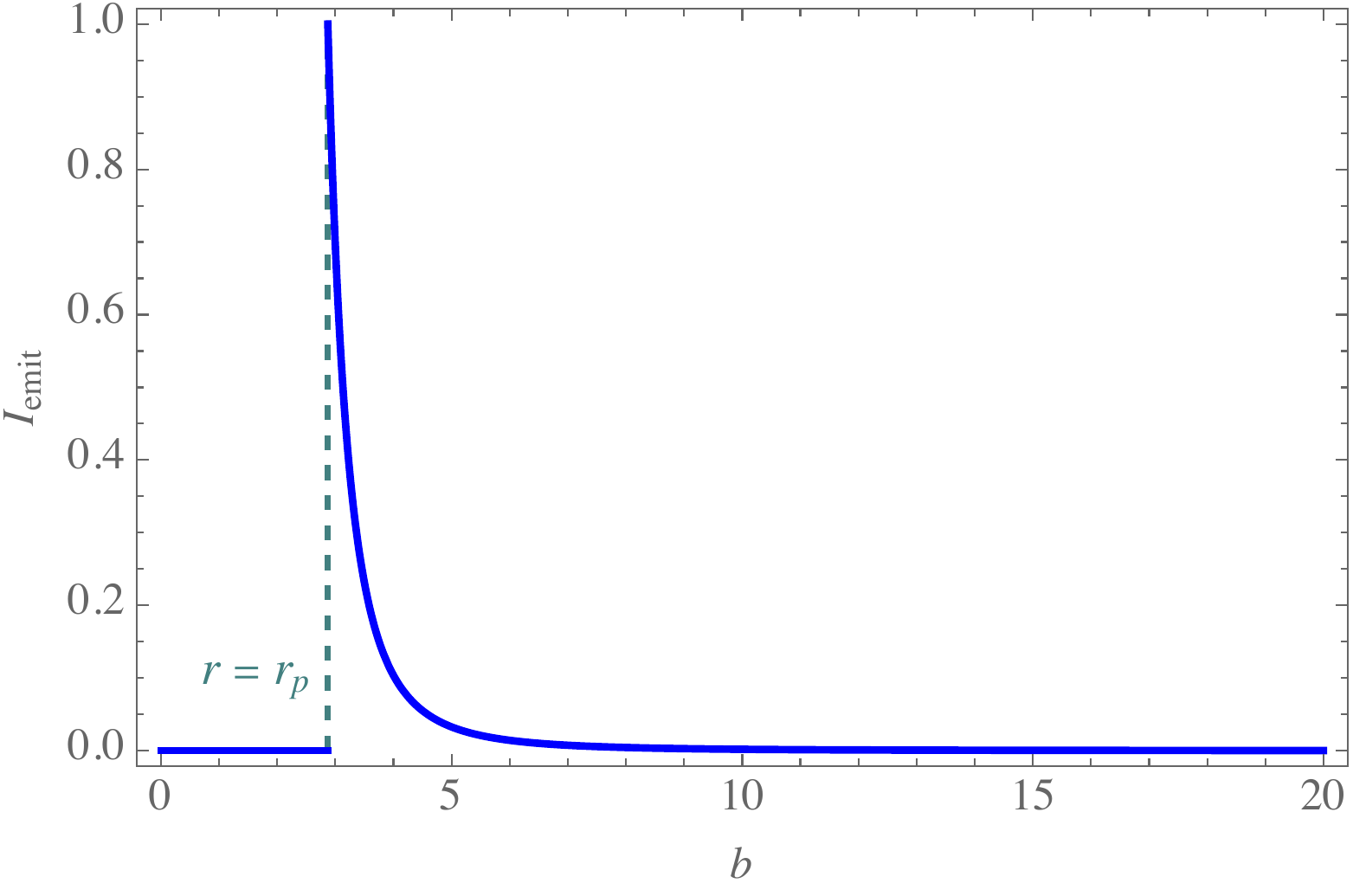}
    \includegraphics[width=5.4cm,height=4.4cm]{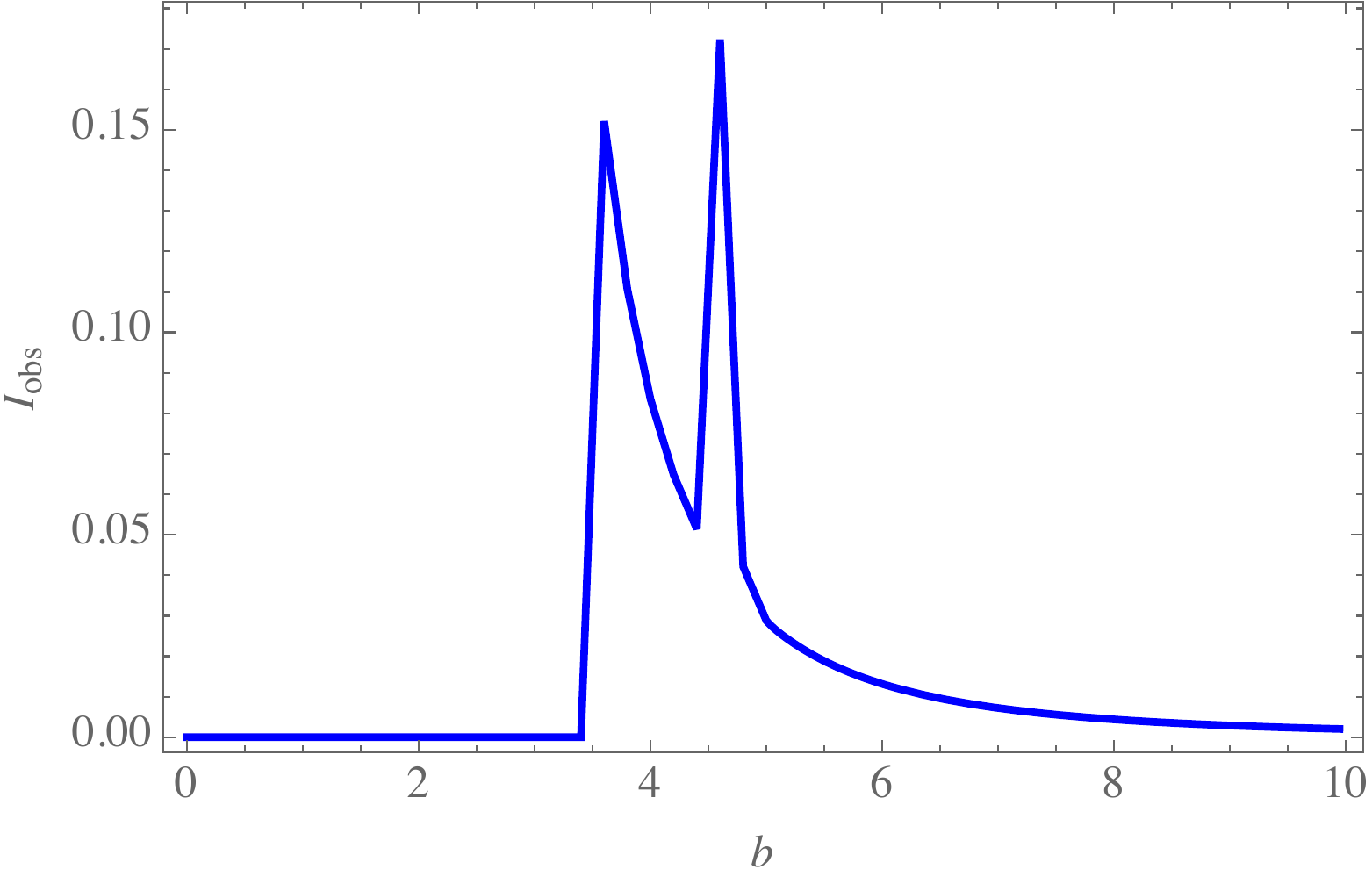}
    \includegraphics[width=5.4cm,height=4.5cm]{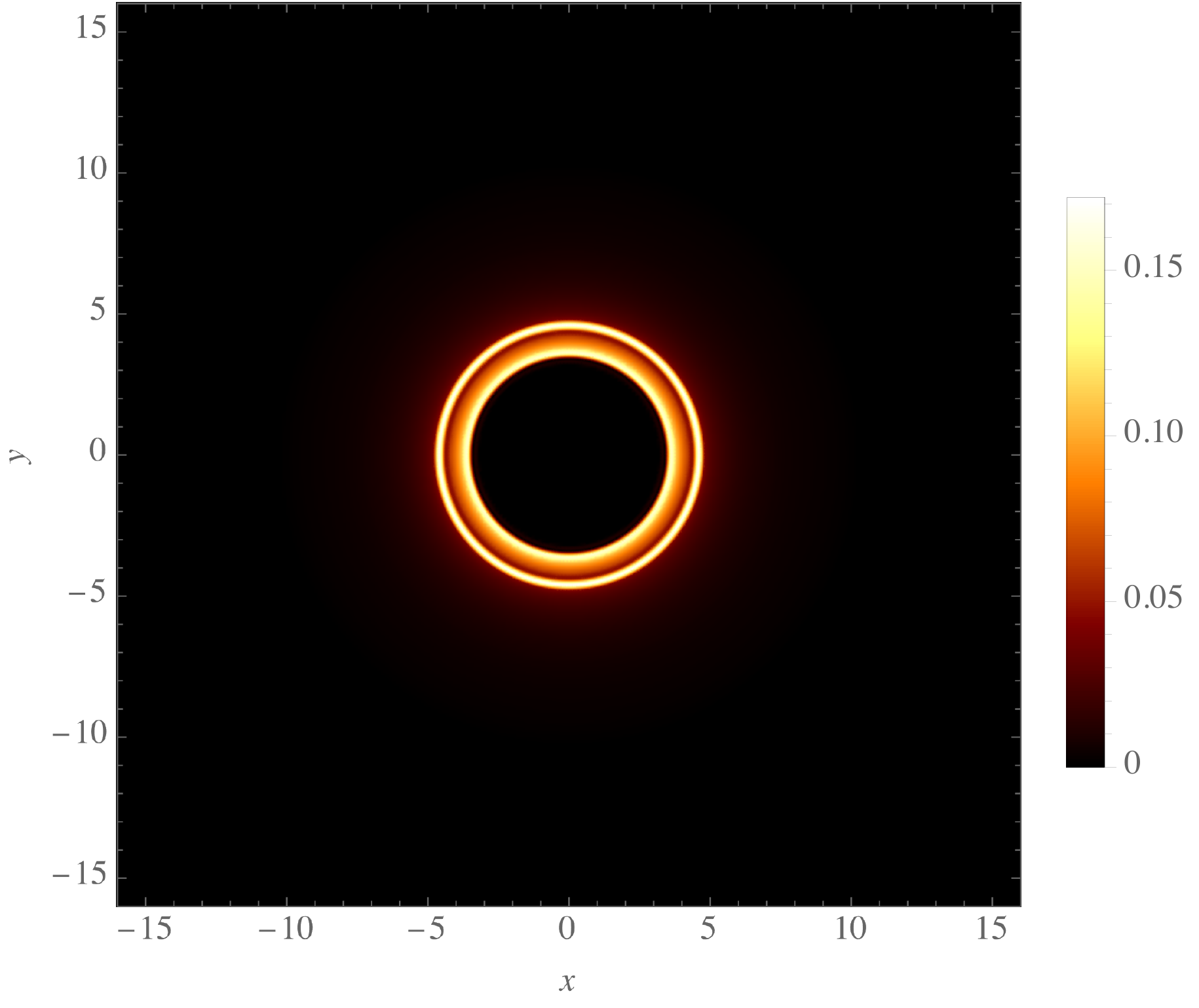}~(b)
    \includegraphics[width=5.4cm,height=4.4cm]{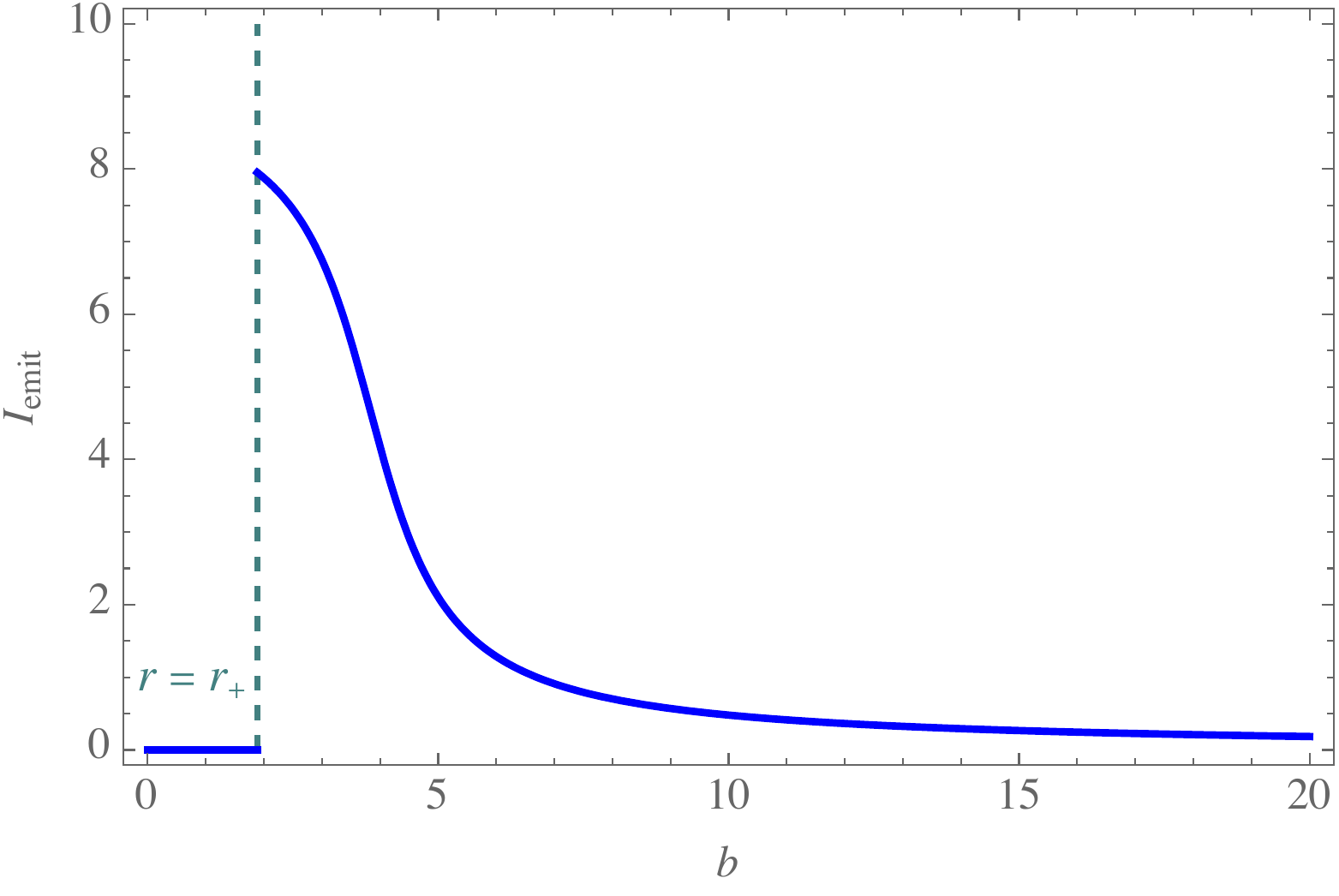}
    \includegraphics[width=5.4cm,height=4.4cm]{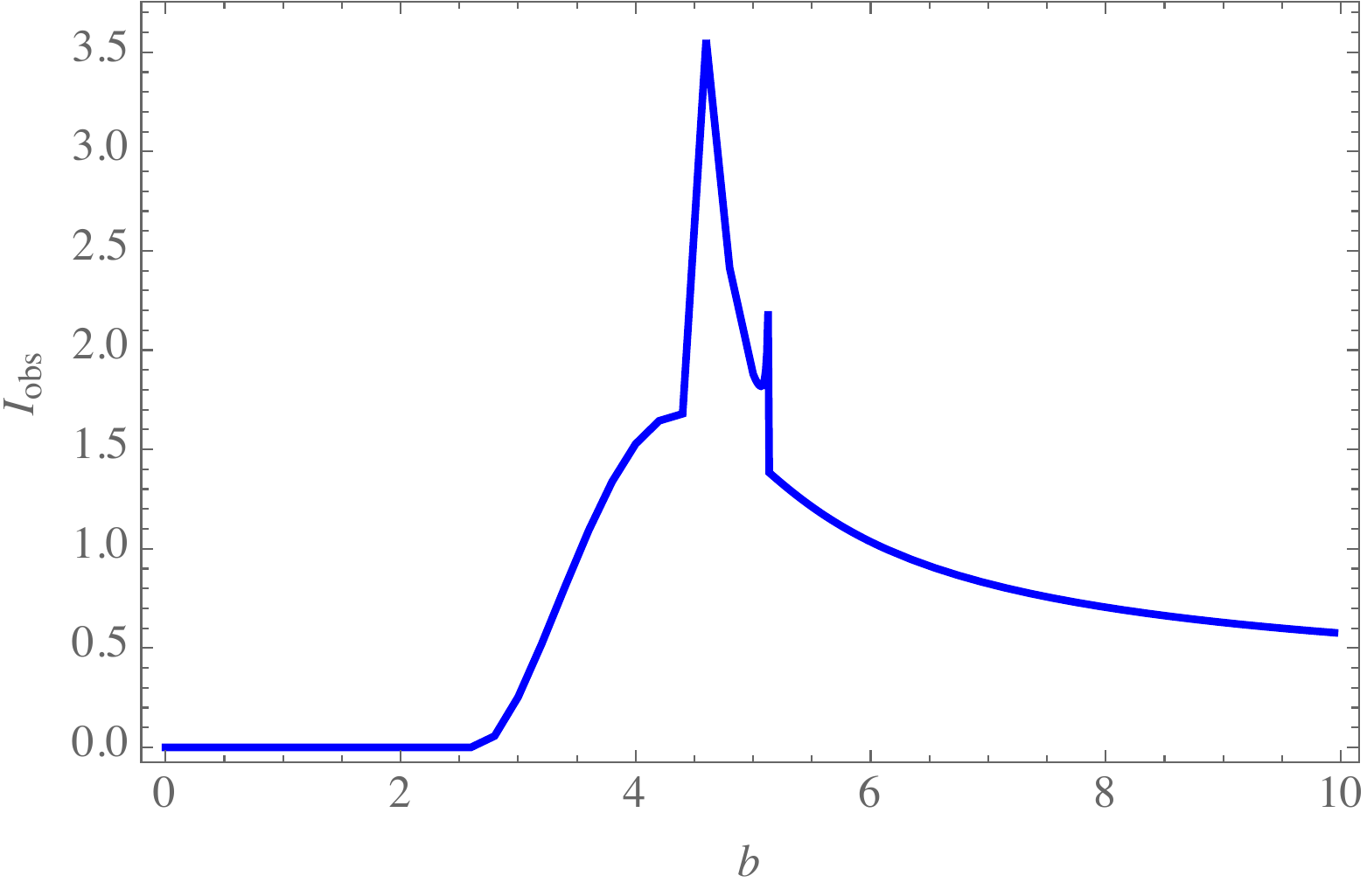}
    \includegraphics[width=5.4cm,height=4.5cm]{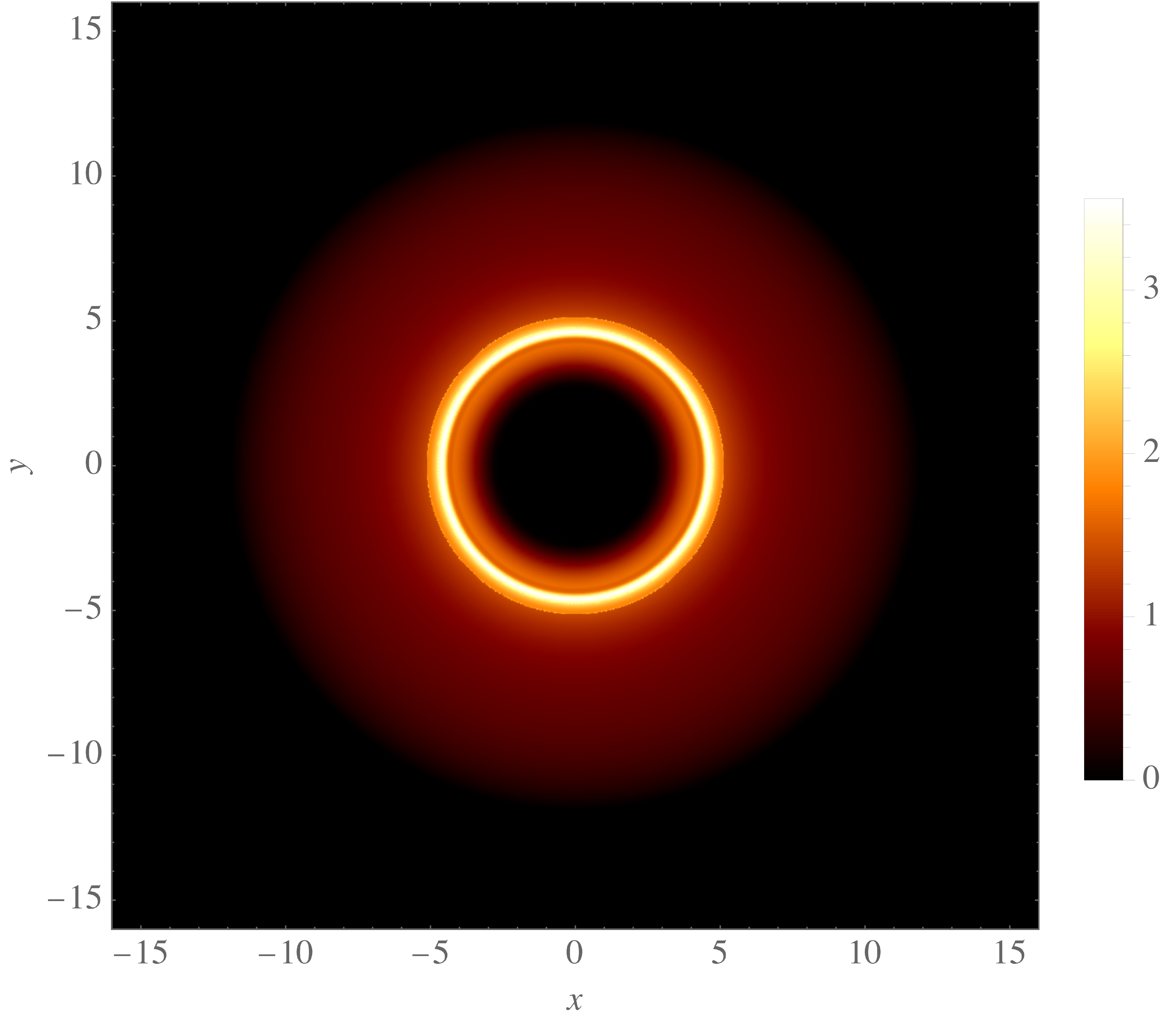}~(c)
    \caption{The case of $\beta=0.031$.}
    \label{fig:ring-beta3}
\end{figure}

\begin{figure}
    \centering
    \includegraphics[width=5.4cm,height=4.4cm]{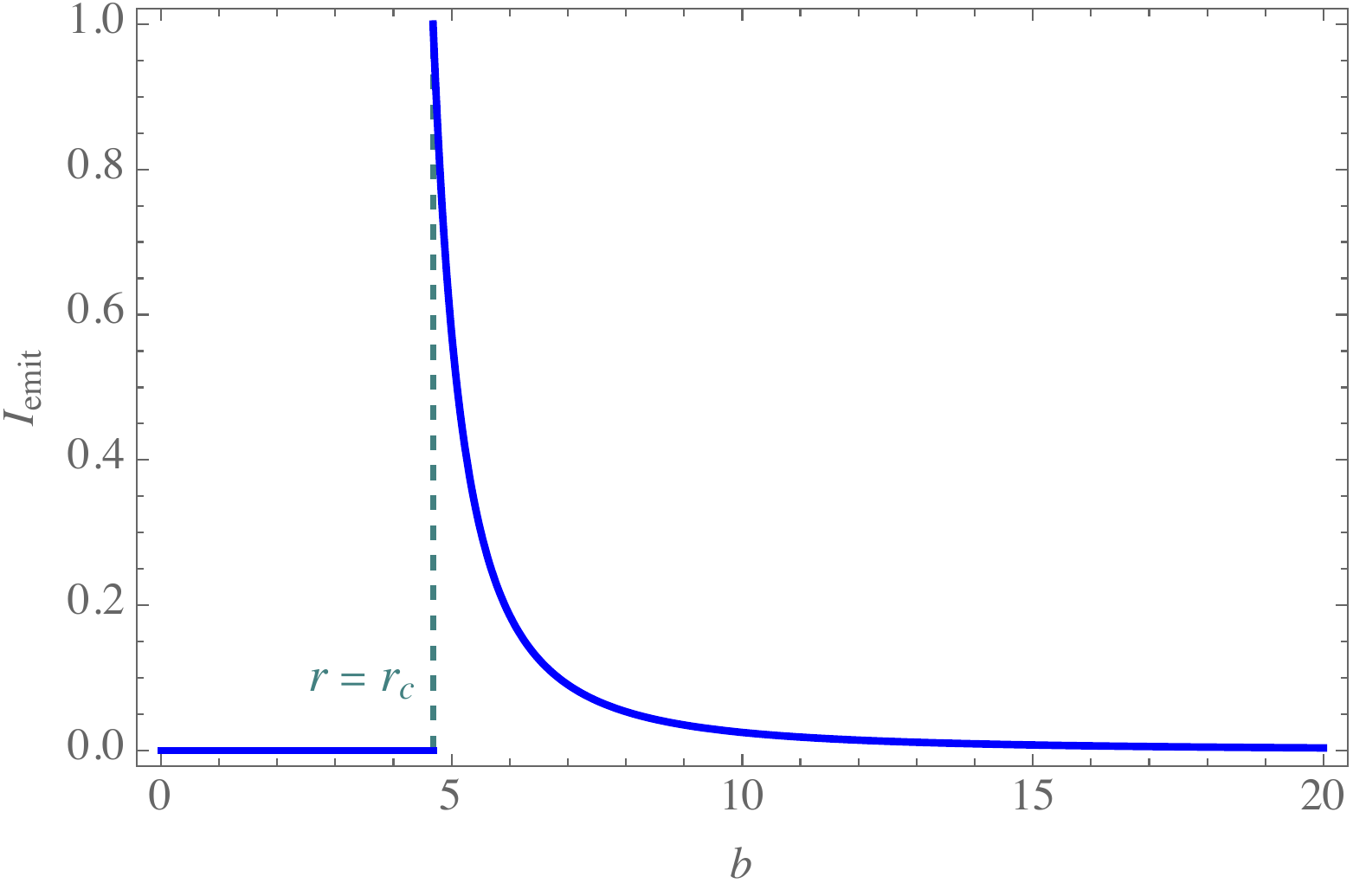}
    \includegraphics[width=5.4cm,height=4.4cm]{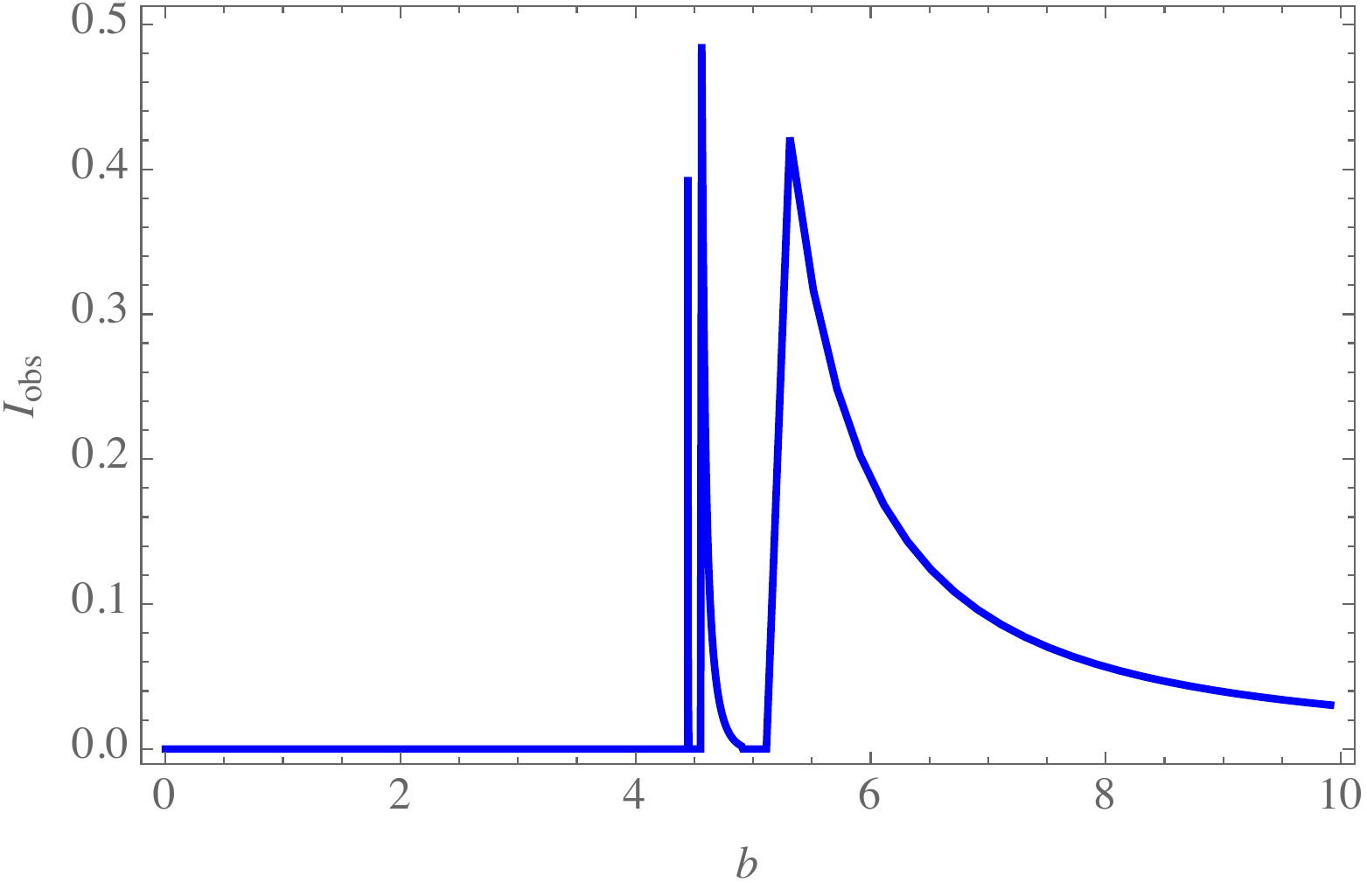}
    \includegraphics[width=5.4cm,height=4.5cm]{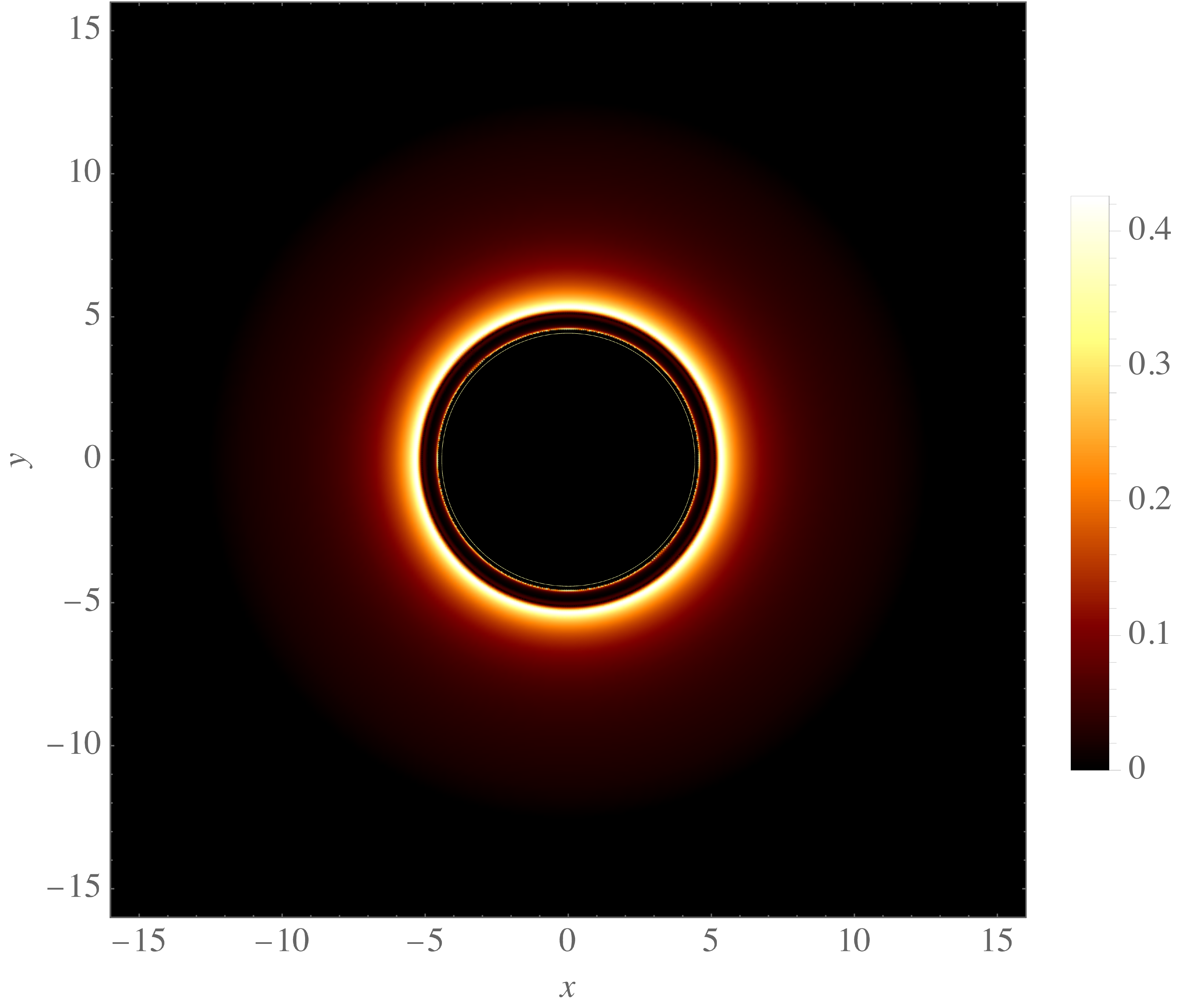}~(a)
    \includegraphics[width=5.4cm,height=4.4cm]{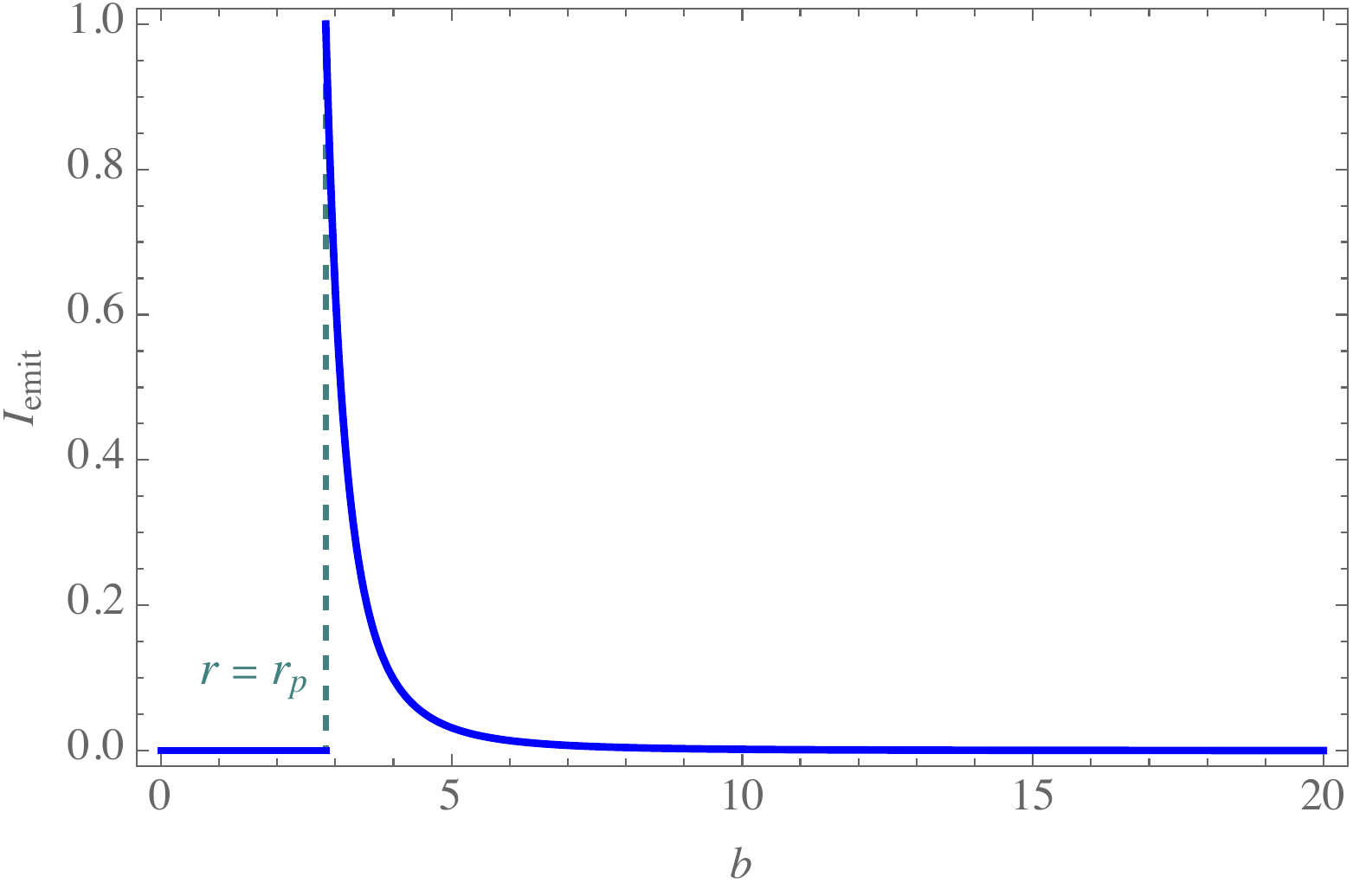}
    \includegraphics[width=5.4cm,height=4.4cm]{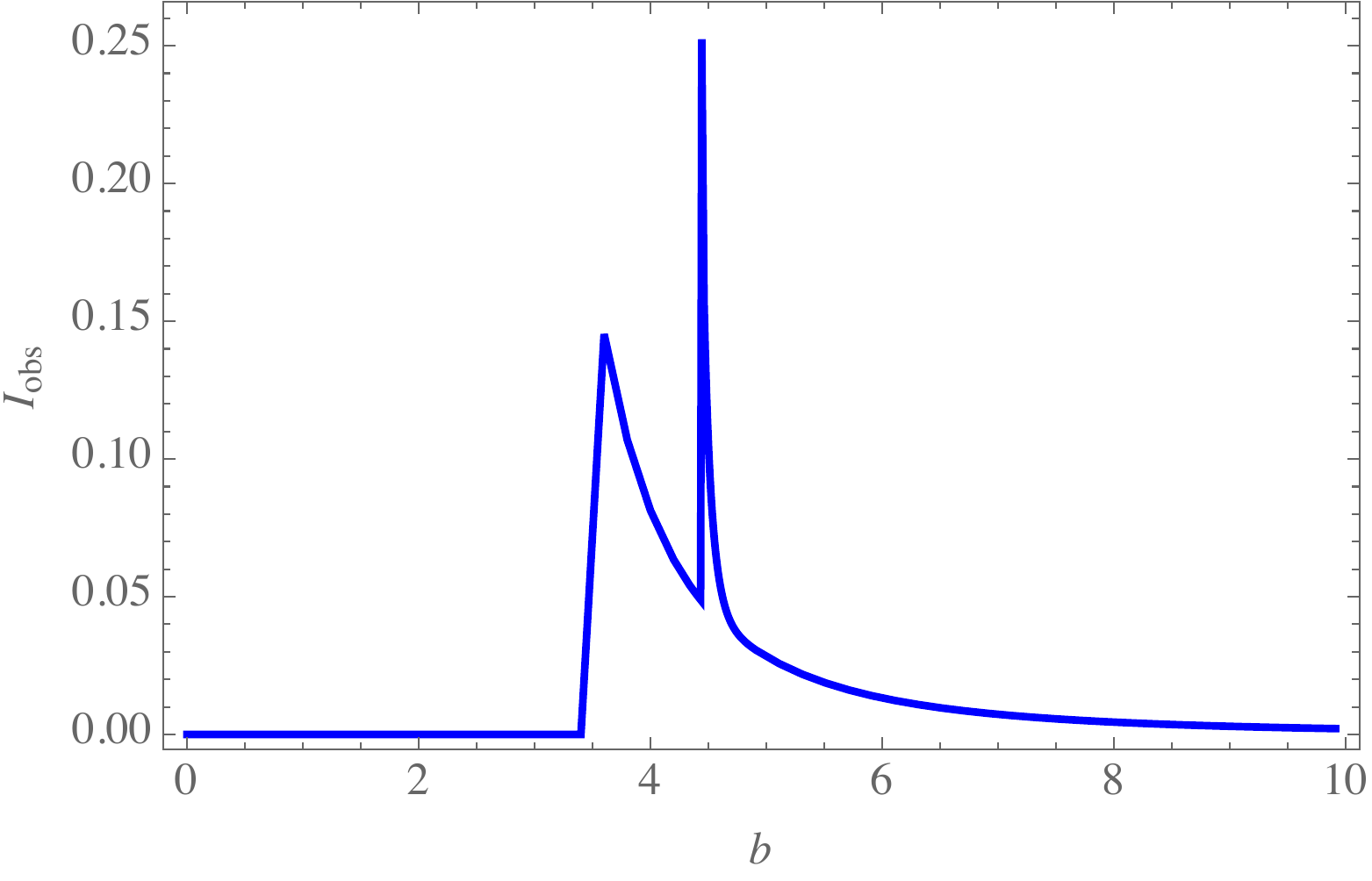}
    \includegraphics[width=5.4cm,height=4.5cm]{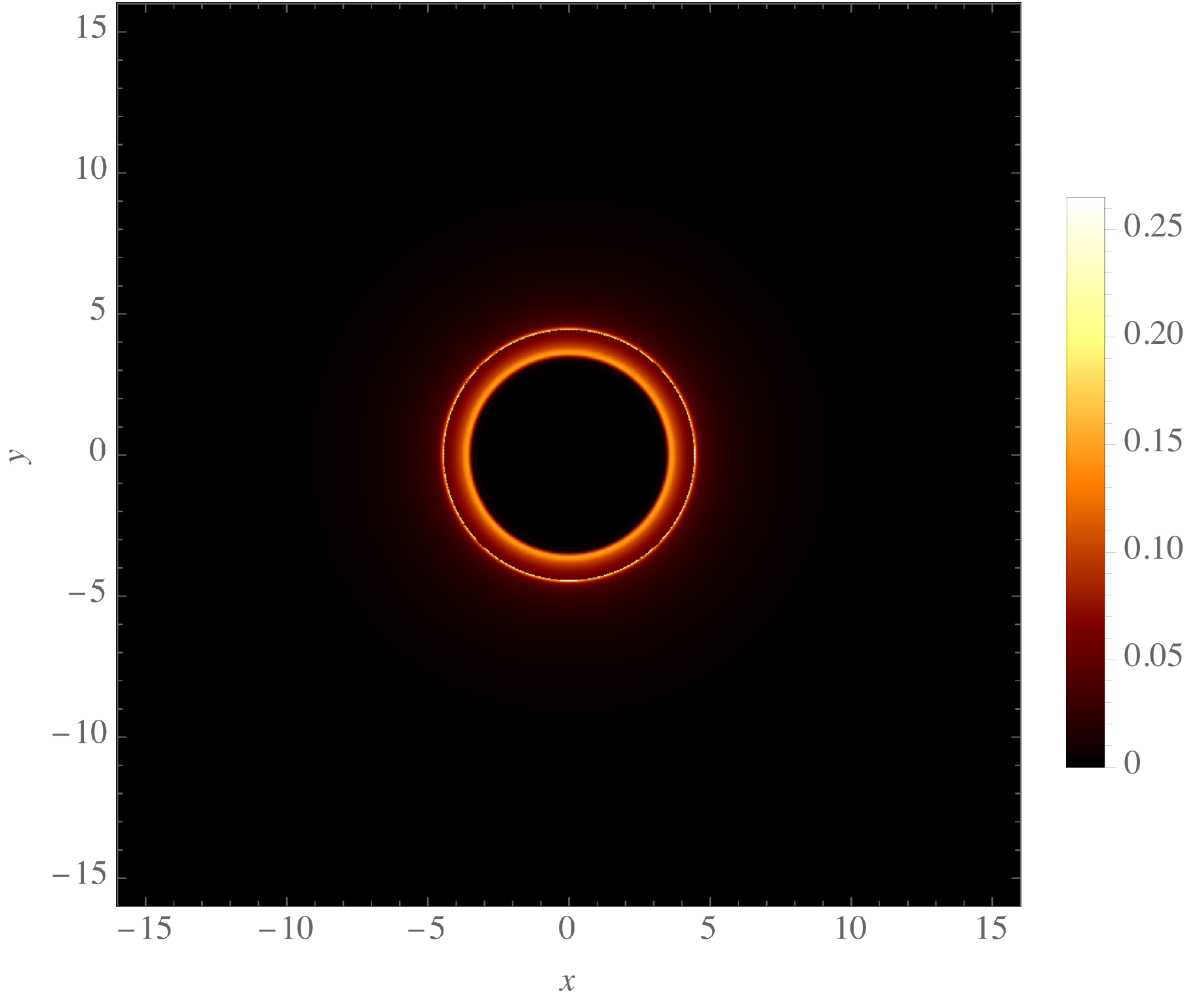}~(b)
    \includegraphics[width=5.4cm,height=4.4cm]{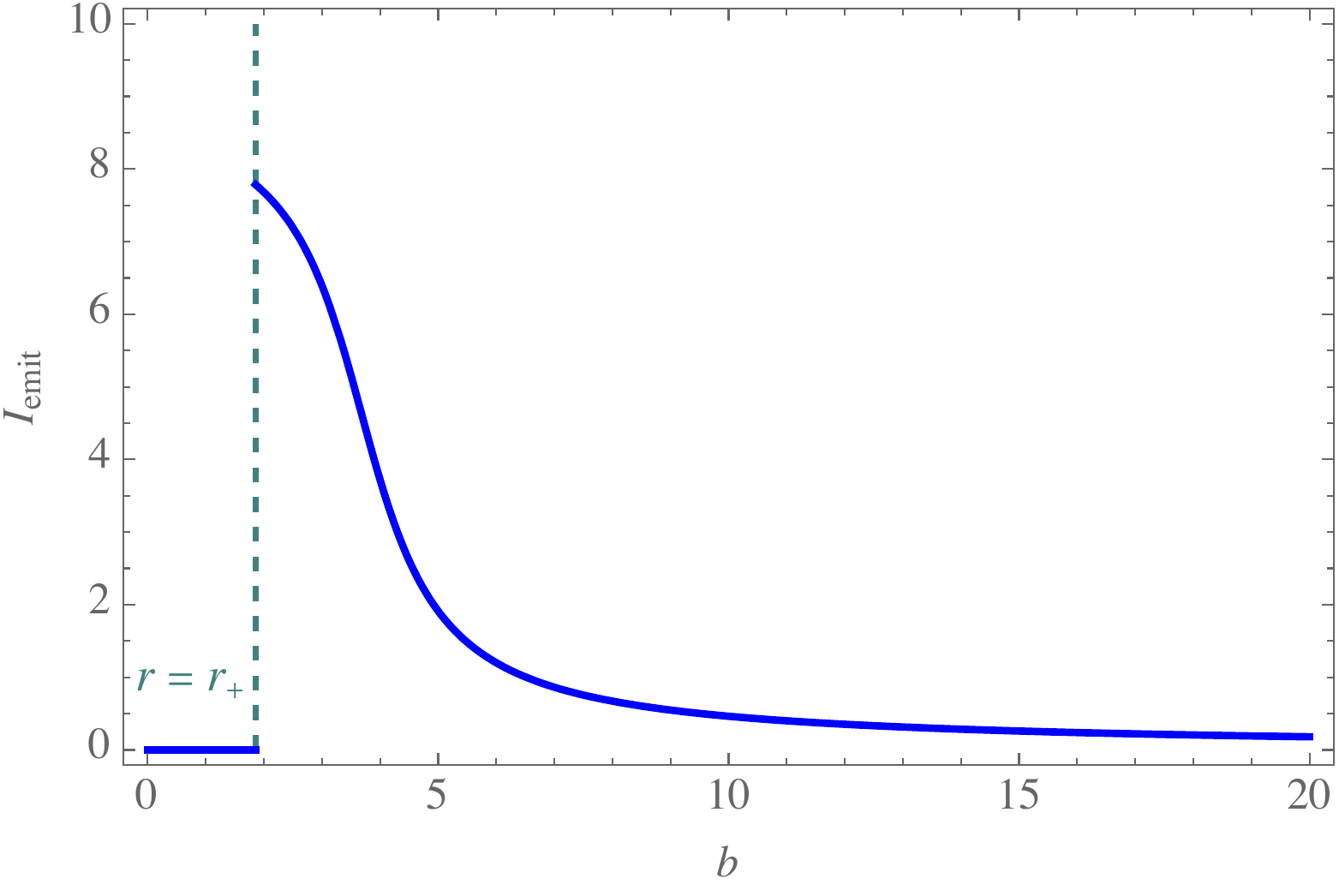}
    \includegraphics[width=5.4cm,height=4.4cm]{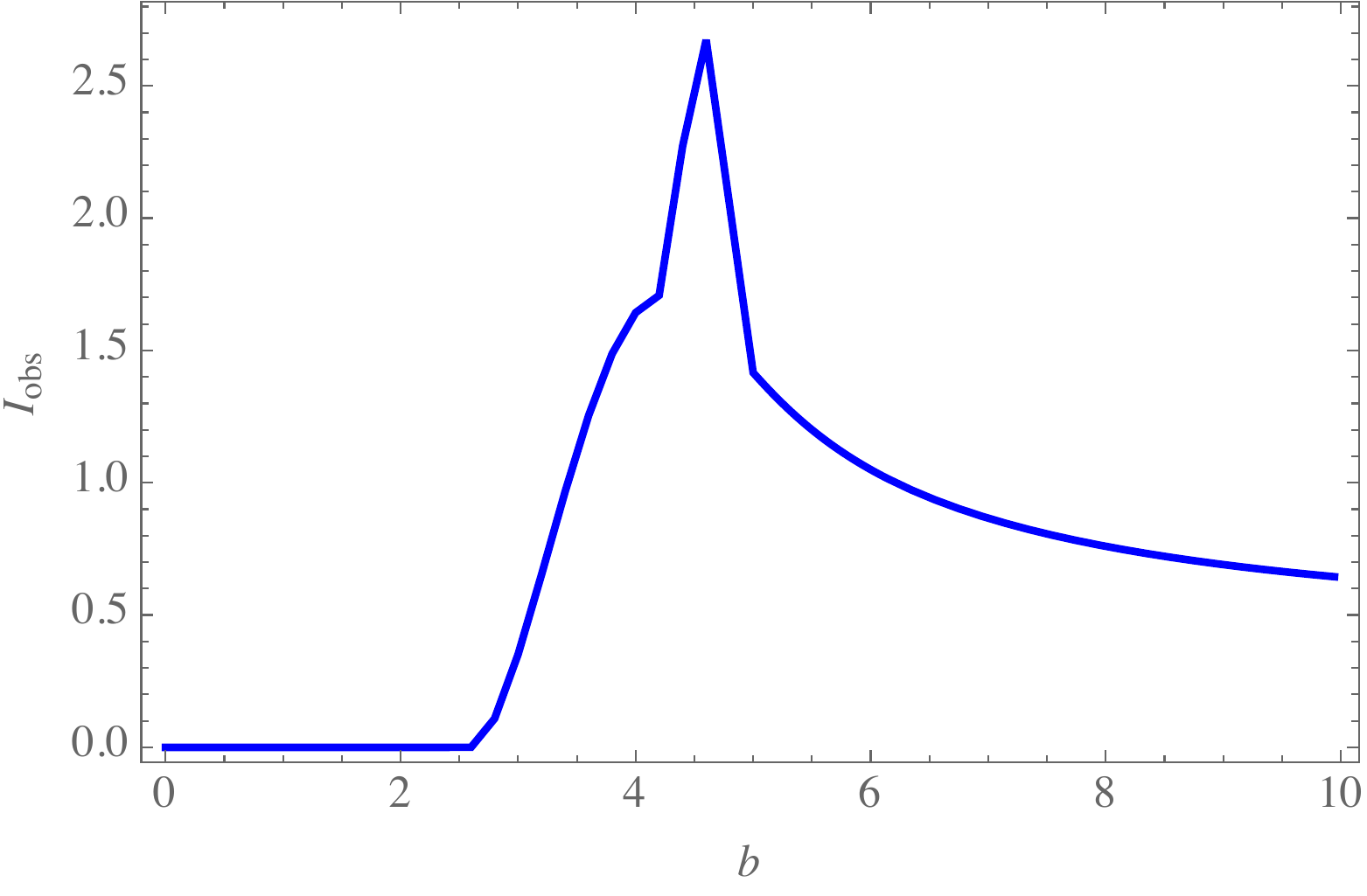}
    \includegraphics[width=5.4cm,height=4.5cm]{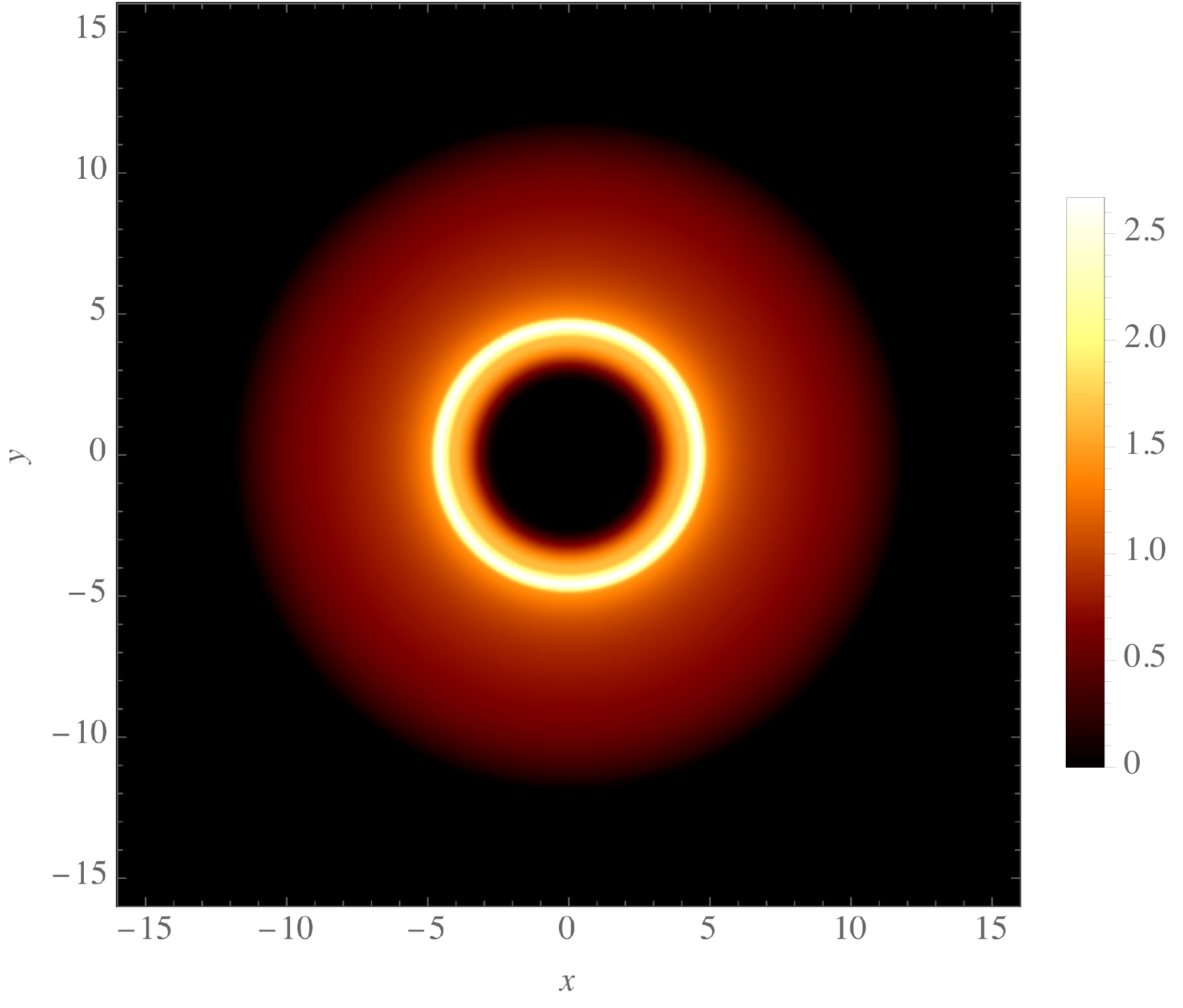}~(c)
    \caption{The case of $\beta=0.041$.}
    \label{fig:ring-beta4}
\end{figure}

From the perspective of a distant observer, the accretion disk serves as the primary light source that illuminates the black hole. The brightness of this source solely depends on the radial coordinate $r$, and as discussed earlier, it can be represented in terms of the emitted intensity $I_\emit$. To explore the observational features of the $f(R)$ black hole further, we consider three toy models for the intensity profile of the thin accretion disk, as described below:

\begin{itemize}

    \item \textbf{Model 1}: In this model, the emission comes from the ISCO, and the intensity profile is given by the decaying function
    \begin{equation}
\displaystyle I_\emit(r) = \begin{cases} 
 {\left[r-(r_c-1)\right]^{-2}} & \text{for $r>r_c$} \\  
 0 & \text{for $r\leq r_c$} 
 \end{cases}.
    \label{eq:Model1}
\end{equation}

    \item \textbf{Model 2}: We assume that the radiation is originated from the photon sphere of the radius $r_p$, and the emission intensity profile is expressed as
    \begin{equation}
\displaystyle I_\emit(r) = \begin{cases} 
 {\left[r-(r_p-1)\right]^{-3}} & \text{for $r>r_p$} \\  
 0 & \text{for $r\leq r_p$} 
 \end{cases}.
    \label{eq:Model2}
\end{equation}

\item \textbf{Model 3}: For the case that the emission starts from the event horizon radius $r_+$, we consider an emission profile in which, the decay is more moderate compared with the last two models, and is given by
    \begin{equation}
\displaystyle I_\emit(r) = \begin{cases} 
 \left[\frac{\pi}{2}-\arctan\left(r-[r_c-1]\right)\right]\left[{\frac{\pi}{2}-\arctan(r_p)}\right]^{-1} & \text{for $r>r_+$} \\  
 0 & \text{for $r\leq r_+$} 
 \end{cases}.
    \label{eq:Model2}
\end{equation}

\end{itemize}
Table \ref{tab:rprcrP} presents the radii of the event horizon, the ISCO, and the photon sphere, serving as the origins for the aforementioned emission models, based on different values of the $\beta$-parameter considered thus far.
\begin{table}[h!]
\centering
 \begin{tabular}{c  c c c} 
 \hline
 
 $\beta$  &  $r_+$ &  $r_c$ &  $r_p$ \\ [0.5ex] 
 \hline\hline
 
 0.0 & $2.00008$ & $6.01981$ & $3.00000$ \\ 
 0.011 & $1.95638$ & $5.31895$ & $2.95031$ \\ 
 0.022 & $1.91959$ & $5.01730$ & $2.90767$ \\ 
 0.031 & $1.88795$ & $4.82752$ & $2.87042$ \\ 
 0.041 & $1.85826$ & $4.68106$ & $2.83497$  \\ [1ex] 
\hline
 \end{tabular}
 \caption{The radii of the origins for the radiation emission models, provided for various values of the $\beta$-parameter.}
 \label{tab:rprcrP}
\end{table}
Note that, the above models have their own specific properties respecting the black hole shadow, and the second model emission profile shows the largest decay. These models, despite being rather idealized, however, can provide useful insights into the light propagation in the exterior of black holes. In Fig.~\ref{fig:ring-beta0}--\ref{fig:ring-beta4}, the observational appearance of the accretion disk around the $f(R)$ black hole has been shown for each of the above models, together with the plots of the emitted and observed intensities for each of the cases.
\begin{figure}
    \centering
    \includegraphics[scale=.4]{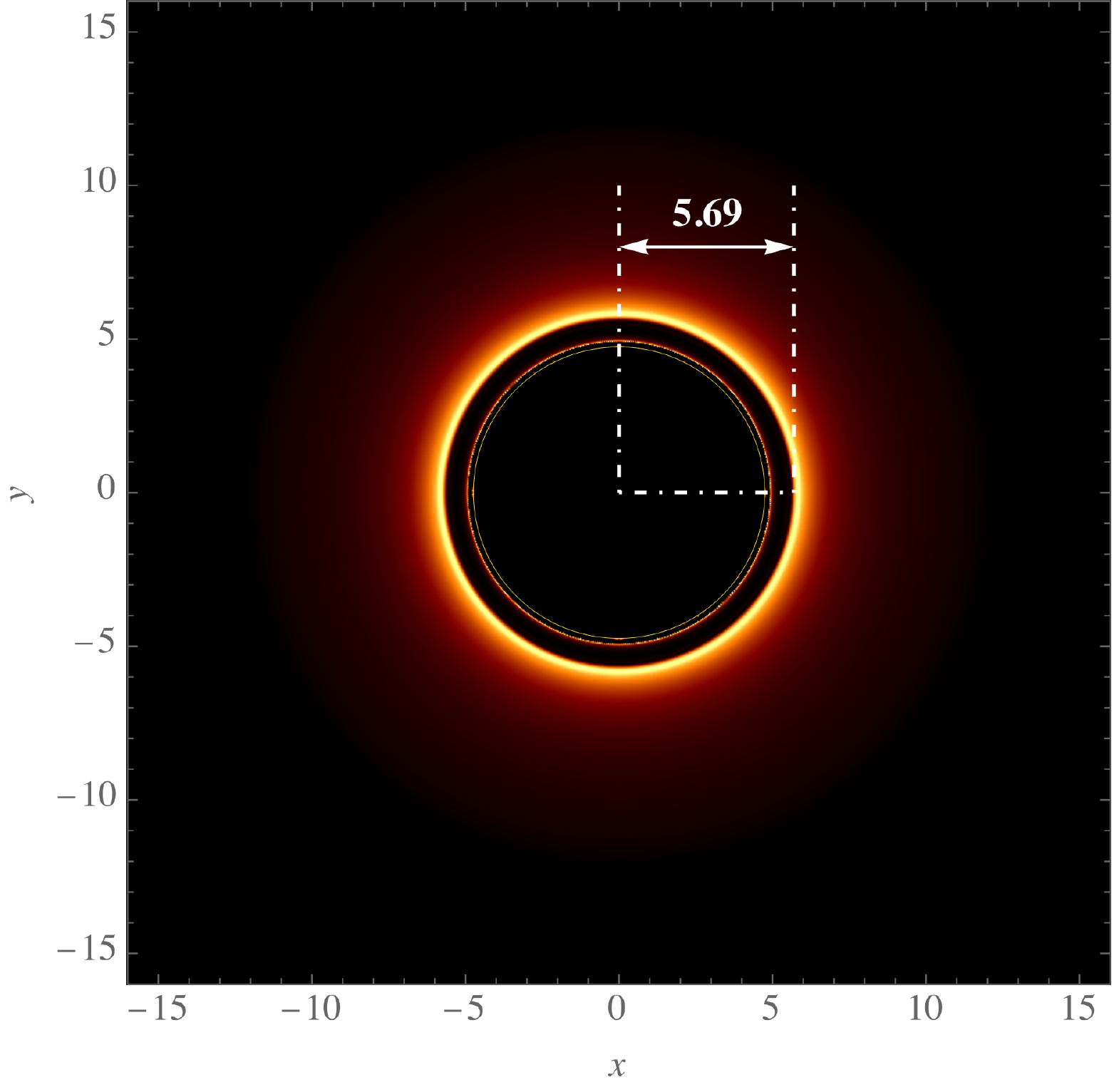}\qquad\qquad
      \includegraphics[scale=0.4]{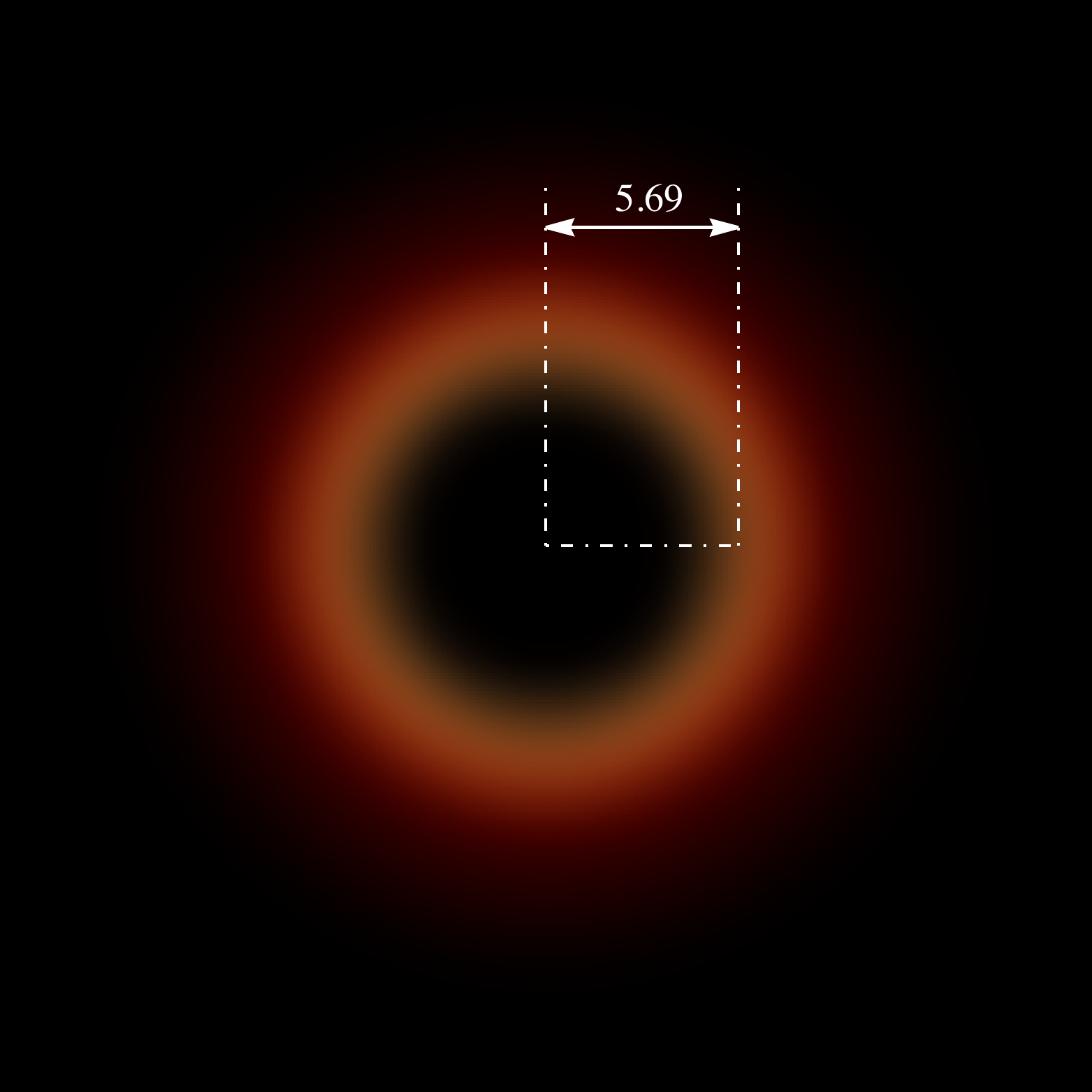}
    \caption{Blurring the shadow in Fig.~\ref{fig:ring-beta2}(a) for $\beta=0.022$ using a Gaussian filter, to emulate the EHT nominal resolutions for the images of M87* and Sgr A*. In the left panel, the starting radius of the direct emission has been shown to be about 5.69, which forms the boundary of the shadow. After applying the Gaussian filter, the lensing and photon rings disappear in the right panel, and hence, the radius of the black hole shadow is estimated as 5.69. This value is much larger than the radius of the photon ring, which is $b_p=4.74$ for the case of $\beta=0.022$.}
    \label{fig:ring-beta2_Gaussian}
\end{figure}

The figures in each row represent, respectively, the emitted and observed intensity, and the shadow of the $f(R)$ black hole for models 1, 2, and 3. In model 1, the emitted intensity shows an asymptotic behavior near $b_p$, decreasing as the radial distance increases and approaching zero. Here, spherical photon orbits occur inside the disk's emission region. The observed intensity exhibits two independent peaks within the lensing ring and photon ring domains. For all values of the $\beta$-parameter, except $\beta=0.031$, the photon ring intensity is smaller than the direct emission, while the lensing ring intensity is always larger\footnote{In the two-dimensional ray-traced shadow images, zooming in reveals the photon ring as a thin circle inside the thicker lensing ring.}. Both peaks have remarkably narrow observational ranges, indicating that, at long distances where the observer is located, the contribution of the lensing and photon rings to the observed intensity is dominated by direct emission. Thus, the observed emission from the black hole in this model is mainly direct emission, as evident from the shadow images. Furthermore, comparing the diagrams for different $\beta$ values reveals that the shadow size decreases with an increase in this parameter, making the Schwarzschild-de Sitter black hole have the largest shadow for this model. In model 2, the emitted intensity peaks at $r_p$ and then sharply drops with increasing radial distance. The observed intensity has two peaks, with the first corresponding to direct emission and the second to a combination of lensing and photon rings. In some cases of $\beta$, both peaks have significant ranges and approximately equal intensities. For most cases of $\beta$, the direct emission dominates the observed intensity, while the rings are strongly demagnified. However, in the exceptional case of $\beta=0.031$, both direct emission and the rings contribute relatively equally to the observed intensity, as evident from the corresponding shadow image. In model 3, the emitted intensity peaks at the event horizon $r_+$ and decreases with radial distance. In this case, direct emission, lensing ring, and photon ring merge and occupy a significant range in the observational domain. Outside the event horizon, there is a smooth uplift in the observed intensity profiles, where direct emission dominates, followed by an intense peak corresponding to the lensing and photon rings. In the case of $\beta=0$ (the Schwarzschild-de Sitter black hole), the observed intensity exhibits a narrow but intense peak for the photon ring, followed by a smaller peak where both rings contribute, forming a wide and bright ring. This bright ring is observed in the intensity profiles and shadow images for all cases of the $\beta$-parameter. Additionally, the brightness of the accretion disk increases with an increase in the $\beta$-parameter. Notably, a thin accretion disk significantly influences the size of the observed black hole shadow. For instance, in Fig. \ref{fig:ring-beta2_Gaussian}, we have applied a Gaussian filter to simulate the angular resolution generated by the EHT on the shadow image presented in Fig. \ref{fig:ring-beta2}(a) as a reference. Calculating the maximums of the observed emission in Eq.~\eqref{eq:Iobs_2} for this case yields an estimated radius of $5.69$ for the direct emission peak, which is significantly larger than the theoretical value ($b_p=4.7445$ for $\beta=0.022$). This difference arises from changes in both the $\beta$-parameter and the disk's emission profile. Therefore, directly inferring the value of $b_p$ from the black hole shadow size makes it challenging to test general relativity using the EHT results.

\subsubsection{Observational signatures of infalling spherical accretion}\label{subsubsec:infalling}

\begin{figure}
    \centering
    \includegraphics[width=7cm]{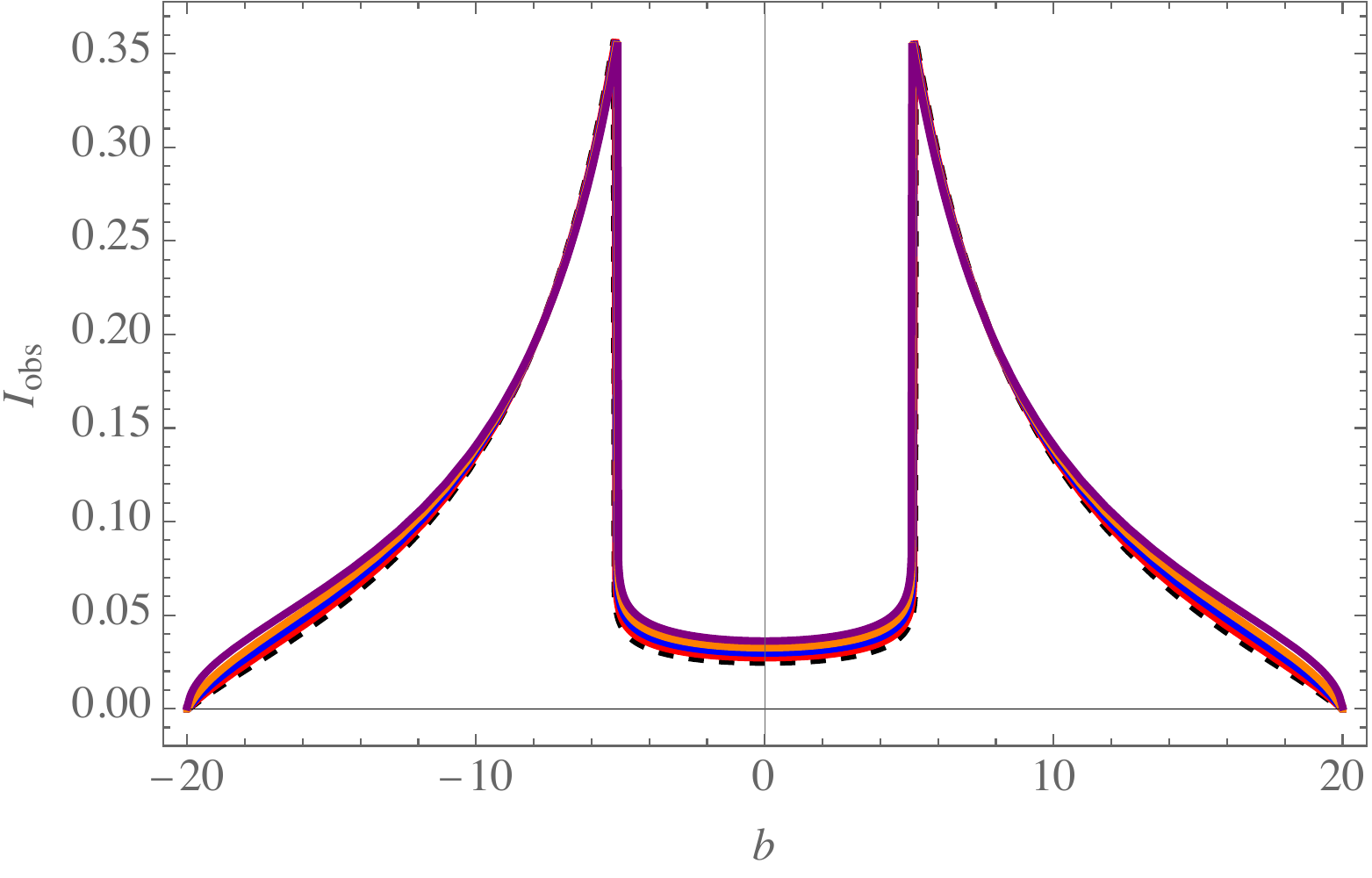}~(a)\qquad\qquad
     \includegraphics[width=7cm]{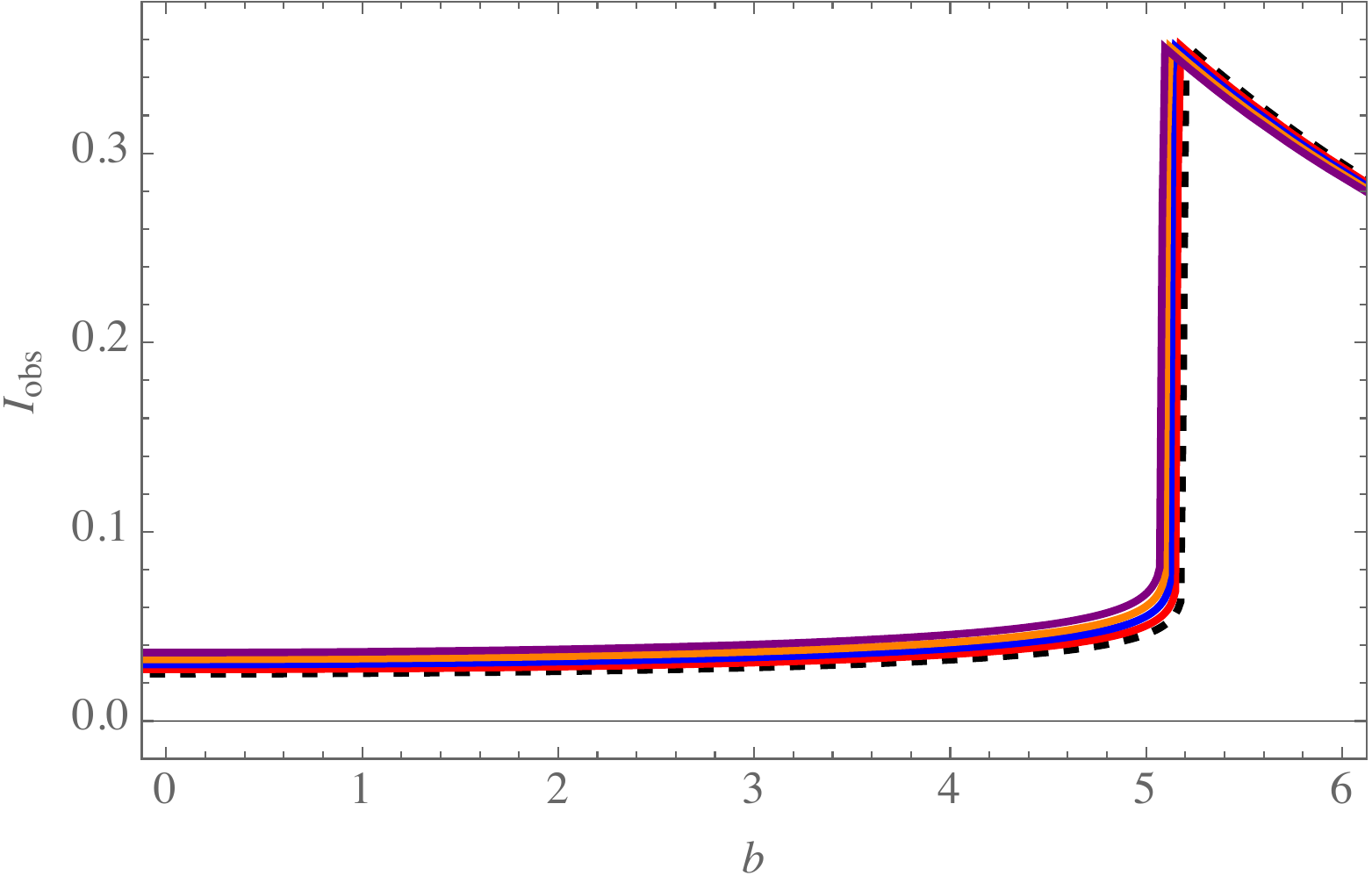}~(b)
    \caption{(a) The $b$-profiles of the observed intensity of the infalling spherical accretion, which from bottom to top correspond to the cases of $\beta = 0, 0.011, 0.022, 0.031$ and $0.041$. (b) The same as panel (a), but showing only a part of the $b$ domain within the positive values.}
    \label{fig:I_obs_infall}
\end{figure}

\begin{figure}[h]
    \centering
    \includegraphics[width=5.4cm]{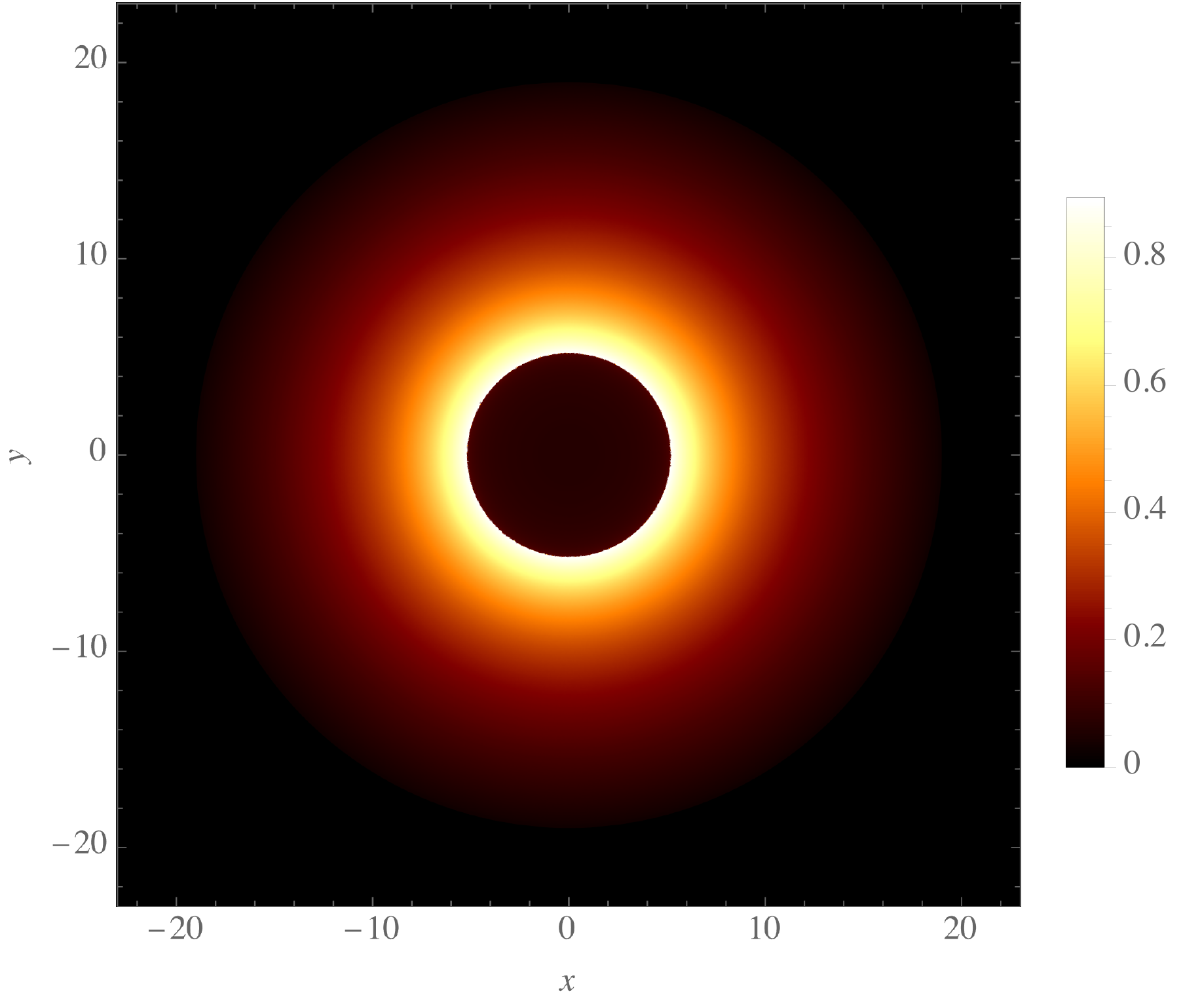}~(a)
     \includegraphics[width=5.4cm]{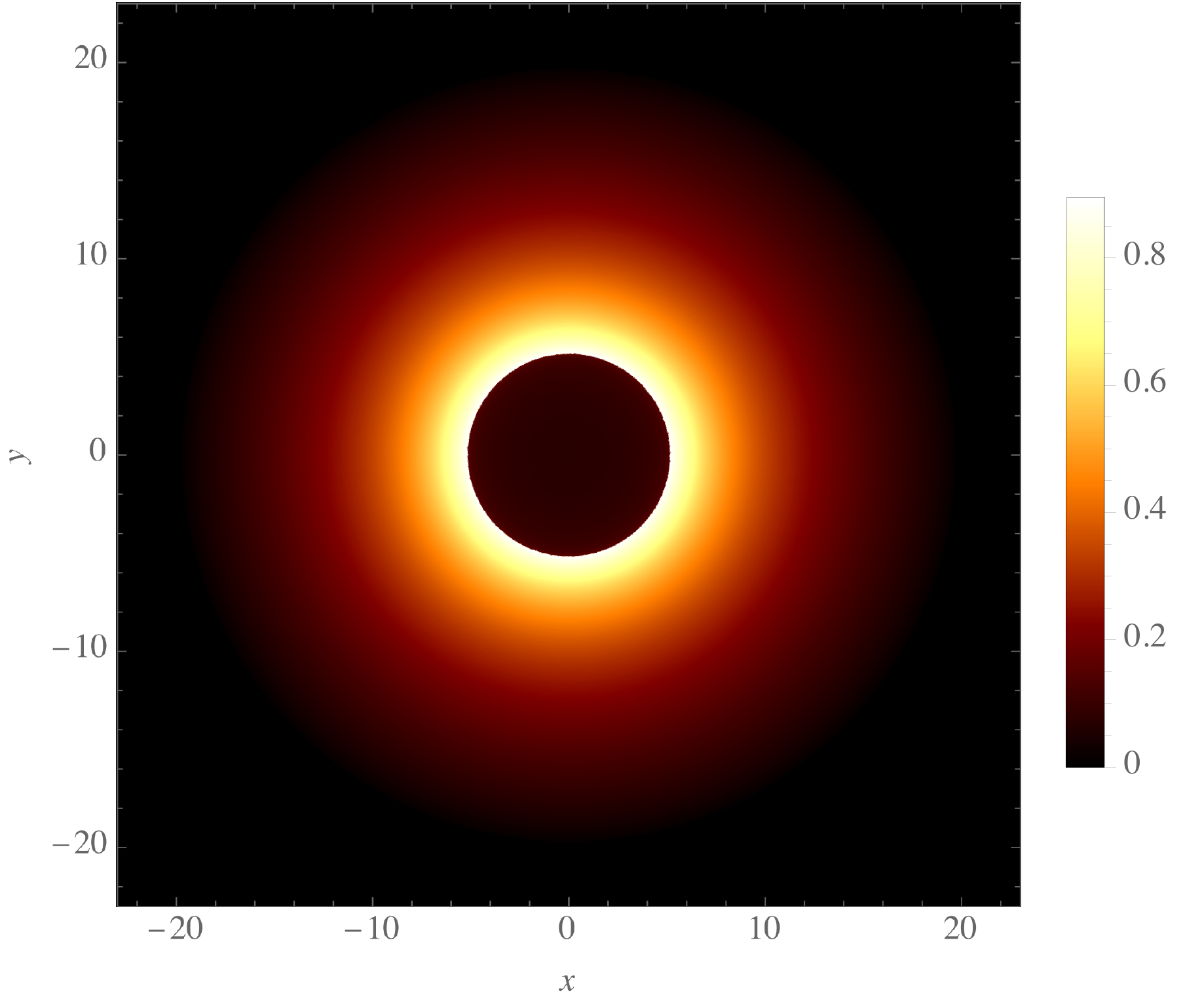}~(b)
     \includegraphics[width=5.4cm]{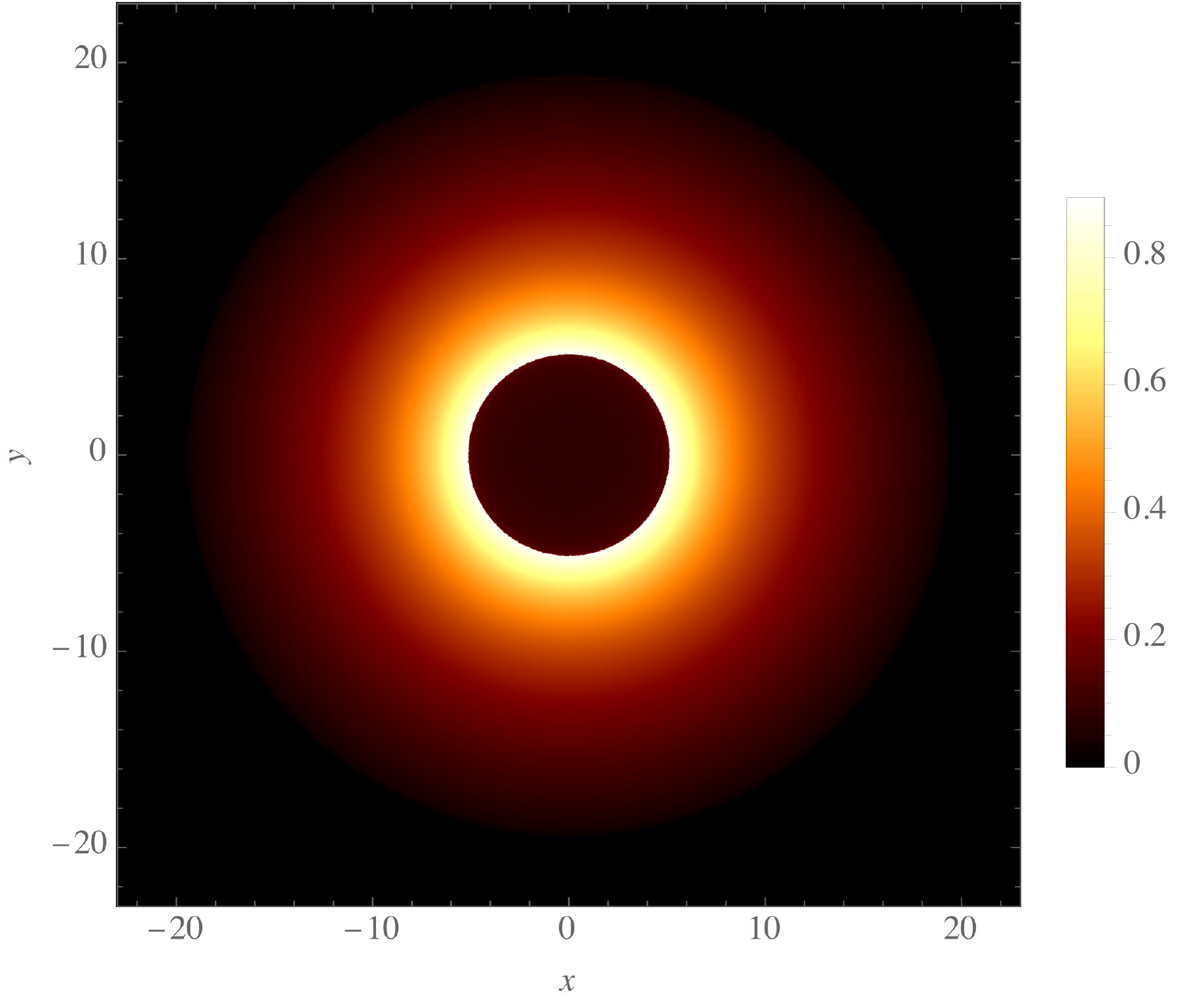}~(c)
     \includegraphics[width=5.4cm]{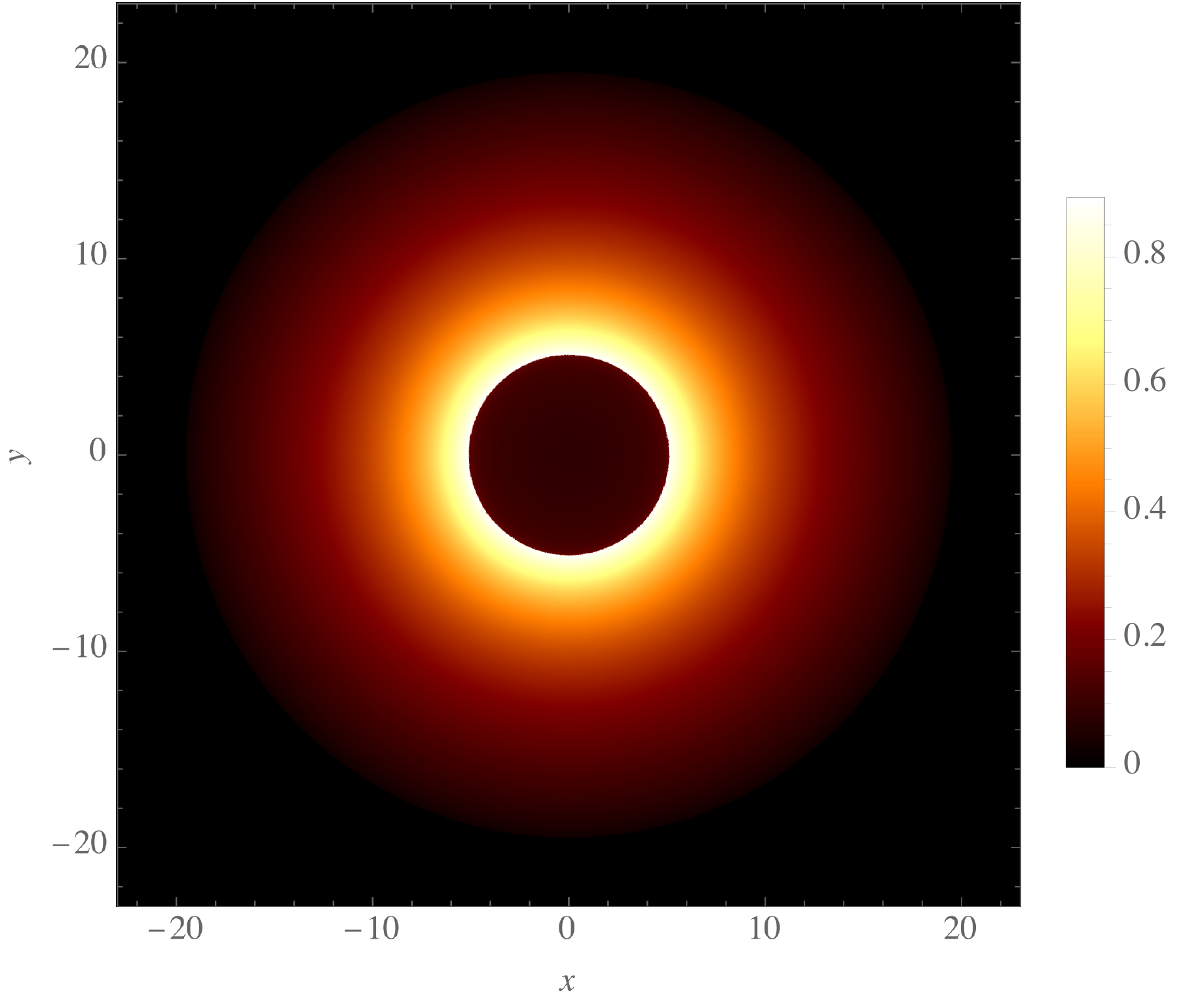}~(d)
     \includegraphics[width=5.4cm]{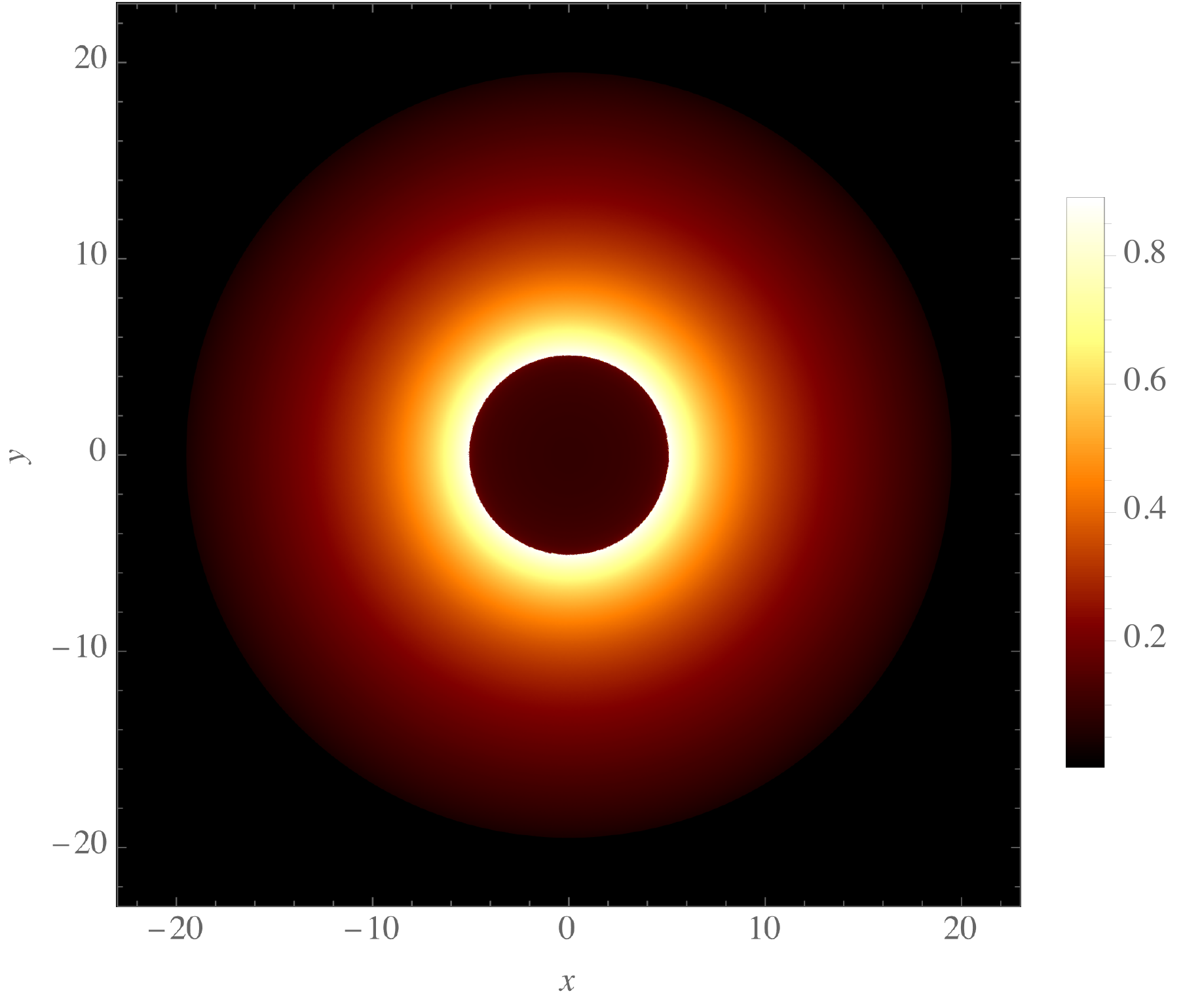}~(e)
    \caption{The images of the disk and the silhouette of the black hole with infalling spherical accretion, given for (a) $\beta = 0$, (b) $\beta = 0.011$, (c) $\beta = 0.022$, (d) $\beta = 0.031$ and (e) $\beta = 0.041$.}
    \label{fig:shadow_infall}
\end{figure}

Here, we investigate the shadow cast of the $f(R)$ black hole while it accretes spherically the radiative gas, constituting its thin emission disk \cite{bambi_can_2013}. In this model, the observed intensity is expressed as 
\begin{equation}
I_\obs=\int_{\bm{\gamma}} \mathscr{R}^3 \mathcal{J}(\nu_\e)\,\ed I_\prop,
    \label{eq:I_obs_infalling_0}
\end{equation}
over the null geodesic congruence $\bm{\gamma}$, in which $\mathscr{R}$ is the redshift factor, $\nu_\e$ is the frequency of emitted photons from the accretion disk, $\ed I_\prop$ is the infinitesimal proper length, and 
\begin{equation}
\mathcal{J}(\nu_\e)\propto\frac{\delta(\nu_\e-\nu_\f)}{r^2},
    \label{eq:j(nu)_0}
\end{equation}
is the permittivity per unit volume in the emitter's rest frame, in which $\nu_\f$ is the monochromatic rest-frame emission frequency, and $\delta$ is the delta function. In this construction, the redshift factor is given by
\begin{equation}
\mathscr{R}=\frac{\Pi_\mu u_\oo^{\mu}}{\Pi_{\nu} u_\e^\nu},
    \label{eq:redshift_factor_0}
\end{equation}
where $\bm{u}_\oo$ and $\bm{u}_\e$ are, respectively, the four-velocities associated with a distant static observer, and the infalling accreting matter. Accordingly, $u_\oo^{\mu}=(1,0,0,0)$, and in the spacetime of the $f(R)$ black hole we can write
\begin{equation}
u_\e^{\mu}=\left(\frac{1}{B(r)},-\sqrt{1-B(r)},0,0\right).
    \label{eq:u_e}
\end{equation}
The $\bm{\Pi}$ covector in Eq.~\eqref{eq:redshift_factor_0} is the four-momentum of the emitted photons from the accretion disk and has the same definition as in Eq.~\eqref{eq:conj_moment}. Since the accretion is supposed to be only in the radial direction, it is then sufficient to recalculate the fraction of the temporal and radial components of $\bm{\Pi}$, which yields \cite{bambi_can_2013}
\begin{equation}
\frac{\Pi_r}{\Pi_t}=\pm\frac{1}{B(r)}\sqrt{1-B(r)\frac{b^2}{r^2}},
    \label{eq:conj_moment_new}
\end{equation}
in which the $\pm$ signs correspond, respectively, to whether the photons approach or recede from the black hole. One can therefore recast the redshift factor as
\begin{eqnarray}
\mathscr{R}&=&\left(
u_\e^t+\frac{\Pi_r}{\Pi_t}u_\e^r
\right)^{-1}\nonumber\\
&=& \left[
\frac{1}{B(r)}\pm\sqrt{
\left(\frac{1}{B(r)}-1\right)\left(\frac{1}{B(r)}-\frac{b^2}{r^2}\right)}\,
\right]^{-1},
    \label{eq:redshift_factor_1}
\end{eqnarray}
and from here, the infinitesimal proper length is obtained as 
\begin{equation}
\ed I_\prop = \Pi_\mu u_\e^\mu\ed\tau = \frac{\Pi_t}{\mathscr{R}|\Pi_r|}\ed r.
    \label{eq:Iprop}
\end{equation}
Accordingly, Eq.~\eqref{eq:I_obs_infalling_0} takes the form
\begin{equation}
I_\obs \propto \int_{\bm{\gamma}} \frac{\mathscr{R}^3}{r^2} \frac{\Pi_t }{|\Pi_r|}\ed r,
    \label{eq:I_obs_infalling_1}
\end{equation}
and the observed intensity is, therefore, obtained by doing the above integration over all the frequencies. Using this method, we can study the brightness of infalling accretion and the shadow of the $f(R)$ black hole. As shown in Fig.~\ref{fig:I_obs_infall}, for all cases of the $\beta$-parameter, by the increase in the impact parameter and by moving away from the origin, the specific observed intensity increases until it reaches a peak around $b_p$, and falls of remarkably in the region $b> b_p$ and goes to zero. On the other hand, by altering the $\beta$-parameter from zero, the peak becomes slightly higher and the bottom line of the profile is lifted by the same value, whereas its width is decreased relevantly. This means that for larger $\beta$, the accretion disk appears brighter to the observer, and the silhouette becomes less dark but smaller in size. Hence, to a distant observer, the Schwarzschild-de Sitter black hole has the darkest and the largest silhouette, and the least bright accretion disk. In Fig.~\ref{fig:shadow_infall}, these effects have been visualized for the $f(R)$ black hole.

\subsection{Discussion}\label{subsec:Discussion}

As mentioned in Section \ref{sec:f(R)_theBH}, the first and second-order linear terms present in the lapse function \eqref{eq:lapse_0} can be interpreted as mimicking the flat galactic rotation curves and the accelerated expansion of the universe, and hence, they can be associated with the dark matter and dark energy components of the $f(R)$ black hole spacetime. In recent years, similar investigations to the one presented in this paper have been conducted on spherically symmetric black holes with linear or nonlinear extra terms, aiming to account for the properties of the dark side of the universe. Notably, the lapse function \eqref{eq:lapse_0} exhibits similarities with quintessential dark fields, which have also been explored in Refs. \cite{zeng_influence_2020,he_shadow_2022} concerning the shadow and observational signatures of infalling spherical accretions. When quintessential dark energy is employed in a black hole spacetime model, the lapse function takes on the form
\begin{equation}
\bar{\mathrm{B}}(r) = 1-\frac{2M}{r}-\frac{q}{r^{3w+1}},
    \label{eq:f(r)-quint}
\end{equation}
where the quintessential and state parameters, are denoted by $q$ and $w$, respectively. For $w=-1$, the resulting lapse function resembles the Schwarzschild-de Sitter black hole spacetime, which has not been addressed in the previously mentioned references. However, in this paper, we consider this case as a baseline to facilitate reliable comparisons between the impacts of different values for the $f(R)$ black hole parameters (see Figs. \ref{fig:ring-beta0} and \ref{fig:shadow_infall}). When $w=-1/3$, the lapse function becomes $\bar{\mathrm{B}}(r)=1-q-2M/r$, corresponding to a Schwazschild black hole associated with cloud of strings \cite{Letelier:1979}. Therefore, the quintessential dark energy field is obtained within the range $-1<w<-1/3$. Another important case is achieved by setting $w=-2/3$, resulting in $\bar{\mathrm{B}}(r)=1-qr-2M/r$, comparable to the lapse function \eqref{eq:lapse_0} for $\Lambda=0$. This specific case has been studied in Ref. \cite{zeng_influence_2020}, where only the quintessential dark energy affects the spacetime, and in Ref.~\cite{he_shadow_2022}, when the cloud of strings is also present. However, the resultant spacetimes for these cases cannot compensate for the flat rotation curves in galactic scales, as this requires a positive contribution of the linear term (i.e., for $q<0$ in $\bar{\mathrm{B}}(r)$), which has not been addressed in the aforementioned references. On the other hand, this aspect is fully considered in our paper, as all studied examples for the $f(R)$ black hole involve a positive $\beta$-parameter. Similar considerations have been explored in Ref. \cite{Li:2021ypw}, where a global monopole black hole in $f(R)$ gravity exhibits a lapse function similar to the Schwarzschild black hole associated with quintessence and cloud of strings, as discussed in Refs. \cite{he_shadow_2022,Fathi:2022a}. However, the aforementioned $f(R)$ black hole contains a negative linear term, and thus cannot compensate for dark matter effects in galactic scales. In contrast, our study benefits from the positive first-order term, allowing for compensation at both large and galactic scales. Furthermore, our investigation also accounts for gravitational effects at large distances, based on the existence of the square term $\Lambda r^2$ in the lapse function. {Therefore, this investigation offers some important new features that are potentially interesting regarding current astrophysical observations}. 

Having studied the analytical structure of light propagation around the $f(R)$ black hole and demonstrated its observational features across various criteria, we can now summarize our results in the next section.


\section{Summary and Conclusion}\label{sec:conclusion}

{The theoretical study of black holes with accretion disks provides a more realistic framework for making reliable comparisons with observational effects.} In this work, we focused on a static spherically symmetric black hole derived from a special $f(R)$ theory of gravity, which is compatible with both small and large-scale structure tests. This black hole features a linear first-order term with a coefficient $\beta$, along with a cosmological constant. While the cosmological constant is meant to compensate for vacuum energy and the accelerated expansion of the universe, the $\beta$-parameter is responsible for mimicking flat galactic rotation curves. {We initially derived the theoretical diameter of the black hole's photon ring and demonstrated its behavior concerning variations in the parameter $\beta$. However, we clarified that this parameter cannot be constrained by comparing the above theoretical diameter with the observed shadow diameters of M87* and Sgr A* since the photon ring is not observed in the images generated by the EHT. Consequently, for the numerical computations throughout the paper, we adopted a specific range for the parameter, setting it between $0$ and $0.05$. This range allows us to incorporate the Schwarzschild-de Sitter black hole scenario and funds a more realistic investigation.} Next, we delved into solving the angular equations of motion for deflecting and critical trajectories, providing exact analytical solutions for each case. These types of orbits play a crucial role in shaping the shadow of black holes when illuminated by an accretion disk. By employing the obtained analytical solutions for deflecting trajectories, we derived the lens equation and calculated the deflection angle, which turned out to be around 10 $\mathrm{\mu as}$ for the assumed values of the $\beta$-parameter. In the latter part of our investigation, we constructed a geometrically and optically thin accretion disk around the black hole using the Novikov-Thorne model and computed the disk's characteristics. We then applied the method introduced in Ref.~\cite{Gralla:2019} to classify the light rings and different types of accretion emission profiles. Based on the number of half orbits $n$ completed by light rays around the black hole, the rings were categorized into lensing rings and photon rings, both of which experienced demagnification due to extreme gravitational lensing. We also determined the impact parameter ranges for each ring, providing their respective thicknesses. By considering three types of disk emission profiles, we observed that an increase in the $\beta$-parameter leads to a decrease in the size of the shadow compared to that of the Schwarzschild-de Sitter black hole. Additionally, the brightness of the rings may vary depending on the value of $\beta$ and the radial position of the direct emission. Furthermore, the accretion disk's brightness is enhanced with an increase in the $\beta$-parameter. {Moreover, we applied a Gaussian filter to a particular case for simulating EHT images, further demonstrating that inferring the black hole size directly from observations remains inconclusive, as we have detailed earlier.} Lastly, we explored a spherically asymmetric infalling accretion scenario for the black hole and obtained the observed intensity profiles for different $\beta$ values. Our analysis demonstrated that as the $\beta$-parameter increases, the black hole becomes brighter in terms of the accretion disk, while the silhouette gradually shrinks. These findings were visually confirmed through appropriate ray-tracing methods. An interesting topic for future exploration could involve examining a plasmic medium characterized by a specific index profile encircling the black hole. This approach allows us to delve into how plasma components might impact the observable characteristics of black holes, serving as more reliable constituents within the accretion disks.

\section*{Acknowledgements} 
We thank the referee for valuable comments, which greatly helped us improve the paper's discussion. We would like to thank Mert Okyay and Ali \"{O}vg\"{u}n for providing some \textit{Mathematica} codes they have used in Ref.~\cite{okyay_nonlinear_2022}. M.F. acknowledges financial support from Vicerrector\'{i}a de Investigaci\'{o}n, Desarrollo e Innovaci\'{o}n - Universidad de Santiago de Chile (USACH), Proyecto DICYT, C\'{o}digo 042331CM$\_$Postdoc.

\appendix

\section{The full expression of $\mathscr{J}(r)$}\label{app:A}

Direct integration of the integral in Eq.~\eqref{eq:flux_0_int} results in
\begin{eqnarray*}
\int\mathcal{L}_c(r)\mathcal{E}'_c(r)\,\ed r \equiv\mathscr{J}(r)\nonumber
&=&\frac{1}{30 \sqrt{2} \beta ^3 \sqrt{\frac{r}{\beta  r^2+2}}}\left\{
\frac{2 \beta  r \left(-30 \Lambda -24 \beta ^2 \Lambda  r^3+2 \beta  r^2 \left(5 \beta ^2+\Lambda \right)+r \left(5 \beta ^2+9 \beta  \Lambda +10 \Lambda \right)\right)}{\beta  r^2+2 r-6}\nonumber\right.\\
&&+\left.\frac{{\beta^{\frac{3}{4}} }}{\sqrt{ (6 \beta +1)} \left(\beta  r^2+2\right)}
\left[
-60  \sqrt{6 \beta +1} \beta ^{\frac{7}{4}}-30  \sqrt{6 \beta +1} \beta ^{\frac{3}{4}} r^2-120 \sqrt{\frac{ (6 \beta +1)}{\sqrt{\beta }}} \Lambda
\right.
\right.\nonumber\\
&&\left.\left.
 -324 \sqrt{\frac{ (6 \beta +1)}{\sqrt{\beta }}} \beta\Lambda -162  \sqrt{6 \beta +1} \beta ^{\frac{7}{4}} \Lambda  r^2-60 \beta  \sqrt{\frac{ (6 \beta +1)}{\sqrt{\beta }}} \Lambda  r^2
\right.\right.\nonumber\\
&&\left.\left.
+60\ 2^{\frac{3}{4}}  \Lambda  \sqrt{(6 \beta +1) r \left(\beta  r^2+2\right)}\, \mathbf{E}\left(\left. \arcsinh\left(\frac{2^{\frac{3}{4}}}{\sqrt{r} {\beta^{\frac{3}{4}} }}\right)\right|-1\right)
\right.\right.\nonumber\\
&&\left.\left.
+ 6\ {2^{\frac{3}{4}}}\beta  (5 \beta +27 \Lambda ) \sqrt{(6 \beta +1) r \left(\beta  r^2+2\right)}\,\mathbf{E}\left(\left. \arcsinh\left(\frac{1}{{2^{\frac{3}{4}}} \sqrt{r} {\beta^{\frac{3}{4}} }}\right)\right|-1\right)
\right.\right.\nonumber\\
&&\left.\left.
-30\ 2^{\frac{3}{4}} \sqrt{r (6 \beta +1) \left(\beta  r^2+2\right)}\, \mathbf{F}\left(\left. \arcsinh\left(\frac{1}{\sqrt[4]{2\beta}\sqrt{r}}\right)\right|-1\right) \beta ^2\right.\right.\nonumber\\
&&\left.\left.-162\ 2^{\frac{3}{4}} \sqrt{r (6 \beta +1) \left(\beta  r^2+2\right)} \Lambda  \,\mathbf{F}\left(\left. \arcsinh\left(\frac{1}{\sqrt[4]{2\beta}\sqrt{r}}\right)\right|-1\right) \beta\right.\right.\nonumber\\
&&\left.\left.-10\ 2^{\frac{3}{4}} \sqrt{r \beta  (6 \beta +1) \left(\beta  r^2+2\right)} \Lambda  \,\mathbf{F}\left(\left. \arcsinh\left(\frac{1}{\sqrt[4]{2\beta}\sqrt{r}}\right)\right|-1\right)\right.\right.\nonumber\\
&&\left.\left.-20\ 2^{\frac{3}{4}} \sqrt{r \beta ^5 (6 \beta +1) \left(\beta  r^2+2\right)} \,\mathbf{F}\left(\left. \arcsinh\left(\frac{1}{\sqrt[4]{2\beta}\sqrt{r}}\right)\right|-1\right)\right.\right.\nonumber\\
&&\left.\left.-30\ 2^{\frac{3}{4}} \sqrt{r^3+\frac{2 r}{\beta }} \Lambda  \,\mathbf{\Pi} \left(\frac{3  \sqrt{2\beta}}{\sqrt{6 \beta +1}+1};\left. \arcsinh\left(\frac{1}{\sqrt[4]{2\beta}\sqrt{r}}\right)\right|-1\right)\right.\right.\nonumber\\
&&\left.\left.-225 \ 2^{\frac{3}{4}} \sqrt{r \beta  \left(\beta  r^2+2\right)} \Lambda  \,\mathbf{\Pi} \left(\frac{3  \sqrt{2\beta}}{\sqrt{6 \beta +1}+1};\left. \arcsinh\left(\frac{1}{\sqrt[4]{2\beta}\sqrt{r}}\right)\right|-1\right)\right.\right.\nonumber\\
&&\left.\left.-270 \ 2^{\frac{3}{4}} \sqrt{r \beta ^3 \left(\beta  r^2+2\right)} \Lambda  \,\mathbf{\Pi} \left(\frac{3  \sqrt{2\beta}}{\sqrt{6 \beta +1}+1};\left. \arcsinh\left(\frac{1}{\sqrt[4]{2\beta}\sqrt{r}}\right)\right|-1\right)\right.\right.\nonumber\\
&&\left.\left.-15 \ 2^{\frac{3}{4}} \sqrt{r \beta ^3 \left(\beta  r^2+2\right)} \,\mathbf{\Pi} \left(\frac{3  \sqrt{2\beta}}{\sqrt{6 \beta +1}+1};\left. \arcsinh\left(\frac{1}{\sqrt[4]{2\beta}\sqrt{r}}\right)\right|-1\right)\right.\right.\nonumber\\
&&\left.\left.-90 \ 2^{\frac{3}{4}} \sqrt{r \beta ^5 \left(\beta  r^2+2\right)} \,\mathbf{\Pi} \left(\frac{3  \sqrt{2\beta}}{\sqrt{6 \beta +1}+1};\left. \arcsinh\left(\frac{1}{\sqrt[4]{2\beta}\sqrt{r}}\right)\right|-1\right)\right.\right.\nonumber\\
&&\left.\left.+30\ 2^{\frac{3}{4}}  \Lambda  \,\mathbf{\Pi} \left(-\frac{3 \sqrt{2\beta}}{\sqrt{6 \beta +1}-1};\left. \arcsinh\left(\frac{1}{\sqrt[4]{2\beta}\sqrt{r}}\right)\right|-1\right) \sqrt{r^3+\frac{2 r}{\beta }}\right.\right.
\end{eqnarray*}
\begin{eqnarray}
&&\left.\left.+30\ 2^{\frac{3}{4}}  \Lambda  \,\mathbf{\Pi} \left(-\frac{3  \sqrt{2\beta}}{\sqrt{6 \beta +1}-1};\left. \arcsinh\left(\frac{1}{\sqrt[4]{2\beta}\sqrt{r}}\right)\right|-1\right) \sqrt{\frac{r (6 \beta +1) \left(\beta  r^2+2\right)}{\beta }}\right.\right.\nonumber\\
&&\left.\left.+30\ 2^{\frac{3}{4}}  \Lambda  \,\mathbf{\Pi} \left(\frac{3  \sqrt{2\beta}}{\sqrt{6 \beta +1}+1};\left. \arcsinh\left(\frac{1}{\sqrt[4]{2\beta}\sqrt{r}}\right)\right|-1\right) \sqrt{\frac{r (6 \beta +1) \left(\beta  r^2+2\right)}{\beta }}\right.\right.\nonumber\\
&&\left.\left.+225\ 2^{\frac{3}{4}}  \Lambda  \,\mathbf{\Pi} \left(-\frac{3  \sqrt{2\beta}}{\sqrt{6 \beta +1}-1};\left. \arcsinh\left(\frac{1}{\sqrt[4]{2\beta}\sqrt{r}}\right)\right|-1\right) \sqrt{r \beta  \left(\beta  r^2+2\right)}\right.\right.\nonumber\\
&&\left.\left.+15\ 2^{\frac{3}{4}}  \,\mathbf{\Pi} \left(-\frac{3  \sqrt{2\beta}}{\sqrt{6 \beta +1}-1};\left. \arcsinh\left(\frac{1}{\sqrt[4]{2\beta}\sqrt{r}}\right)\right|-1\right) \sqrt{r \beta ^3 \left(\beta  r^2+2\right)}\right.\right.\nonumber\\
&&\left.\left.+270\ 2^{\frac{3}{4}}  \Lambda  \,\mathbf{\Pi} \left(-\frac{3  \sqrt{2\beta}}{\sqrt{6 \beta +1}-1};\left. \arcsinh\left(\frac{1}{\sqrt[4]{2\beta}\sqrt{r}}\right)\right|-1\right) \sqrt{r \beta ^3 \left(\beta  r^2+2\right)}\right.\right.\nonumber\\
&&\left.\left.+90\ 2^{\frac{3}{4}}  \,\mathbf{\Pi} \left(-\frac{3  \sqrt{2\beta}}{\sqrt{6 \beta +1}-1};\left. \arcsinh\left(\frac{1}{\sqrt[4]{2\beta}\sqrt{r}}\right)\right|-1\right) \sqrt{r \beta ^5 \left(\beta  r^2+2\right)}\right.\right.\nonumber\\
&&\left.\left.+60\ 2^{\frac{1}{4}} \Lambda  \,\mathbf{F}\left(\left. \arcsinh\left(\frac{{2^{\frac{1}{4}}}}{\sqrt{r} \beta^{\frac{1}{4}}}\right)\right|-1\right) \sqrt{r (6 \beta +1) \left(\beta  r^2+2\right)}\right.\right.\nonumber\\
&&\left.\left.+135\ 2^{\frac{3}{4}}  \Lambda  \,\mathbf{\Pi} \left(-\frac{3  \sqrt{2\beta}}{\sqrt{6 \beta +1}-1};\left. \arcsinh\left(\frac{1}{\sqrt[4]{2\beta}\sqrt{r}}\right)\right|-1\right) \sqrt{r \beta  (6 \beta +1) \left(\beta  r^2+2\right)}\right.\right.\nonumber\\
&&\left.\left.+135\ 2^{\frac{3}{4}}  \Lambda  \,\mathbf{\Pi} \left(\frac{3  \sqrt{2\beta}}{\sqrt{6 \beta +1}+1};\left. \arcsinh\left(\frac{1}{\sqrt[4]{2\beta}\sqrt{r}}\right)\right|-1\right) \sqrt{r \beta  (6 \beta +1) \left(\beta  r^2+2\right)}\right.\right.\nonumber\\
&&\left.\left.+15\ 2^{\frac{3}{4}}  \,\mathbf{\Pi} \left(-\frac{3  \sqrt{2\beta}}{\sqrt{6 \beta +1}-1};\left. \arcsinh\left(\frac{1}{\sqrt[4]{2\beta}\sqrt{r}}\right)\right|-1\right) \sqrt{r \beta ^3 (6 \beta +1) \left(\beta  r^2+2\right)}\right.\right.\nonumber\\
&&\left.\left.+15\ 2^{\frac{3}{4}}  \,\mathbf{\Pi} \left(\frac{3  \sqrt{2\beta}}{\sqrt{6 \beta +1}+1};\left. \arcsinh\left(\frac{1}{\sqrt[4]{2\beta}\sqrt{r}}\right)\right|-1\right) \sqrt{r \beta ^3 (6 \beta +1) \left(\beta  r^2+2\right)}\right.\right.\nonumber\\
&&\left.\left.+60\ 2^{\frac{3}{4}}  \,\mathbf{\Pi} \left(-\frac{3  \sqrt{2\beta}}{\sqrt{6 \beta +1}-1};\left. \arcsinh\left(\frac{1}{\sqrt[4]{2\beta}\sqrt{r}}\right)\right|-1\right) \sqrt{r \beta ^5 (6 \beta +1) \left(\beta  r^2+2\right)}\right.\right.\nonumber\\
&&\left.\left.+60\ 2^{\frac{3}{4}}  \,\mathbf{\Pi} \left(\frac{3  \sqrt{2\beta}}{\sqrt{6 \beta +1}+1};\left. \arcsinh\left(\frac{1}{\sqrt[4]{2\beta}\sqrt{r}}\right)\right|-1\right) \sqrt{r \beta ^5 (6 \beta +1) \left(\beta  r^2+2\right)}
\right]\right\},
\end{eqnarray}
where $\mathbf{F}(\varphi|\mathfrak{m})$, $\mathbf{E}(\varphi|\mathfrak{m})$ and $\mathbf{\Pi}(\mathfrak{n};\varphi|\mathfrak{m})$ are, respectively, the incomplete elliptic integrals of the first, second, and third kind of argument $\varphi$, modulus $\mathfrak{m}$ and characteristic $\mathfrak{n}$ \cite{byrd_handbook_1971}. Note that the above expression does not cover the case of $\beta=0$, so the corresponding profile has to be obtained by doing numerical integration of Eq.~\eqref{eq:flux_0_int}.

\bibliographystyle{ieeetr}
\bibliography{biblio_v1.bib}

\end{document}